\documentclass[preprint,a4paper,authoryear,11pt]{elsarticle}
\usepackage[left=1.5cm,right=1.5cm,top=2cm, bottom=3cm]{geometry}
\usepackage[onehalfspacing]{setspace}
\usepackage{credits}
\usepackage{pgf,tikz}
\usepackage{pgfplots}       
\usepackage{pgfplotstable}
\usetikzlibrary {arrows.meta}
\usetikzlibrary{patterns}
\usetikzlibrary{intersections}
\usepgfplotslibrary{colormaps,external, fillbetween}
\usepgfplotslibrary{groupplots}
\usepgfplotslibrary{statistics}
\usetikzlibrary{calc}
\usepgfplotslibrary{colorbrewer}
\usetikzlibrary{fit,calc}
\usepackage{pgf-pie}
\usetikzlibrary{spy}
\usetikzlibrary{backgrounds}
\usetikzlibrary{decorations}
\usepackage{tikzviolinplots}
\usepackage{scontents}
\usepackage{xcolor}
\usepackage{amsfonts}
\usepackage{xspace}
\usepackage{amssymb}

\usepackage[table]{xcolor}

\usepackage{enumitem}
\setlist{nolistsep,leftmargin=*}

\usepackage{bm}

\usepackage{hyperref}       
\usepackage{booktabs}

\usepackage{array,multirow,graphicx}
\graphicspath{{imgs/}}

\usepackage{amsmath}
\makeatletter
\newcommand\niton{\mathrel{\m@th\mathpalette\canc@l\owns}}
\newcommand\canc@l[2]{{\ooalign{$\hfil#1/\mkern1mu\hfil$\crcr$#1#2$}}}
\makeatother

\usepackage{enumitem}
\usepackage{float}
\usepackage[T1]{fontenc}

\usepackage[parfill]{parskip} 

\newcommand{\NP}{NP-}
\newcommand{\TSP}{TSP}
\newcommand{\VRP}{VRP}

\usepackage{etoolbox}

\usepackage{algorithm}
\usepackage{algpseudocode}
\usepackage{rotating}
\usepackage{tabularx}

\newcommand{\FILOX}{FILO2$^x$\xspace}

\makeatletter
\makeatletter
\g@addto@macro\bfseries{\boldmath}
\makeatother

\makeatletter
\AfterEndEnvironment{algorithm}{\let\@algcomment\relax}
\AtEndEnvironment{algorithm}{\kern2pt\hrule\relax\vskip3pt\@algcomment}
\let\@algcomment\relax
\newcommand\algcomment[1]{\def\@algcomment{\footnotesize#1}}

\renewcommand\fs@ruled{\def\@fs@cfont{\bfseries}\let\@fs@capt\floatc@ruled
  \def\@fs@pre{\hrule height.8pt depth0pt \kern2pt}%
  \def\@fs@post{}%
  \def\@fs@mid{\kern2pt\hrule\kern2pt}%
  \let\@fs@iftopcapt\iftrue}
\makeatother

\algnewcommand{\IfThen}[2]{
	\State \algorithmicif\ #1\ \algorithmicthen\ #2}
\algnewcommand{\LineComment}[1]{\State \(\triangleright\) #1}

\algnewcommand{\InlineFor}[2]{
	\State \algorithmicfor\ #1\ \algorithmicendfor\ #2}

\usepackage{siunitx}
\usepackage[framemethod=tikz]{mdframed}
\usetikzlibrary{calc}
\usepackage{fourier-orns}
\tikzset{ warningsymbol/.style={draw=red, fill=white, scale=1, overlay} }
\mdfdefinestyle{warning}{%
 hidealllines=true,leftline=true,
 skipabove=1pt,skipbelow=1pt,
 innertopmargin=0.4em,%
 innerbottommargin=0.4em,%
 innerrightmargin=0.7em,%
 rightmargin=0.7em,%
 innerleftmargin=1.7em,%
 leftmargin=0.7em,%
 middlelinewidth=.2em,%
 linecolor=red,%
 fontcolor=red,%
 firstextra={\path let \p1=(P), \p2=(O) in ($(\x2,0)+0.5*(0,\y1)$) 
                           node[warningsymbol] {\danger};},%
 secondextra={\path let \p1=(P), \p2=(O) in ($(\x2,0)+0.5*(0,\y1)$) 
                           node[warningsymbol] {\danger};},%
 middleextra={\path let \p1=(P), \p2=(O) in ($(\x2,0)+0.5*(0,\y1)$) 
                           node[warningsymbol] {\danger};},%
 singleextra={\path let \p1=(P), \p2=(O) in ($(\x2,0)+0.5*(0,\y1)$) 
                           node[warningsymbol] {\danger};},%
}
\newmdenv[style=warning]{Warning}

\usepackage{mathtools, nccmath}

\usepackage{subfig}
\pgfplotsset{compat=newest}

\usepackage{natbib}
 \bibpunct[, ]{(}{)}{,}{a}{}{,}%
 \def\bibsep{\smallskipamount}%
\journal{XXX}

\usepackage{multirow}
\usepackage{lscape}
\usepackage{adjustbox}
\usepackage{graphicx}
\usepackage{longtable}

\usepackage{import}
\usepackage{xifthen}
\usepackage{pdfpages}
\usepackage{transparent}

\begin{document}

\begin{frontmatter}

\title{Asynchronous Cooperative Optimization of a Capacitated Vehicle Routing Problem Solution}


\author[1]{Luca Accorsi}
\ead{accorsi@google.com}

\author[3]{Demetrio Lagan\`{a}}
\ead{demetrio.lagana@unical.it}
%
\author[2]{Federico Michelotto}
\ead{federico.michelotto2@unibo.it}
\author[3]{Roberto Musmanno}
\ead{roberto.musmanno@unical.it}
\author[2,4]{Daniele Vigo}
\ead{daniele.vigo@unibo.it}
\address[1]{Google, Erika-Mann-Stra{\ss}e 33, Munich, 80636, Bavaria, Germany}
\address[3]{DIMEG, University of Calabria, Via Pietro Bucci,  87036 - Rende (CS), Italy}
\address[2]{DEI ``G. Marconi'', University of Bologna, Viale del Risorgimento 2, 40136 - Bologna, Italy}
\address[4]{CIRI-ICT, University of Bologna, via Quinto Bucci, 336, 47521 - Cesena, Italy}

\date{June 2024}

\begin{abstract}
We propose a parallel shared-memory schema to cooperatively optimize the solution of a Capacitated Vehicle Routing Problem instance with minimal synchronization effort and without the need for an explicit decomposition.
To this end, we design \FILOX as a single-trajectory parallel adaptation of the FILO2 algorithm originally proposed for extremely large-scale instances and described in \citet{accorsi2024}.
Using the locality of the FILO2 optimization applications, in \FILOX several possibly unrelated solution areas are concurrently asynchronously optimized.
The overall search trajectory emerges as an iteration-based parallelism obtained by the simultaneous optimization of the same underlying solution performed by several solvers.
Despite the high efficiency exhibited by the single-threaded FILO2 algorithm, the computational results show that, by better exploiting the available computing resources, \FILOX can greatly enhance the resolution time compared to
the original approach, still maintaining a similar final solution quality for instances ranging from hundreds to hundreds of thousands customers.
\end{abstract}


\end{frontmatter}

\section{Introduction}
\label{sec:introduction}
The Vehicle Routing Problem (\VRP) \citep{dantzig1959, Toth2002, Toth2014} is an \NP hard
combinatorial optimization problem that is widely studied in the field of operations research, both for its notorious difficulty and its wide range of applications in freight and people transportation. The \VRP \ aims to find the least-cost routes that visit exactly once each customer in a given set. The \VRP \ has different variants depending on the objectives and constraints. The most studied is the Capacitated Vehicle Routing Problem (CVRP) \citep{Toth2002, Toth2014}, in which routes are performed by a fleet of homogeneous vehicles with limited capacity.
The CVRP is formally defined on an undirected graph $G = \left(V,E\right)$, where $V$ is the set of vertices and $E = \left\{\left(i,j\right) \in V \times V: i < j\right\}$ is the set of edges. The vertex set $V$ is partitioned into $V = \left\{0\right\} \cup V'$, where vertex 0 represents the depot and $V' = \left\{1, \ldots, N\right\}$ is the vertex set associated with $N$ customers. A nonnegative cost $c_{ij}$ is associated with each edge $\left(i,j\right) \in E$. Moreover, we assume that the cost matrix $\textbf{C} = \left\{c_{ij}, (i,j) \in E\right\}$ satisfies the triangle inequality. 
For a subset $V' \subseteq V$, we identify with $\mathcal{N}_i^k(V')$ the ordered list of the $k$ nearest neighbor vertices $j \in V'$ of vertex $i$ with respect to the cost matrix $\textbf{C}$. If $k = |V'|$, we omit the apex $k$, that is, $\mathcal{N}_i(V')$.
Each customer $i \in V'$ requires an integer quantity $q_i > 0$ of goods from the depot, and $q_0 = 0$. An unlimited fleet of homogeneous vehicles, each with capacity $Q$, is located at the depot to serve the customers. A CVRP solution $S = \left\{r_1, \ldots, r_h, \ldots, r_{|S|}\right\}$ is composed of  $|S|$  Hamiltonian circuits, called \textit{routes} $r_h = \left\{i_0, i_1, \ldots, i_p, i_{p + 1}\right\}$, starting from the depot (i.e., $i_0 = 0$), visiting a subset of customers $i_1 , \ldots, i_p \in V'$, and returning to the depot (i.e., $i_{p + 1} = 0$).  A solution $S$ is feasible for the CVRP   if all customers are visited exactly once and none of the vehicles exceed their capacity. The cost of a route is defined as the total cost of the edges connecting the visited customers: $c_{r^h} = \sum^p_{j = 0}c_{i_ji_{j + 1}}$, while the cost $c_S$ of a solution $S$ is given by the sum of the costs of the routes of $S$: $c_S = \sum^{|S|}_{u = 1}c_{r^u}$. The objective of the CVRP is to find a feasible solution with  a minimum total cost. 

The exact methods for the CVRP are based on branch-and-bound, branch-and-cut, branch-and-price, and dynamic programming \citep{Toth2014} and, in the worst-case, they all require  exponential time with respect to the input size (number of customers) to obtain an optimal solution. The high time complexity of exact methods makes them unsuitable for solving large-scale CVRP instances. To address this challenge, heuristic algorithms are employed, offering satisfactory solutions in a reasonable amount of time. Many heuristic algorithms consume a considerable amount of CPU time in the improvement phase. This results in a considerable degradation in performance when solving very large-scale instances. 

In this work, we focus on reducing the computing time required to solve very small- and large-scale CVRP instances by implementing a scalable parallel computing approach. In general, parallel computing
is a methodology for solving a specific problem by distributing the computational workload among multiple concurrent processes (solvers) on a set of processors \citep{crainic2010}. The objective is to distribute the computational load among different processors to enhance performance
when the size of the problem increases. The workload to be distributed may concern the algorithm, the search space, or the problem structure. In the context of algorithmic workload distribution, the implementation of \textit{functional parallelism} is a key objective. This approach involves the distribution of computations that require the most CPU time across available processors that execute parallel tasks utilizing the same data, with each task being allocated to a specific processor. These processors collaborate with one another, if required, to ensure the efficient utilization of the resources. The distribution of workload in terms of search space is achieved through the utilization of \textit{data parallelism}, which involves the decomposition of the problem domain into smaller portions, with each parallel processor undertaking the processing of a specific portion. Finally, the distribution of the workload determined by the problem structure entails the decomposition of the problem into a series of subproblems. The subsequent tasks are either concerned with addressing these subproblems or with integrating their solutions into the original problem. In general, it is possible to define \textit{fine-} and \textit{coarse-grained} parallelisms in accordance with the small and large size of the tasks resulting from the decomposition of the problem. In the existing literature, two main strategies for the partitioning of search space are identified: \textit{domain decomposition} and \textit{multisearch}. Domain decomposition is explicit, while multisearch is implicit. 
More precisely, the domain decomposition strategy is characterized by explicit partitioning of the search space, whereas the multisearch strategy involves implicit segmentation through independent thread explorations of the search space. In the context of the multisearch method, a prevalent issue pertains to the non-overlapping exploration of the search space, a challenge that can be mitigated by employing different search strategies. From an algorithmic perspective, the most elementary form of parallelism arises from the concurrent execution of inner loop iterations with synchronization components. However, it should be noted that this approach can result in substantial delays associated with the communication overhead required by the synchronization.
The success of parallel/distributed metaheuristic for the \VRP \ is based on the Cooperative Search (CS) paradigm \citep{alba2005, crainic2005b, crainic2005c, talbi2006, crainic2008, crainic2008b, crainic2010}.
Cooperative search uses multiple solution methods and cooperation mechanisms to share information asynchronously and generate new data. Cooperative-search strategies have several key aspects: $i)$ the solution methods engaged in cooperation (including the same method with different parameter settings or populations); $ii)$ the information shared (e.g., the best solution and status updates); $iii)$ the way of sharing information and global search control; and $iv)$ how the exchanged information is used globally and how each process uses the received information.
In the context of parallel algorithms based on CS strategies for the CVRP, meta-heuristics such as Tabu Search (TS), Genetic and Evolutionary Algorithms (GEAs), Memetic Algorithms (MAs), Ant Colony Algorithms (ACOs) and Simulated Annealing (SA) methods have been shown to be particularly effective.  The rationale behind the employment of these metaheuristics as a single-processor local search algorithm is primarily due to their ability to efficiently explore the search space while escaping from local minima, and to represent the CVRP solution as a set of paths or routes, which allows for easy updating through information exchange between processors that can optimize subsets of routes. Nevertheless, these metaheuristics are inherently limited in their ability to scale as the size of the instances increases, especially when a CVRP solution comprises a very large number of routes and the number of processors is either quite limited or quite large in relation to the number of routes. In the latter cases, particular attention is required for the mechanism by which information is shared and the subset of routes is updated. This inherent limitation shows up in the parallel algorithms, resulting in suboptimal efficiency and consequent degradation in the quality of the solutions obtained when these algorithms are used to solve very large CVRP instances. The present work aims to partially address this gap.
Our work contributes to the literature on parallel algorithms for the CVRP on the following grounds:
\begin{itemize}[noitemsep]
    \item we propose a parallel algorithm based on a shared-memory CS strategy whose execution results in a single-trajectory parallel adaptation of the FILO2 algorithm \citep{accorsi2024} that is executed by several solvers to simultaneously optimize localized solution areas;
    \item the proposed approach results in the coordinated optimization of a shared solution, without the need of decomposition nor the imposition of rigid boundaries on the solvers. Synchronization among solvers is minimized by using data redundancy and a message-passing communication;
    \item the computational results demonstrate the effectiveness of the parallel algorithm, which achieves results comparable with those obtained by the sequential FILO2 algorithm, but in a fraction of time, on a wide range of instance sizes.
\end{itemize}
The paper is organized as follows.
In Section~\ref{sec:literature_review}, we position our work with respect to the literature, while Section \ref{sec:filo2-and-filo} provides the background of the FILO2 and FILO algorithms. We describe our solution approach in Section~\ref{sec:solution_approach}. In Section~\ref{sec:computational-results}, we present the computational performance of our solution approach. Section~\ref{sec:analysis} presents a computational analysis of the algorithm components. Finally, Section~\ref{sec:conclusions} concludes our work.


\section{Literature Review}
\label{sec:literature_review}


Parallel algorithms for the Traveling Salesman Problem (\TSP) and \VRP have been researched for almost three decades. As this work focuses primarily on the CVRP, this section provides an overview of the main contributions to parallel and distributed algorithms for \VRP. 

The algorithms that achieved the best performance in parallel/distributed computing for solving very large-scale instances of the \VRP \ are based on decomposition techniques, as outlined in \cite{kindervater1986, gendron1994, alba2005, talbi2006, crainic2006, crainic2010}. Decomposition-based methods are based on three subsequent stages. The first one is the {\em Decomposition} stage, which splits the original problem into sub-problems. In the  {\em Integration} stage,  the results of these sub-problems are combined to create solutions to the original problem. Finally, in the last  stage the evolution of the complete method is defined. It is clearly essential that these mechanisms are explicitly adapted to align with the specific heuristic strategies adopted in the solution of sub-problems.
In 2005, \cite{crainic2005} presented a classification system for parallel metaheuristic strategies. The authors introduced three dimensions for controlling the global process and managing data. The first dimension is \textit{Search Control Cardinality}, which delineates whether a singular (1C; 1-control) or multiple (pC; p-control) processes oversee the global search. The second dimension is \textit{Search Control and Communication}, which describes the data exchange between processes: namely, \textit{Rigid Synchronization} (\textit{RS}), \textit{Knowledge Synchronization} (\textit{KS}), \textit{Collegial} (\textit{C}), and \textit{Knowledge Collegial} (\textit{KC}). The third dimension is the \textit{Search Differentiation}, which concerns how searches are initialized  and which search method is used. Four cases have been identified in this respect: \textit{SPSS} (\textit{Same Initial Point/Population, Same Search Strategy}), \textit{SPDS} (\textit{Same Initial Point/Population, Different Search Strategy}), \textit{MPSS} (\textit{Multiple Initial Point/Population, Same Search Strategy}), and \textit{MPDS} (\textit{Multiple Initial Point/Population, Different Search Strategy}). Within this nomenclature, the following parallel search strategies can be defined. \textit{Low-Level 1-Control Parallelization Strategies} are used by exact or heuristic methods. These strategies are based on a rigid synchronization between the master process and other processes, and are implemented in the traditional master-slave parallel programming model, where communication between slave processes is prohibited with the objective of accelerating the search process. The \textit{Neighborhood-based, 1-Control/Rigid Synchronization/Same Initial Point, Same Search Strategy} approach enables the parallelization of the neighborhood search between processes. A neighborhood solution can be generated and evaluated independently, facilitating parallelization. This can be achieved through the classical master-slave mechanism. The Independent Multi-Search (IMS) is a p-Control parallelization strategy that involves multiple searches on the entire search space, starting from the same or different initial solutions, with the objective of selecting the best solution at the end. The IMS is a pC/RS class with multiple starting points/populations that implements a MPSS search strategy. The Cooperative Search strategies \citep{crainic2010} advance IMS methods by allowing data sharing throughout the search. The data exchange mechanism helps achieve better results than standalone multisearch techniques. The data-sharing cooperation mechanism specifies how different search threads interact.

Search processes may exchange data directly or indirectly, typically through a \textit{blackboard repository}. Blackboard repositories store information from all processes and can be used as a source of information accessible to all processes. Cooperation can be classified as synchronous or asynchronous. In synchronous cooperation, referred to as \textit{p-Control Knowledge Synchronization} (\textit{pC/KS}) strategy, processes share knowledge about solutions at a pre-determined time. The \textit{p-Control Collegial} (\textit{pC/C}) strategy is a fully distributed one, with cooperation actions initiated independently by search processes. When the \textit{pC/C} communication strategy is used to modify the knowledge of specific processes, it results in a \textit{p-Control Knowledge Collegial} (\textit{pC/KC}). In accordance with the classification system outlined above, the following analysis examines the most recent contributions to the scientific literature concerning CVRP. 

\cite{Doerner2004} explores the application of Ant Colony Optimization (ACO) for solving the CVRP and evaluates three ACO paradigms: Rank-Based Ant System (ASrank), Max-Min Ant System (MMAS), and Ant Colony System (ACS).
The approach can be classified as a \textit{pC/RS} strategy due to the structured communication between processors and synchronized pheromone updates. The computational analysis is based on 14 standard \VRP \ benchmark instances taken from \cite{christofides1979}. These instances comprise between 50 and 199 customers, with the configurations including both random customer locations on a plane and locations included in clusters.
A comparison is conducted between the three ACO paradigms (ASRank, MMAS and ACS) by measuring the relative percentage deviation from the best known solutions. An analysis of several algorithm settings is conducted in terms of the average and worst results over five runs from the best known solutions. The average deviation of ASRank, MMAS, and ACS is 0.97, 0.80, and 1.22, respectively, when using the best setting. In order to analyze the effectiveness of the parallel algorithm, the speedup and efficiency are measured for varying numbers of processors. The speedup is defined as $S_p = \frac{T_{seq}}{T_p}$, where $S_p$ is the speedup obtained on $p$ processors, $T_{seq}$ is the measured sequential execution time and $T_p$ is the parallel execution time on $p$ processors. The implementation of this speedup measure, using eight processors, has been demonstrated to result in a sixfold enhancement in efficiency for problem instances involving 150 customers.\\

\cite{Kalatzantonakis2005} investigates the impact of cooperation strategies in parallel General Variable Neighborhood Search (GVNS) algorithms for solving the CVRP. Three parallelization models are proposed:
\begin{itemize}
\item non-cooperative model (NCM): \textit{pC/RS/SPSS}, featuring rigid synchronization with no solution exchange among parallel processes;
\item managed cooperative model (MCM): \textit{pC/C/SPSS}, enabling asynchronous solution sharing through a centralized manager, promoting intensification;
\item parameterized cooperative model (PCM): \textit{pC/KC/SPSS}, with asynchronous and selective solution sharing that adapts dynamically based on search progression, optimizing both intensification and diversification.
\end{itemize}
All computational tests are carried out on 76 instances of the CVRPLIB library available at \url{http://vrp.atd-lab.inf.puc-rio.br}, with the number of customers varying from 32 to 360. 
In order to ensure a fair comparison and evaluation of the performance and quality of the solutions, all experiments are repeated ten times, and the average gap values of each parallel model are reported compared to the best-known solution values. Specifically, NCM presents an average gap of 0.837\%, MCM of 1.260\%, and PCM of 0.729\%, thus demonstrating that PCM is the best approach, producing high-quality solutions.
All parallel algorithms successfully achieve the optimum value for 14 out of 65 instances. No analysis in terms of speedup is presented, and none of the existing datasets, including  larger instances, is examined in the computational testing.

\cite{ Barbucha2011} investigates the solution of CVRP using a genetic algorithm based on a team of heterogeneous cooperating agents within a multi-agent system framework. The proposed approach employs an asynchronous team paradigm, where agents represent various local search heuristics and collaborate to iteratively refine a shared population of solutions stored in global memory.
The methodology corresponds to a \textit{pC/KC} strategy because of its asynchronous cooperation and the reliance on shared memory for solution refinement. Computational experiments are carried out on 5 test instances of the \cite{christofides1979} containing 50$-$199 customers and only the capacity constraint. The quality of the solutions is measured in terms of relative deviation from the best-known solution. The team of heterogeneous cooperative agents achieved an average deviation of 0.87\%, compared to the best-known solution on all instances. No information on speedups is provided in the paper.

\cite{Barbucha2014} presents a hybrid approach to solve the CVRP by integrating cooperative multiple neighborhood search with a multi-agent system (MAS) paradigm.
This approach can be classified as a \textit{pC/KC} strategy due to its distributed processes, asynchronous communication, and reliance on shared memory for solution exchange.
The experiments are carried out on CVRP instances derived from \cite{christofides1979}, with a number of customers ranging from 50 to 199. The population size is set at 30 individuals and the algorithm terminates after 180 seconds of execution  elapsed without improvement. The quality of the solutions is measured as Mean Relative Error (MRE) when compared to the optimal or best known solution, as reported by \cite{laporte2000}.
The MRE (in \%) from the best known solution is reported, calculated separately for cases where a single agent or a team of optimizing agents is involved in the search process and ranges between 2.27\% and 3.57\%. The work does not present any analysis in terms of speedup performance.

A parallel micro genetic algorithm for solving the CVRP is proposed in \cite{Borcinova2018}. The approach employs a coarse-grained parallelization strategy where each node executes a micro genetic algorithm on a small population, and synchronization occurs through a master node that collects the best solutions from all nodes and broadcasts the best global solution.
The methodology is a \textit{pC/RS} strategy due to its structured master-slave communication model, where nodes synchronize at defined intervals to exchange and update the global best solution.
The experiments make use of 16 problems drawn from the sets (A and P) of benchmark problems  by \cite{augerat1995}. The instances vary in size, with number of customers (including the depot) ranging from 16 to 101.
The quality of the solutions is assessed by computing the relative deviation of the solution value from the known optimal costs.
The average relative deviation from the optimal solution values, when employing the random generation of the initial solution, is 0.14, and is reduced to 0.07 when utilizing the parameterized Clarke-Wright algorithm. However, no information is provided regarding the potential speedups achieved with the parallel implementation.

In \cite{ Abdelatti2020} an innovative GPU-accelerated genetic algorithm for solving the CVRP is illustrated. By leveraging NVIDIA's CUDA architecture, the algorithm achieves high-speed parallel computations. It incorporates 2-opt local search and advanced population management techniques to enhance solution quality and prevent premature convergence.
The methodology aligns with a \textit{pC/KC} classification.
The performance of the algorithm is evaluated using 14 common CVRP benchmarks, selected from various sources \citep{augerat1995,christofides1969,christofides1981,fisher1994,gillet1976,feiyue2005}. These benchmarks range in size from 16 to 76 vertices. To assess the performance of the GPU-based algorithm, the authors used a CPU-only based implementation of their algorithm.  Each algorithm is executed ten times on each benchmark problem, and averaged results are reported. The quality of the solutions obtained is evaluated by computing the percentage of the gap, defined as $\left(\frac{s_0 - s^*}{s^*}\right) \times 100$, where $s_0$ is the best solution obtained by the algorithm in 10 runs and $s^*$ is the best known solution from the literature. 
The speedups of the GPU version over the CPU range from 52.7 times for the 21-node problem to 454.4 times for the 76-node problem.
The gap obtained on instances with a number of customers between 60 and 76 ranges from 7.86\% to 10.23\%. The evaluation of the proposed algorithm for larger problems is lacking.

\cite{Yelmewad2021} propose a GPU-based parallelization of Local Search Heuristic (LHS) algorithms to efficiently solve the CVRP for large-scale instances. The parallelization is implemented at two levels: route-level and customer-level designs. The route-level approach assigns each route to a GPU thread, while the customer-level approach assigns a thread to each customer, significantly increasing computational parallelism.
The approach can be classified as a \textit{pC/KC} strategy. This classification reflects the asynchronous nature of the parallelization, where GPU threads independently explore solution neighborhoods and periodically update global memory to share improved solutions. The experimental analysis is carried out on the following CVRP data sets: the M set of \cite{christofides1979} instances, \cite{golden1998} instances, \cite{feiyue2005} instances, \cite{kytojoki2007} instances, \cite{uchoa2017} instances, and the \cite{arnold2019xxl} instance set. The instance sizes thus range from 101 to 30,000 customers, exhibiting a wide spectrum of complexity and magnitude. 
On the \cite{uchoa2017} data set, the route-level parallel strategy achieves a speedup that is 1.11 times faster in the worst-case, 4.40 times faster in the best-case and 2.17x on average compared to the sequential version for the same data set. No information on gap values compared with the best known solutions is provided. When compared with the sequential version on the \cite{uchoa2017} data set, the customer-level parallel strategy is 1.99 times faster in the worst-case, 9.83 times on average, and 22.97 times in the best-case. Considering the sequential version on the \cite{arnold2019xxl} instances, the customer-level parallel strategy obtains speedups of 29.9, 83.52, and 147.20 times in the worst, average, and best-case, respectively. Finally, the speedup achieved in the worst, average, and best-case is 1.99, 4.69, and 9.03 times, respectively, compared with the route level parallel design. No gap value with respect to the best known solutions is reported since the LHS used in parallel strategies is not able to produce competitive results compared with the state of the art method.

\cite{Muniasamy2023} introduces ParMDS, an efficient shared-memory parallel method for solving the CVRP. ParMDS employs a novel combination of Minimum Spanning Tree and Depth-First Search heuristics with randomization to generate candidate solutions quickly.
The method is classified as a \textit{pC/KC} strategy because of its asynchronous parallelization and solution sharing through memory. The ParMDS is tested on 130 instances of the CVRP from the CVRPLIB library, with sizes ranging from 76 to 30,000 customers and vehicle capacities between 3 and 2210. These instances include different sets, including the $\mathbb{X}$ and $\mathbb{B}$ datasets proposed by \cite{uchoa2017} and \citet{arnold2019xxl}.
The quality of the solution is measured by the \% gap. The ParMDS achieves a geometric mean deviation of 11.85\% from the best known solutions related to two largest input instances per type, thereby significantly improving on the two baselines that implement a metaheuristic algorithm for VRP on GPU in CUDA, and a genetic algorithm for VRP on GPU, respectively.
ParMDS, which is based on shared parallelization via OpenMP, demonstrates a significant speedup compared to baseline GPU implementations, varying between 36× and 1189×. However, the comparison with the BKSs is limited to very few large instances, and the quality of solutions compared to the BKSs turns out to be rather poor.

\section{Background: The FILO2 and FILO Algorithms} \label{sec:filo2-and-filo}

The parallel solution approach we propose, named \FILOX, builds upon the FILO2 algorithm introduced by \citet{accorsi2024}, an efficient randomized metaheuristic designed to solve extremely large-scale CVRP instances in relatively short computing times.
To facilitate the description of \FILOX, this section first provides an overview of the main components of the FILO2 algorithm. Interested readers are referred to the original paper for further details.

FILO2 is an evolution of the FILO algorithm originally proposed in \citet{accorsi2021} for the efficient resolution of large-scale CVRP instances with several thousand customers.
FILO2 shares the same structure as FILO but replaces some computationally expensive procedures and memory-demanding data structures to make the resolution of extremely large-scale instances with up to one million customers approachable with ordinary computer systems.

The main strength of the FILO and FILO2 approaches consists in the iterative optimization of localized solution portions through the application of a sophisticated local search engine making use of heuristic acceleration and pruning techniques.
Several local search operators such as the CROSS exchange (see \citet{taillard1997tabu}), the 2-opt (see \citet{Reinelt1994TheTS}), and the ejection chain (see \citet{GLOVER1996223}) are applied in a variable neighborhood descent fashion (VND, see \citet{MLADENOVIC19971097}). 
Their neighborhoods are efficiently explored using the static move descriptors (SMD) framework, originally proposed by \citet{zachariadis2010} and \citet{BEEK20181}, which cleverly exploits the locality of local search changes to avoid the re-evaluation of unrelated local search moves.
In addition, neighborhoods are granular (see \citet{toth2003}) and their size is dynamically adjusted with a vertex-wise sparsification factors to better control local search strength and intensify it to specific areas that require it the most.
Finally, the optimization is localized to a limited set of vertices stored in a \textit{selective vertex cache} to concentrate the search on recently modified solution areas.
By combining these techniques, FILO2 and FILO exhibit a computing time that grows linearly with instance size and state-of-the-art results on well-studied literature datasets.

Algorithm \ref{algorithm:filo} outlines the high-level structure of the FILO and FILO2 approaches.
\begin{algorithm}
	\footnotesize
	\caption{High-level structure of FILO and FILO2}\label{algorithm:filo}
	\algcomment{Input parameters: instance $\mathcal{I}$, seed $s$}
	\begin{algorithmic}[1]
		\Procedure{filo}{$\mathcal{I}, s$}
		\State $\mathcal{R} \gets \Call{RandomEngine}{s}$
		\State $S \gets \Call{Construction}{ }$
		\State $k \gets \Call{GreedyRoutesEstimate}{\mathcal{I}}$
		\IfThen{$|S| > k$}{$S \gets \Call{RouteMin}{S, \mathcal{R}}$}
		\State $S \gets \Call{CoreOpt}{S, \mathcal{R}}$
        \State \Return $S$
		\EndProcedure
	\end{algorithmic}
\end{algorithm}
First, the construction phase builds an initial feasible solution using a variation of the savings algorithm proposed by \citet{clarke1964}. 
Then, the improvement phase refines this solution through the optional application of a route minimization procedure, followed by the core optimization procedure.
The following paragraphs will provide more details on these two key phases of the FILO algorithms.

\subsection{Instance Preprocessing} \label{sec:filo2-and-filo:ds-initialization}
Before optimization can start, some data structures must be initialized. In particular, the instance under examination is preprocessed and the cost matrix $\bm{C}$ and the neighbor lists $\mathcal{N}_i(V), i \in V$ are defined. These data structures will be extensively used during all the algorithm procedures.

\subsection{Construction Phase} \label{sec:filo2-and-filo:construction}
The initial feasible solution is built using an adaptation of the savings algorithm proposed by \citet{arnold2019xxl} to solve large-scale instances.
In particular, rather than computing the savings score for all pairs of customers, which would require memory and computing time growing quadratically with the number of customers, only a limited number of pairs are considered.
More precisely, for each customer $i$, savings are computed only for pairs $(i, j)$ where $j$ is among the $n_{cw} = 100$ nearest customers to $i$.

\subsection{Improvement Phase} \label{sec:filo2-and-filo:improvement}
The initial solution generated by the construction phase is then possibly improved during the improvement phase.
First, a route minimization procedure is potentially applied if the number of routes of the initial solution is greater than a heuristic estimate obtained by a greedy first-fit solution of the bin-packing problem associated with the CVRP instance.
Then, the core optimization procedure, which represents the core part of the algorithm, is applied to the resulting solution for a predefined number of iterations.

Both the route minimization and the core optimization procedures are based on the iterated local search paradigm (see \citet{Lourenco2003}) and work by reoptimizing, through the local search engine mentioned in Section \ref{sec:filo2-and-filo}, a localized area disrupted by ruin-and-recreate applications.
The route minimization procedure may evaluate partial solutions to more effectively compact routes. In contrast, the core optimization procedure exclusively explores the feasible space, employing a simulated annealing acceptance rule (see \citet{Kirkpatrick671}) to achieve an effective diversification.

\subsubsection{The Route Minimization Procedure} \label{sec:filo2-and-filo:improvement:routemin}
The route minimization procedure tries to compact the initial solution, but still favors quality over a lower number of routes.
Algorithm \ref{algorithm:routemin} shows the pseudo-code of the procedure.
\begin{algorithm}[!ht]
	\footnotesize
	\caption{Route minimization procedure}
	\algcomment{Input parameters: instance $\mathcal{I}$, solution $S^*$, route estimate $k$, random engine $\mathcal{R}$, uniform real distribution $U(0, 1)$}
	\label{algorithm:routemin}
	\begin{algorithmic}[1]
		\Procedure{RouteMin}{$\mathcal{I}, S^*, k, \mathcal{R}$}
        \State $S \gets S^*, \mathcal{P} \gets 1, L \gets [\;]$ \label{algorithm:routemin:1}
		\For{$n \gets 1 \text{ to } \Delta_{RM}$}
		    \State $(r_i, r_j) \gets \Call{PickPairOfRoutes}{S, \mathcal{R}}$ \label{algorithm:routemin:2}
		    \State $S \gets S \smallsetminus r_i \smallsetminus r_j$ \label{algorithm:routemin:3}
			\State $L \gets [L, \Call{CustomersOf}{r_i, r_j}]$
		    \State $\bar{L} = [\;]$
		    \For{$i \in \Call{Ordered}{L, \mathcal{I}, \mathcal{R}}$} \label{algorithm:routemin:4}
		        \State $p \gets \Call{BestInsertionPositionInExistingRoutes}{i,S}$ \label{algorithm:routemin:bestpos}
		        \If{$p \not= none$}
                    \State $S \gets \Call{Insert}{i, p, S}$
                \Else
                    \If{$|S| < k \lor U(0, 1) > \mathcal{P}$} \label{algorithm:routemin:7}
                        \State $S \gets \Call{BuildSingleCustomerRoute}{i, S}$ \label{algorithm:routemin:singlecustomerroute}
                    \Else
                        \State $\bar{L} \gets [\bar{L}, i]$ \label{algorithm:routemin:addtolist}
                    \EndIf \label{algorithm:routemin:8}
		        \EndIf
		    \EndFor \label{algorithm:routemin:5}
		    \State $L \gets \bar{L}$
		    \State $S \gets \Call{LocalSearch}{S, \mathcal{R}}$ \label{algorithm:routemin:6}

		    \If{$|L| = 0 \land (\Call{Cost}{S} < \Call{Cost}{S^*} \lor (\Call{Cost}{S} = \Call{Cost}{S^*} \land |S| < |S^*|))$} \label{algorithm:routemin:9}
                \State $S^* \gets S$
                \IfThen{$|S^*| \leq k$}{ \Return $S^{*}$}
		    \EndIf \label{algorithm:routemin:10}
		    \State $\mathcal{P} \gets \Call{Decrease}{\mathcal{P}}$ \label{algorithm:routemin:11}
            \IfThen{$\Call{Cost}{S} > \Call{Cost}{S^*}$}{$S \gets S^*$} \label{algorithm:routemin:reverttobest}
		\EndFor
		\State \Return $S^{*}$
		\EndProcedure
	\end{algorithmic}
\end{algorithm}

First, the current reference solution $S$ is set to be equal to the current best-found solution $S^*$ generated by the construction phase (line \ref{algorithm:routemin:1}).
The main loop is then executed for at most $\Delta_{RM}$ iterations and consists of
(a) removing from $S$ the customers belonging to a selected pair of routes $(r_i, r_j)$ (lines \ref{algorithm:routemin:2} and \ref{algorithm:routemin:3}),
(b) sorting any unserved customers either randomly or based on their increasing demand,
(c) potentially greedily inserting any removed customer in the position of $S$ that minimizes the solution cost increase (lines \ref{algorithm:routemin:4} to \ref{algorithm:routemin:5}),
and finally, (d) applying local search to $S$ (line \ref{algorithm:routemin:6}).

The main peculiarity of this procedure consists, during step (c), of probabilistically leaving unserved customers that cannot be inserted into the existing routes because this would violate their load (lines \ref{algorithm:routemin:7} to \ref{algorithm:routemin:8}).
When this happens, such customers can be left unserved for the current iteration with a probability drawn from a uniform real distribution $U(0, 1)$ that decreases exponentially over the iterations (line \ref{algorithm:routemin:11}). These customers will be considered in the subsequent iteration in steps (b) and (c).

The solution obtained after the local search application in step (d) will then replace $S^*$ if it serves all customers and has a better cost or has the same cost but a lower number of routes (line \ref{algorithm:routemin:9}).
Indeed, the main goal of the procedure, which follows the CVRP objective function, is still to find good quality solutions rather than only reducing the number of routes. However, on average, the route minimization procedure proved to be effective in both quickly improving and compacting initial solutions.

\subsubsection{The Core Optimization Procedure}
The core optimization procedure is the main tool for improving the quality of the solution. 
Algorithm \ref{algorithm:coreopt} shows the pseudocode of the procedure.
\begin{algorithm}[!ht]
	\footnotesize
	\caption{Core optimization procedure}
	\label{algorithm:coreopt}
	\algcomment{Input parameters: solution $S^*$, random engine $\mathcal{R}$, uniform real distribution $U(0, 1)$}
	\begin{algorithmic}[1]
		\Procedure{CoreOpt}{$S^*, \mathcal{R}$}
        \State $S \gets S^*, \mathcal{T} \gets \mathcal{T}_0$ \label{algorithm:coreopt:1}
		\State $\bm{\omega} \gets (\omega_1, \omega_2, \ldots, \omega_{|V_c|}), \omega_i \gets \omega_{base} \; \forall i \in V_c$ 
		\State $\bm{\gamma} \gets (\gamma_0, \gamma_1, \ldots, \gamma_{|V_c|}), \gamma_i \gets \gamma_{base} \; \forall i \in V$ \label{algorithm:coreopt:8}
		\For{$n \gets 1 \text{ to } \Delta_{CO}$} \label{algorithm:coreopt:forbegins}
		    \State $S' \gets S$ \label{algorithm:coreopt:2}
			\State $S' \gets \Call{Shaking}{S', \mathcal{R}, \bm{\omega}}$
            \State $S' \gets \Call{LocalSearch}{S', \mathcal{R}}$ \label{algorithm:coreopt:3}
            \If{$\Call{Cost}{S'} < \Call{Cost}{S^*}$}
                \State $S^* \gets S'$ \label{algorithm:coreopt:4}
                \State $\Call{ResetSparsificationFactors}{\bm{\gamma}}$ \label{algorithm:coreopt:6}
            \Else
                \State $\Call{UpdateSparsificationFactors}{\bm{\gamma}}$ \label{algorithm:coreopt:7}
            \EndIf
            \State $\Call{UpdateShakingParameters}{\bm{\omega}, S', S, \mathcal{R}}$ \label{algorithm:coreopt:update-omega}
            \IfThen{$\Call{Cost}{S'} < \Call{Cost}{S} - \mathcal{T} \log(U(0, 1))$}{$S \gets S'$} \label{algorithm:coreopt:5}
            \State $\mathcal{T} \gets c \cdot \mathcal{T}$
		\EndFor \label{algorithm:coreopt:forends}
		\State \Return $S^{*}$
		\EndProcedure
	\end{algorithmic}
\end{algorithm}

Initially, the current reference solution $S$ is set equal to the current best-found solution $S*$, generated by the construction phase, possibly followed by the route minimization procedure (line \ref{algorithm:coreopt:1}). 
Then, the main loop is performed for $\Delta_{CO}$ iterations.
At each iteration, a neighbor solution $S'$ is generated by applying a shaking and a local search procedure to the current reference solution $S$ (lines \ref{algorithm:coreopt:2} to \ref{algorithm:coreopt:3}).
The generated neighbor $S'$ will replace the best-found solution $S^*$ whenever the former has a better quality (line \ref{algorithm:coreopt:4}).
In addition, $S'$ may replace the current reference solution $S$ according to a standard simulated annealing criterion (line \ref{algorithm:coreopt:5}).

The ruin-and-recreate-based shaking procedure first removes a number of $\omega_i$ customers identified by a random walk on the graph underlying the solution,  starting from a randomly selected seed customer $i$.
The removed customers are then sorted by increasing or decreasing distance from the depot, decreasing demand value, or randomly shuffled, and sequentially reinserted in the position that minimizes their insertion cost.
The parameters $\bm{\omega} = (\omega_1, \ldots, \omega_{|V_c|})$ define the ruin intensity per customer.
They are initially set at a value $\omega_{base} = \lceil \ln{|V|} \rceil$ and iteratively updated for the customers involved in the ruin step, based on the quality of the resulting neighbor solution $S'$ with respect to the initial source solution $S$. 
In particular, $\omega_i$ for customer $i$ involved in the ruin step are increased if $S$ and $S'$ have a similar cost, they are decreased if $S'$ has a much higher cost than $S$, and finally they are randomly changed otherwise.
This allows for an iterative fine-tuning of the ruin intensity based on the search trajectory and on the instance and solution structure.

Another set of parameters $\bm{\gamma} = (\gamma_0, \ldots, \gamma_{|V_c|})$ defines the local search intensification level.
The value $\gamma_i \in [0, 1], i \in V$ determines a vertex-specific sparsification level ruling the number of move generators used during a local search application involving the vertex $i$.
In particular, granular neighborhoods are defined by the set of move generators $T = \cup_{i \in V} \{(i, j), (j, i) \in E: j \in \mathcal{N}^{n_{gs}}_i(V \smallsetminus \{i\})\}$, which identifies the arcs that connect $i$ to its $n_{gs}$ neighbors $j$, for every vertex $i$.
The parameters $\gamma_i$ allow the selection of a fraction of the least cost move generators for every vertex $i$.

The core optimization procedure begins with $\gamma_i = \gamma_{base}, i \in V$ (line \ref{algorithm:coreopt:8}), with $\gamma_{base}$ being the base sparsification factor such that at least $n_{gs} \cdot \gamma_{base}$ move generators are always considered for every vertex.
Whenever the best known solution $S^*$ is replaced by a better neighbor solution $S'$, the parameters $\gamma_i$ for the vertices $i$ involved in the last local search application are reset to $\gamma_{base}$ (line \ref{algorithm:coreopt:6}).
If, on the other hand, a vertex $i$ is involved in a certain number of consecutive unsuccessful local search applications, the value of $\gamma_i$ is set to $\gamma_i = min\{\gamma_i \cdot \lambda, 1\}$, where $\lambda$ is an increment factor (line \ref{algorithm:coreopt:7}).

\subsection{Design Highlights}
In the following, we highlight some FILO and FILO2 design choices that are key ingredients of the proposed parallel approach.

\subsubsection{Localized Optimization Pattern} \label{sec:localized-optimization-pattern}
The key principle of the FILO approaches is the localization of the optimization performed during the iterations of the improvement procedures.
This is achieved by employing a technique called selective vertex caching, which allows the local search step to mainly re-optimize the area disrupted by the ruin-and-recreate application. To this end, a cache tracks the vertices involved in the disruption process.
At the beginning of every improvement procedure iteration, the cache is empty, as no action has been performed yet. 
The shaking application causes some vertices involved in the ruin-and-recreate procedure to be inserted into the cache.
Subsequent local search applications are then only started from move generators having at least one of the endpoints in such a cache.
Moreover, whenever a change is successfully applied to a solution (for example, due to the execution of an improving move during the exploration of a neighborhood), related vertices are inserted into the cache.

The cache has a limited size of $C$ vertices, and it is managed with a least-recently changed eviction policy to constantly keep the optimization focused on more recently changed areas.
This technique, which is also the basis for the design of the proposed parallel algorithm \FILOX, is the main factor that localizes the iterations of the improvement procedures and therefore makes them extremely efficient and largely independent of the actual size of the instance.

\subsubsection{Incremental Solution Synchronization} \label{sec:filo2-and-filo:sol-sync}
One of the critical changes introduced in FILO2 to manage extremely large-scale instances is related to the management of solution copies.
In particular, due to the localized optimization pattern described in Section \ref{sec:localized-optimization-pattern}, the difference between a reference solution $S$ and a neighboring solution $S'$ is usually negligible as the size of the instance increases.
Thus, in purely practical terms, performing a full linear-time solution copy (see, for example, lines \ref{algorithm:coreopt:2} and \ref{algorithm:coreopt:5} of Algorithm \ref{algorithm:coreopt}) may be incredibly expensive and largely unnecessary.
For this reason, FILO2 replaces the copy operation with an incremental synchronization between solutions.

The construction phase generates the best-found solution $S^*$. An identical reference solution $S$ is then obtained by performing the only complete copy executed by the entire algorithm.
Any subsequent change to the current reference solution $S$ is stored in the so-called do- and undo-lists.
Possible changes to a solution can in practice be reduced to the following actions: insertion and removal of vertices, creation of one-customer routes, and deletion of empty routes.

The undo-list is used to restore a previous state of the reference solution by undoing part of the stored actions. 
This is particularly helpful in avoiding the explicit creation of an additional neighbor solution $S'$.
In fact, the ruin, recreate and local search steps can be applied to the reference solution $S$, and later undone if the resulting neighbor is not accepted according to the chosen acceptance criteria.
Similarly, the do-list is used to update the best-found solution $S^*$ by applying all the actions that separate it from the current reference solution.

Synchronization between solutions is clearly no longer a linear-time procedure, but it is rather tied to the specific optimization pattern of the algorithm.

\subsubsection{Reduced Memory Footprint} \label{sec:filo2-and-filo:reduced-memory-footprint}
Another important difference introduced in FILO2 to scale to extremely large-scale instances consists in limiting the memory occupation of its data structures.
More specifically, the cost matrix $\bm{C}$ is never explicitly stored, but the costs are computed on demand or retrieved from other existing data structures. 
For example, solutions keep track of the cost of their edges and, similarly, move generators provide the cost associated with their underlying edges to local search operators.
Additionally, neighbor lists are limited to $n_{nn}$ vertices that are identified by using a $kd$-tree built on top of vertex coordinates; see \citet{bentley1990k}.

\section{The \FILOX Solution Approach} \label{sec:solution_approach}
In this section we describe \FILOX, our proposed parallel solution approach for a timely resolution of small- to extremely large-scale CVRP instances.
As will be better detailed in the following, the additional $x$ superscript in the name denotes the usage of $x$ FILO2-like algorithms, or simply \textit{solvers}, to seamlessly optimize a common reference solution.

The main methodological contribution of our research consists in defining an efficient schema to allow an effective collaboration among the individual solvers aimed at collectively optimize a shared solution.

\subsection{The Parallel Schema} \label{sec:parallel-schema}

\FILOX represents the family of parallel solution approaches using $x$ FILO2-based solvers to cooperatively and asynchronously optimize the same underlying solution.
Ultimately, \FILOX results in the coordinated optimization of a shared solution, without the imposition of any guidance or rigid boundaries on the solvers.
In particular, this is achieved by using data redundancy and message-passing communication for coordination while minimizing the need for inefficient synchronization.

Figure \ref{fig:FILO2x.highlevel} shows a high-level representation of the proposed parallel schema.
\begin{figure}[htpb]
\centering
\caption{High-level \FILOX schema. Different filling patterns are used to associated candidate changes to the respective generator solver. Queues highlight how solvers may have not all applied the same amount of received changes, but changes in all queues are received in the same order.}
\label{fig:FILO2x.highlevel}
\begin{tikzpicture}[scale=0.75, transform shape]

\draw[pattern=north west lines, rounded corners] (0,0) rectangle (2,1) node[fill=white, pos=.5] {\textsc{Solver}};
\draw[->, arrows = {-Latex[width=5pt, length=5pt]}] (2.25,0.5) -- (3.75,0.5);
\draw[pattern=north west lines] (4,0.25) rectangle (4.5,0.75)  node[below, pos=.5, yshift=-10pt] {\textsc{Change}};
\draw[->, arrows = {-Latex[width=5pt, length=5pt]}] (4.75, 0.5) -- (6.75,1.25);

\draw[fill=white] (-4.35,0.15) rectangle (-1.90,0.85) node[below, pos=0.5, yshift=-10pt] {\textsc{Queue}};
\draw[pattern=vertical lines] (-3.70,0.25) rectangle (-3.20,0.75);
\draw[pattern=north west lines] (-3.10,0.25) rectangle (-2.60,0.75);
\draw[pattern=crosshatch dots] (-2.5,0.25) rectangle (-2,0.75);
\draw[->, arrows = {-Latex[width=5pt, length=5pt]}] (-1.75,0.5) -- (-0.25,0.5);

\draw[pattern=north east lines, rounded corners] (0,1.5) rectangle (2,2.5) node[fill=white, pos=.5] {\textsc{Solver}};
\draw[->, arrows = {-Latex[width=5pt, length=5pt]}] (2.25, 2) -- node[below] {\scriptsize generate} (3.75,2);
\draw[pattern=north east lines] (4,1.75) rectangle (4.5,2.25);
\draw[->, arrows = {-Latex[width=5pt, length=5pt]}] (4.75, 2) -- (6.75,2.25);

\draw[fill=white] (-4.35,1.65) rectangle (-1.90,2.35);
\draw[pattern=vertical lines] (-2.5,1.75) rectangle (-2,2.25);
\draw[->, arrows = {-Latex[width=5pt, length=5pt]}] (-1.75,2.0) -- node[below] {\scriptsize sync} (-0.25,2.0);

\draw[pattern=vertical lines, rounded corners] (0,3) rectangle (2,4) node[fill=white, pos=.5] {\textsc{Solver}};
\draw[->, arrows = {-Latex[width=5pt, length=5pt]}] (2.25, 3.5) -- (3.75,3.5);
\draw[pattern=vertical lines] (4,3.25) rectangle (4.5,3.75);
\draw[->, arrows = {-Latex[width=5pt, length=5pt]}] (4.75, 3.5) -- (6.75,3.25);

\draw[fill=white] (-4.35,3.15) rectangle (-1.90,3.85);
\draw[pattern=vertical lines] (-3.70,3.25) rectangle (-3.20,3.75);
\draw[pattern=north west lines] (-3.10,3.25) rectangle (-2.6,3.75);
\draw[pattern=crosshatch dots] (-2.5,3.25) rectangle (-2,3.75);
\draw[->, arrows = {-Latex[width=5pt, length=5pt]}] (-1.75,3.5) -- (-0.25,3.5);

\draw[pattern=crosshatch dots, rounded corners] (0,4.5) rectangle (2,5.5) node[fill=white, pos=.5] {\textsc{Solver}};
\draw[->, arrows = {-Latex[width=5pt, length=5pt]}] (2.25, 5) -- (3.75,5);
\draw[pattern=crosshatch dots] (4,4.75) rectangle (4.5,5.25);
\draw[->, arrows = {-Latex[width=5pt, length=5pt]}] (4.75, 5) -- node[above, rotate=-22] {\scriptsize append} (6.75,4.25);

\draw[fill=white] (-4.35,4.65) rectangle (-1.90,5.35);
\draw[pattern=vertical lines] (-3.10,4.75) rectangle (-2.6,5.25);
\draw[pattern=north west lines] (-2.5,4.75) rectangle (-2,5.25);
\draw[->, arrows = {-Latex[width=5pt, length=5pt]}] (-1.75,5) -- (-0.25,5);


\draw[fill=white] (7.0,2.4) rectangle (9.45,3.1) node[below, pos=0.5, yshift=-10pt] {\textsc{Queue}};
\draw[pattern=vertical lines] (8.85,2.5) rectangle (9.35,3);
\draw[pattern=north west lines] (8.25,2.5) rectangle (8.75,3);

\draw[->, arrows = {-Latex[width=5pt, length=5pt]}] (9.65, 2.75) -- node[below] {\scriptsize get} (11, 2.75);

\draw[fill=white, rounded corners] (11.25,2.25) rectangle (14, 3.25) node[pos=.5] {\textsc{Dispatcher}};

\draw[-] (14.25, 2.75) -- (15, 2.75) -- (15, 6) -- node[above] {\scriptsize broadcast} (-7, 6) -- (-7, 2.75);
\draw[->, arrows = {-Latex[width=5pt, length=5pt]}] (-7, 2.75) -- (-6, 2.75);

\draw[-] (-5.5, 0.5) -- (-5.5, 5);
\draw[preaction={fill, white}, pattern=crosshatch dots] (-5.75,2.5) rectangle (-5.25,3);
\draw[->, arrows = {-Latex[width=5pt, length=5pt]}] (-5.5, 0.5) -- (-4.5, 0.5);
\draw[->, arrows = {-Latex[width=5pt, length=5pt]}] (-5.5, 2) -- (-4.5, 2);
\draw[->, arrows = {-Latex[width=5pt, length=5pt]}] (-5.5, 3.5) -- (-4.5, 3.5);
\draw[->, arrows = {-Latex[width=5pt, length=5pt]}] (-5.5, 5) -- node[above, pos=.5]{\scriptsize append} (-4.5, 5);

\end{tikzpicture}
\end{figure}
Initially, the desired number of solvers is created and initialized with the solution obtained through the construction phase and the greedy estimate of the required number of routes $k$; see Sections \ref{sec:filo2-and-filo:construction} and \ref{sec:filo2-and-filo:improvement}.
Both the construction phase and the route estimation procedure do not require any solver-specific data structures, and as a design choice, they are performed before solvers are created. 
A different but equivalent choice could have been to execute them independently by all solvers.

As soon as they are created, the solvers asynchronously start to optimize their solution by running first the route minimization procedure described in Section \ref{sec:filo2-and-filo:improvement:routemin} if required, followed by a parallel version of the core optimization procedure.

The route minimization procedure does require solver-specific data structures for the local search execution and it is thus performed by every solver independently. All solvers will however traverse the same search trajectory and converge to the same final solution as they are using a common seed for their pseudo-random number generator.

At every iteration of the parallel core optimization procedure, a candidate change is generated from the application of the ruin, recreate, and local search steps to the current reference solution.
Candidate changes are sent to a centralized dispatcher component, which keeps them sorted according to their arrival time.
The dispatcher actively and continuously retrieves the next available candidate change and broadcasts it to all available solvers.
Candidate changes are evaluated by the solvers and possibly implemented to update their reference solution.

\subsection{Memory Model}
The \FILOX algorithm employs a multithreaded shared memory architecture in which the solvers and the dispatcher are backed by a logical computer thread having its own independent execution flow.

Candidate changes generated by the solvers are stored into a shared thread-safe dispatcher queue data structure keeping them ordered according to a first-in first-out policy.
The queue includes both improving and non-improving candidate changes.
The latter are relevant for the simulated annealing acceptance criteria used in the core optimization phase (see line \ref{algorithm:coreopt:5} of Algorithm \ref{algorithm:coreopt}).

The insertion of a new change into the queue is performed by the generating solver by first gaining exclusive access to the data structure through standard mutex locking mechanisms.
Similarly, the removal of the least recent candidate change is performed by the dispatcher once access to the queue has been locked for the solvers.
The dispatcher continuously retrieves the next available candidate change from the queue and broadcasts it to all solvers.
When the queue is empty, the dispatcher suspends its execution, waiting for more changes.

Similarly, every solver manages a queue that stores in chronological order the candidate changes sent by the dispatcher.
As for the dispatcher queue, exclusive access to the solver queue is required when the dispatcher adds a new change or when the solver retrieves the least recently available one.

\subsection{Parallel Core Optimization Procedure}
Every solver behaves as the single-threaded FILO2 algorithm described in Section \ref{sec:filo2-and-filo}.
However, in addition, during the core optimization procedure, it leverages the work performed by other solvers to simulate the execution of the iterations they have already completed.

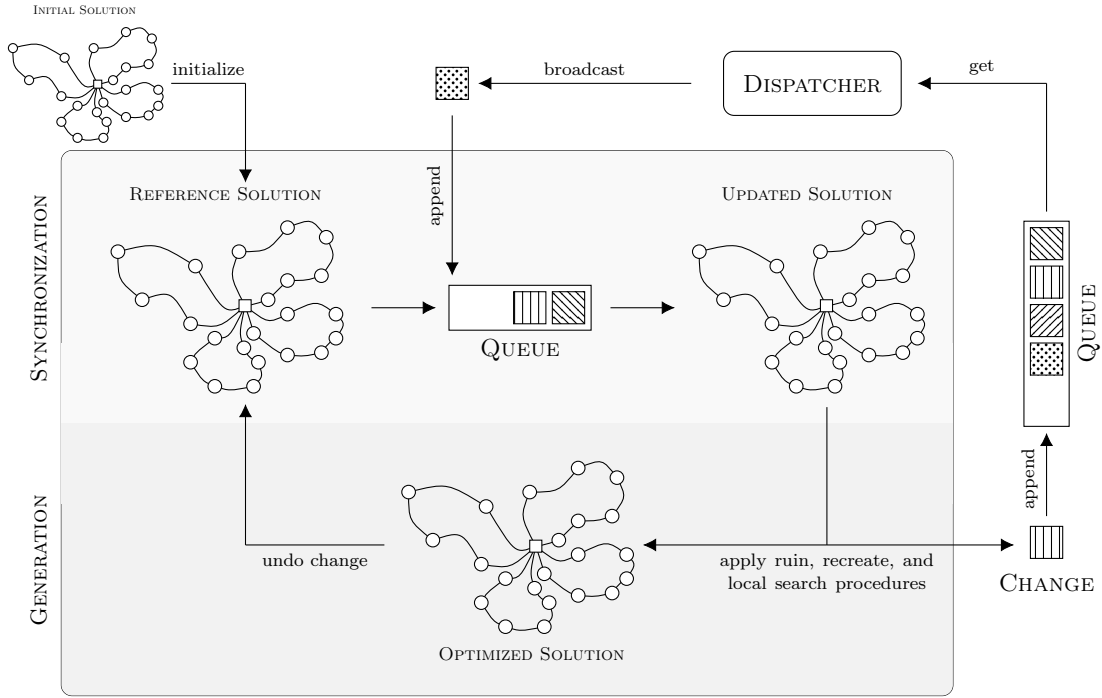
\begin{figure}
\centering
\caption{Component activities during the parallel core optimization procedure. The solver is depicted by the gray box with the top lighter part representing the synchronization stage and the bottom darker one illustrating the generation stage.}
\label{fig:FILO2x.solver}
\newcommand\DrawSolution[6]{%

\begin{scope}[scale=#3]

\draw (10.125+#1, 10.375+#2) .. controls (9.6+#1,9.9+#2) .. (9+#1,10+#2);
\draw (9+#1, 10+#2) .. controls (8.6+#1,10.7+#2) .. (8+#1,10.5+#2);
\draw (8+#1, 10.5+#2) .. controls (7.5+#1,10.9+#2) .. (7.5+#1,11.5+#2);
\draw (7.5+#1, 11.5+#2) .. controls (8.3+#1,11.7+#2) .. (9.1+#1,11.2+#2);
\draw (9.1+#1, 11.2+#2) .. controls (9.7+#1,10.2+#2) .. (10.125+#1,10.375+#2);

\draw (10.125+#1, 10.375+#2) .. controls (9.8+#1,11+#2) .. (10+#1,11.5+#2);
\draw (10+#1, 11.5+#2) .. controls (10.6+#1,11.5+#2) .. (11+#1,12+#2);
\draw (11+#1, 12+#2) .. controls (11.5+#1,12.2+#2) .. (11.8+#1,11.8+#2);
\draw (11.8+#1, 11.8+#2) .. controls (12+#1,11.4+#2) .. (11.7+#1,11+#2);
\draw (11.7+#1, 11+#2) .. controls (11.4+#1,10.9+#2) .. (11+#1,10.9+#2);
\draw (11+#1, 10.9+#2) .. controls (10.8+#1,10.7+#2) .. (10.6+#1,10.5+#2);
\draw (10.6+#1, 10.5+#2) .. controls (10.3+#1,10.4+#2) .. (10.125+#1,10.375+#2);

\draw (10.125+#1, 10.375+#2) .. controls (10.6+#1,10+#2) .. (11+#1,10.2+#2);
\draw (11+#1,10.2+#2) .. controls (11.6+#1,10.5+#2) .. (12+#1,10.2+#2);
\draw (12+#1,10.2+#2) .. controls (12.2+#1,10+#2) .. (12+#1,9.7+#2);
\draw (12+#1,9.7+#2) .. controls (11.85+#1,9.6+#2) .. (11.7+#1,9.4+#2);
\draw (11.7+#1,9.4+#2) .. controls (11.35+#1,9.32+#2) .. (11+#1,9.35+#2);
\draw (11+#1,9.35+#2) .. controls (10.8+#1,9.4+#2) .. (10.3+#1,9.8+#2);
\draw (10.3+#1,9.8+#2) .. controls (10.25+#1,10+#2) .. (10.125+#1, 10.375+#2);

\draw (10.125+#1, 10.375+#2) .. controls (10+#1,9.8+#2) .. (10.1+#1,9.5+#2);
\draw (10.1+#1,9.5+#2) .. controls (10.2+#1,9.3+#2) .. (10.4+#1,9.2+#2);
\draw (10.4+#1,9.2+#2) .. controls (10.5+#1,8.9+#2) .. (10.3+#1,8.7+#2);
\draw (10.3+#1,8.7+#2) .. controls (9.8+#1,8.6+#2) .. (9.5+#1,8.7+#2);
\draw (9.5+#1,8.7+#2) .. controls (9.25+#1,8.8+#2) .. (9+#1,9.2+#2);
\draw (9+#1,9.2+#2) .. controls (9.1+#1,9.4+#2) .. (9.5+#1,9.7+#2);
\draw (9.5+#1,9.7+#2) .. controls (9.8+#1,9.7+#2) .. (10.125+#1, 10.375+#2);

\draw[fill=white] (10+#1, 10.25+#2) rectangle (10.25+#1,10.50+#2) node[above, xshift=#6pt, yshift=#5pt] {\textsc{#4}};
\draw[fill=white] (9+#1,10+#2) circle (4pt);
\draw[fill=white] (8+#1,10.5+#2) circle (4pt);
\draw[fill=white] (7.5+#1,11.5+#2) circle (4pt);
\draw[fill=white] (9.1+#1,11.2+#2) circle (4pt);

\draw[fill=white] (10+#1,11.5+#2) circle (4pt);
\draw[fill=white] (11+#1,12+#2) circle (4pt);
\draw[fill=white] (11.8+#1,11.8+#2) circle (4pt);
\draw[fill=white] (11.7+#1,11+#2) circle (4pt);
\draw[fill=white] (11+#1,10.9+#2) circle (4pt);
\draw[fill=white] (10.6+#1,10.5+#2) circle (4pt);

\draw[fill=white] (11+#1,10.2+#2) circle (4pt);
\draw[fill=white] (12+#1,10.2+#2) circle (4pt);
\draw[fill=white] (12+#1,9.7+#2) circle (4pt);
\draw[fill=white] (11.7+#1,9.4+#2) circle (4pt);
\draw[fill=white] (11+#1,9.35+#2) circle (4pt);
\draw[fill=white] (10.3+#1,9.8+#2) circle (4pt);

\draw[fill=white] (10.1+#1,9.5+#2) circle (4pt);
\draw[fill=white] (10.4+#1,9.2+#2) circle (4pt);
\draw[fill=white] (10.3+#1,8.7+#2) circle (4pt);
\draw[fill=white] (9.5+#1,8.7+#2) circle (4pt);
\draw[fill=white] (9+#1,9.2+#2) circle (4pt);
\draw[fill=white] (9.5+#1,9.7+#2) circle (4pt);

\end{scope}
}

\newcommand\DrawQueue[2]{
\draw[fill=white] (9+#1,9.9+#2) rectangle (11.2+#1,10.6+#2) node[below, pos=.5, yshift=-10pt] {\textsc{Queue}};
\draw[pattern=vertical lines] (10+#1,10+#2) rectangle (10.5+#1,10.5+#2);
\draw[pattern=north west lines] (10.6+#1,10+#2) rectangle (11.1+#1,10.5+#2);
\draw[->, arrows = {-Latex[width=5pt, length=5pt]}] (11.5+#1,10.25+#2) -- (12.5+#1,10.25+#2);
\draw[->, arrows = {-Latex[width=5pt, length=5pt]}] (7.8+#1,10.25+#2) -- (8.8+#1,10.25+#2);
}

\newcommand\DrawHeap[2] {
\draw[pattern=crosshatch dots] (8+#1,3.5+#2) rectangle (8.5+#1,4+#2);
\draw[-] (8.25+#1, 3.5+#2) -- (7.75+#1, 3+#2);
\draw[-] (8.25+#1, 3.5+#2) -- (8.75+#1, 3+#2);
\draw[pattern=vertical lines] (7.5+#1,2.5+#2) rectangle (8+#1,3+#2);
\draw[-] (7.75+#1, 2.5+#2) -- (7.25+#1, 2+#2);
\draw[-] (7.75+#1, 2.5+#2) -- (8.25+#1, 2+#2);
\draw[pattern=north west lines] (8.5+#1,2.5+#2) rectangle (9+#1,3+#2);
\draw[-] (8.75+#1, 2.5+#2) -- (9.25+#1, 2+#2);
\draw[pattern=crosshatch dots] (8+#1,1.5+#2) rectangle (8.5+#1,2+#2) node[below, pos=.5, yshift=-10pt] {\textsc{Heap}};
\draw[pattern=north west lines] (7+#1,1.5+#2) rectangle (7.5+#1,2+#2);
\draw[pattern=north east lines] (9+#1,1.5+#2) rectangle (9.5+#1,2+#2);
}

\newcommand\DrawVerticalQueue[2] {
\draw[fill=white] (7.9+#1,4.1+#2) rectangle (8.6+#1,0.9+#2) node[below, rotate=90, pos=.5, yshift=-10pt] {\textsc{Queue}};

\draw[pattern=north west lines] (8+#1,3.5+#2) rectangle (8.5+#1,4+#2);

\draw[pattern=vertical lines] (8+#1,3.4+#2) rectangle (8.5+#1,2.9+#2);
\draw[pattern=north east lines] (8+#1,2.8+#2) rectangle (8.5+#1,2.3+#2);

\draw[pattern=crosshatch dots] (8+#1,2.2+#2) rectangle (8.5+#1,1.7+#2);
}

\begin{tikzpicture}[scale=0.85, transform shape]

\draw[rounded corners] (4,9.2) rectangle (17.8, 17.7);

\fill[fill=gray!5] (4,13.45) rectangle (17.8, 14.7);
\fill[fill=gray!5, rounded corners] (4,13.45) rectangle (17.8, 17.7) node[pos=0, rotate=90, xshift=60pt, yshift=10pt]{\small \textsc{Synchronization}};

\fill[fill=gray!10] (4,12.2) rectangle (17.8, 13.45);
\fill[fill=gray!10, rounded corners] (4,9.2) rectangle (17.8, 13.45) node[pos=0, rotate=90, xshift=60pt, yshift=10pt]{\small \textsc{Generation}};

\DrawSolution{-1}{27.1}{0.5}{Initial Solution}{55}{-15}
\draw[-] (5.7, 18.75) -- node[pos=0.5, above] {\scriptsize initialize} (6.85, 18.75);
\draw[->, arrows = {-Latex[width=5pt, length=5pt]}] (6.85, 18.75) -- (6.85, 17.2);

\DrawSolution{-1}{10}{0.75}{Reference Solution}{55}{-15}
\DrawSolution{11}{10}{0.75}{Updated Solution}{55}{-15}
\DrawQueue{1}{5}

\draw[-] (15.85, 13.7) -- (15.85, 11.55);
\draw[->, arrows = {-Latex[width=5pt, length=5pt]}] (15.85, 11.55) -- (13, 11.55);
\draw[->, arrows = {-Latex[width=5pt, length=5pt]}] (15.85, 11.55) -- node[below, pos=0, text width=3.5cm, text centered, execute at begin node=\setlength{\baselineskip}{10pt}]{\scriptsize apply ruin, recreate, and local search procedures} (18.75, 11.55);

\draw[-] (9, 11.55) -- (6.85, 11.55) node[below, pos=0.5]{\scriptsize undo change};
\draw[->, arrows = {-Latex[width=5pt, length=5pt]}] (6.85, 11.55) -- (6.85, 13.75);

\DrawVerticalQueue{11}{12.5}

\draw[->, arrows = {-Latex[width=5pt, length=5pt]}] (8.25+11, -0.50+12.5) -- node[pos=0.45, rotate=90, above]{\scriptsize append} (8.25+11, 0.8+12.5);
\draw[pattern=vertical lines] (8+11,-1.16+12.5) rectangle (8.5+11,-0.66+12.5) node[below, pos=0.5, yshift=-10pt]{\textsc{Change}};

\draw[-] (8.25+11, 4.25+12.5) -- (8.25+11, 6.25+12.5);
\draw[->, arrows = {-Latex[width=5pt, length=5pt]}] (8.25+11, 6.25+12.5) -- node[pos=0.5, above]{\scriptsize get} (5.25+12, 6.25+12.5);

\draw[->, arrows = {-Latex[width=5pt, length=5pt]}] (1.75+12, 6.25+12.5) -- node[pos=0.5, above]{\scriptsize broadcast} (-0.55+11, 6.25+12.5);

\draw[pattern=crosshatch dots] (-1.20+11,6+12.5) rectangle (-0.70+11,6.5+12.5);

\draw[->, arrows = {-Latex[width=5pt, length=5pt]}] (-0.95+11, 5.75+12.5) -- node[pos=0.5, rotate=90, above]{\scriptsize append} (-0.95+11, 3.25+12.5);

\DrawSolution{5}{5}{0.75}{Optimized Solution}{-75}{-6}

\draw[fill=white, rounded corners] (14.25,18.25) rectangle (17, 19.25) node[pos=.5] {\textsc{Dispatcher}};

\end{tikzpicture}
\end{figure}
As depicted in Figure \ref{fig:FILO2x.solver}, once initialized with the solution obtained through the construction phase and possibly the route minimization procedure, a solver performs the parallel core optimization procedure through a sequence of synchronization and generation stages.

During the synchronization stage, some candidate changes in the solver queue are applied to its reference solution $S$ to generate an updated reference solution $S'$.
Note that, due to the concurrent and asynchronous nature of the algorithm, candidate changes keep flowing constantly to every solver queue. 
The update is thus limited to the sequence of changes available when the synchronization stage begins and does not include any additional received ones.

Each candidate change describes the effects of an iteration of the core optimization procedure on a reference solution through a list of actions; see Section \ref{sec:filo2-and-filo:sol-sync}.
More precisely, a change contains the list of actions associated with the ruin, recreate, and local search steps performed by some solver.
Because a change may have been generated by a different solver on an arbitrarily different reference solution, the application of the associated actions to the current solver reference solution may produce an infeasible neighbor solution.
For example, it may either generate a solution with route load constraints violations or even create customer sub-tours.
Thus, before being applied, a candidate change is checked to verify whether it produces a feasible solution, and it is discarded otherwise.

Whenever a candidate change generates a feasible solution when applied to the current solver reference solution, an entire procedure iteration of Algorithm \ref{algorithm:coreopt} (lines \ref{algorithm:coreopt:forbegins} to \ref{algorithm:coreopt:forends}) is simulated with the ruin, recreate, and local search steps defined by the actions belonging to the change.
The simulation includes the update of the algorithm parameters (lines \ref{algorithm:coreopt:6}, \ref{algorithm:coreopt:7}, and \ref{algorithm:coreopt:update-omega})
and the possible update of the best-found and reference solutions (lines \ref{algorithm:coreopt:4} and \ref{algorithm:coreopt:5}).


The generation stage consists of applying the ruin, recreate, and local search procedures to the updated reference solution $S'$, resulting from the synchronization stage, to generate an optimized reference solution $S''$.
The sequence of actions required to convert $S'$ into $S''$ is recorded using the technique described in Section \ref{sec:filo2-and-filo:sol-sync} and used to define a candidate change that is sent to the dispatcher.

Finally, the optimized solution $S''$ is reverted to $S'$ by applying in reverse order the inverse actions that compose the generated candidate change (see Section \ref{sec:filo2-and-filo:sol-sync}), essentially bringing the current solver reference solution back to the same search trajectory traversed by all other solvers.
The solver then returns to the synchronization stage with $S = S'$ as the next reference solution.

The pseudo-random number generator of each solver is initialized with a different seed. Therefore, since the ruin, recreate, and local search steps contain some randomization, each solver may produce different candidate changes during the generation stage. 
However, during the synchronization stage, a randomized update of the $\bm{\omega}$ parameters may occur (see line \ref{algorithm:coreopt:update-omega} of Algorithm \ref{algorithm:coreopt}).
To ensure that all solvers update these parameters consistently, a different pseudo-random number generator is instead initialized with the same seed for all the solvers. 

\subsection{Synchronization Mechanism}
The dispatcher is the key component that guarantees the synchronization among solvers.
Indeed, by collecting and relaying all candidate changes, it ensures all solvers receive them in the same absolute order, thus implicitly fixing a shared and unique search trajectory.
Because solvers process received changes in the same order, they will always make the same decisions during the synchronization stage, eventually converging to the same final solution.

However, solvers may differ in the number of changes processed at a given time because of when they are scheduled by the operating system and due to their unique generation stage.
At a specific time, they may thus be at different points on the same search trajectory, therefore causing their reference solutions to be possibly considerably different.
This may induce the generation of candidate changes that, when applied by other solvers, might not produce feasible solutions.
However, intuitively, as the size of the instance grows, given the localized optimization pattern used by the algorithm, the likelihood of defining these inapplicable changes decreases greatly, since different solvers will probably optimize different, and possibly disjoint, areas of the solution.




\subsection{Parallel Instance Preprocessing}
The initial instance preprocessing mentioned in Sections \ref{sec:filo2-and-filo:ds-initialization} and \ref{sec:filo2-and-filo:reduced-memory-footprint}, which mainly consists of identifying $n_{nn}$ neighbors for every vertex $i \in V$, can be very time consuming for extremely large-scale instances.
In \FILOX, once the $kd$-tree is built on top of the instance coordinates, the neighbors of the vertices are retrieved in parallel by employing $x$ different threads. To make comparisons easier, we decided to link the number of threads employed during this preprocessing to the number of solvers used during the optimization. However, in principle, a different and greater number could profitably be employed.

\section{Computational Results} \label{sec:computational-results}
The computational testing has the main objective of validating the strengths and limits of the proposed parallel approach.
In particular, we assess whether \FILOX is capable of finding solutions having a quality comparable to those obtained by FILO2, but in a fraction of time.
To this end, we test the proposed algorithm on the well-known $\mathbb{X}$ instances proposed by \citet{uchoa2017}, which are the current standard reference instances for the CVRP.
In addition, we evaluate \FILOX on the very large-scale $\mathbb{B}$ instances presented by \citet{arnold2019xxl} and on the extremely large-scale $\mathbb{I}$ instances introduced in \citet{accorsi2024}.
Afterwards, in Section \ref{sec:analysis}, we provide more insight into the behavior of the algorithm.

\subsection{Implementation and Experimental Environment}
The proposed \FILOX algorithm was implemented in C++ and compiled using g++ 11.4.
The experiments were executed on a 64-bit GNU/Linux Ubuntu 22.04 workstation with an Intel Xeon Gold 6254 CPU, running at 3.1 GHz, and equipped with 384 GB of RAM.
The source code for \FILOX is available at \url{https://github.com/acco93/filo2x}.

For the experiments, we considered a standard version of \FILOX and a longer version, called \FILOX (long), performing $10^5$ and $10^6$ core optimization iterations, respectively.
Given that \FILOX is a randomized algorithm, to get a sense of the variability between executions, we performed a symbolic number of ten runs for each problem instance. All results and analyses described in the following sections refer to the average behavior of the algorithm.

\FILOX was compared to the original sequential algorithm FILO2, available at \url{https://github.com/acco93/filo2}. The results of FILO2 presented in this study were obtained by running the algorithm on the same machine described above. In addition, for reference, we include the results of state-of-the-art algorithms on the different studied datasets.

The best known solution values (BKS) for the $\mathbb{X}$ and $\mathbb{B}$ instances were taken from \citet{cvrplib} at the time of writing, whereas the BKS for the $\mathbb{I}$ instances come from \citet{accorsi2024}.
The gaps are computed as $100 \cdot (z - \text{BKS}) / \text{BKS}$ where $z$ is the final solution value obtained by the algorithms, and the computing time is always reported in seconds.
Finally, to ensure conciseness, in the following we opted to report only aggregated results. More details are included in the Appendix.

\subsection{Parameter Tuning}
\begin{table}[]
    \renewcommand{\arraystretch}{1.3}
	\footnotesize
	\centering
	\caption{Inherited main FILO2 parameters. \label{table:inherited-parameters}}
	\begin{tabular}{p{2.5cm} lp{12cm}}
    \rowcolor{gray!20} \multicolumn{3}{l}{Scalability} \\
    \rowcolor{gray!5} $n_{nn} = 1500$        && Maximum number of nearest neighbors that are computed, stored and used during the algorithm execution.\\
    \rowcolor{gray!5} $C = 50$  && Maximum capacity of the selective vertex cache.\\
    \rowcolor{gray!20} \multicolumn{3}{l}{Initial solution definition (construction phase and route minimization procedure)} \\
	\rowcolor{gray!5} $n_{cw} = 100$        && Number of neighbors considered in the savings computation.\\
	\rowcolor{gray!5} $\Delta_{RM} = 10^3$ && Maximum number of route minimization iterations.\\
	\rowcolor{gray!20} \multicolumn{3}{l}{Granular neighborhood}\\
	\rowcolor{gray!5} $n_{gs} = 25$        && Number of neighbors considered for the definition of granular neighborhoods.\\
	\rowcolor{gray!5} $\gamma_{base} = 0.25$ && Base sparsification factor, i.e., at least $n_{gs} \cdot \gamma_{base}$ move generators are always considered for every vertex.\\
	\rowcolor{gray!5} $\delta = 0.5$ && Reduction factor used in the definition of the fraction of non-improving iterations performed before increasing a sparsification parameter.\\
	\rowcolor{gray!5} $\lambda = 2$ && Sparsification increment factor.\\
	\rowcolor{gray!20} \multicolumn{3}{l}{Core optimization procedure} \\
	\rowcolor{gray!5} $\Delta_{CO} = 10^5, 10^6$ && Number of core optimization iterations for short and long runs, respectively.\\
	\rowcolor{gray!5} $\mathcal{T}_0$, $\mathcal{T}_f$ && Initial and final simulated annealing temperatures.\\
	\rowcolor{gray!5} $\omega_{base} = \lceil \ln{|V|} \rceil$ && Initial shaking intensity.\\
	\end{tabular}
\end{table}
To minimize behavioral differences between sequential and parallel versions, \FILOX inherits all parameters and their values from the FILO2 sequential algorithm.
Table \ref{table:inherited-parameters} summarizes the inherited parameters, their value, and their meaning.
In addition, \FILOX uses a single supplementary parameter $x$ defining the number of solvers used during optimization.
Whenever $x = 1$, \FILOX behaves similarly to FILO2. 
When $x > 1$, \FILOX results in an emerging behavior that can be assimilated into a FILO2 approach in which multiple iterations of the core optimization procedure are executed in parallel.
We tested values for $x$ ranging from $2$ to $10$, as these values were found to provide the best performance on the specific machine and problem instances used in this study.

\subsection{Testing on the \texorpdfstring{$\mathbb{X}$}{X} Instances}
The $\mathbb{X}$ dataset, proposed by \citet{uchoa2017}, contains hundred instances with a number of customers ranging from 100 to 1000 modeling a wide range of customer demands and layout distributions.
The $\mathbb{X}$ instances are the current reference benchmark literature dataset for the CVRP.
Optimal or near-optimal solutions are known for most of the instances, and the best performing algorithms on this dataset usually differ in the order of a few percentage gap decimal points.

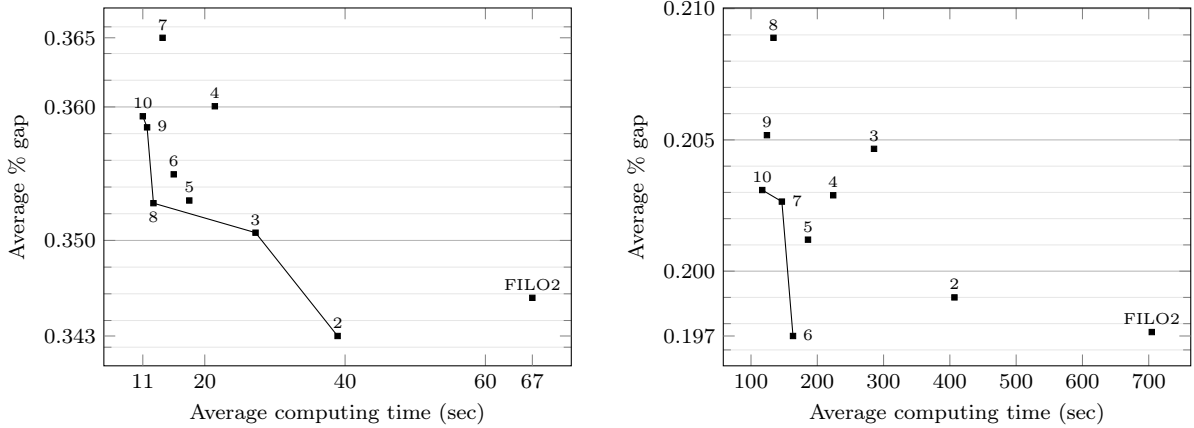
\begin{figure}
\centering
\caption{Performance charts of \FILOX and FILO2 on the $\mathbb{X}$ dataset when performing standard (left) and long (right) runs. The individual numbers identify how many solvers have been used to run the \FILOX algorithm.}
\begin{tikzpicture}[]
\scriptsize

\begin{groupplot}[
    group style={group size= 2 by 1, horizontal sep=2cm},
    height=180pt,
    width=220pt,
    ymajorgrids=true,
    yminorgrids=true,
    minor y tick num=4,
    minor grid style={line width=.01pt,draw=black!10},
    major grid style={line width=.01pt,draw=black!30},
]

    \nextgroupplot[
	title={},
	xlabel={Average computing time (sec)},
	ylabel={Average \% gap},
        extra y ticks={0.36518, 0.34284},
        extra y tick labels={0.365, 0.343},
        extra y tick style={grid=none},
        extra x ticks={11.168, 66.693},
        extra x tick labels={11, 67},
        extra x tick style={grid=none},
        yticklabel=\pgfkeys{/pgf/number format/.cd,fixed,precision=3,zerofill}\pgfmathprintnumber{\tick}
]

\addplot[] coordinates {

(11.168,0.3593)
(11.777,0.35847)
(12.683,0.35279)
(27.235,0.35058)
(38.932,0.34284)

};

\addplot[only marks, mark=square*, mark size=1pt, point meta=explicit symbolic, nodes near coords, text width=3cm, align=center, font=\tiny\linespread{0.8}\selectfont] coordinates {

(66.693,0.3457)[FILO2]
(38.932,0.34284)[2]
(27.235,0.35058)[3]
(21.435,0.36005)[4]
(17.789,0.35299)[5]
(15.577,0.35495)[6]
(13.985,0.36518)[7]
(11.168,0.3593)[10]

};

\addplot[only marks, mark=square*, mark size=1pt, point meta=explicit symbolic, nodes near coords, text width=3cm, align=center, every node near coord/.append style={yshift=-0.35cm}, font=\tiny\linespread{0.8}\selectfont] coordinates {
(12.683,0.35279)[8]
};


\addplot[only marks, mark=square*, mark size=1pt, point meta=explicit symbolic, nodes near coords, text width=3cm, align=center, every node near coord/.append style={xshift=+0.20cm, yshift=-0.16cm}, font=\tiny\linespread{0.8}\selectfont] coordinates {
(11.777,0.35847)[9]
};

\nextgroupplot[
	title={},	
	xlabel={Average computing time (sec)},
	ylabel={Average \% gap},
        extra y ticks={0.19753},
        extra y tick labels={0.197},
        extra y tick style={grid=none},
    yticklabel=\pgfkeys{/pgf/number format/.cd,fixed,precision=3,zerofill}\pgfmathprintnumber{\tick}
    ]

\addplot[] coordinates {

(116.963,0.20309)
(146.555,0.20265)
(163.365,0.19753)

};

\addplot[only marks, mark=square*, mark size=1pt, point meta=explicit symbolic, nodes near coords, text width=3cm, align=center, font=\tiny\linespread{0.8}\selectfont] coordinates {

(704.609,0.19768)[FILO2]
(406.893,0.199)[2]
(285.581,0.20466)[3]
(223.993,0.20289)[4]
(185.977,0.2012)[5]
(133.904,0.20889)[8]
(124.053,0.20518)[9]
(116.963,0.20309)[10]

};



\addplot[only marks, mark=square*, mark size=1pt, point meta=explicit symbolic, nodes near coords, text width=3cm, align=center, every node near coord/.append style={xshift=+0.20cm, yshift=-0.16cm}, font=\tiny\linespread{0.8}\selectfont] coordinates {

(163.365,0.19753)[6]
(146.555,0.20265)[7]

};

\end{groupplot}

	\end{tikzpicture}
\label{fig:FILO2x.performance.x}
\end{figure}
Although these small to large instances might not be the ideal candidates for parallel optimization, Figure \ref{fig:FILO2x.performance.x} shows that the proposed parallel approach obtains competitive results in terms of final solution quality and an almost linear speedup when using up to five solvers.
For experiments using more than five solvers, the computing time converges to approximately 10 seconds for short runs and 100 seconds for longer ones, indicating that for these instances and hardware configuration, more solvers cannot provide substantial speedup benefits.

The graphs show small differences in the average percentage gap obtained by varying the number of solvers. However, statistical analyses of these results, detailed in \ref{appendix:x}, show that the final solution quality obtained by the parallel versions is similar to that obtained by the original sequential algorithm.

Finally, to better frame the results obtained by \FILOX, consider the hybrid genetic search (HGS) and the slack induction by string removals (SISR) algorithms proposed by \citet{vidal22} and \citet{sisr}, respectively, which are two of the most effective sequential algorithms for the CVRP. HGS and SISR obtain an average gap of 0.11\% and 0.20\%, respectively, when executed for $|V_c| \times 240 / 100$ seconds on the authors machines; see \citet{vidal22}.

\subsection{Testing on the \texorpdfstring{$\mathbb{B}$}{B} Instances}
\begin{figure}
\centering
\caption{Performance charts of \FILOX and FILO2 on the $\mathbb{B}$ dataset when performing standard (left) and long (right) runs. The individual numbers identify how many solvers have been used to run the \FILOX algorithm.}
\begin{tikzpicture}[]
\scriptsize

\begin{groupplot}[
    group style={group size= 2 by 1, horizontal sep=2cm},
    height=180pt,
    width=220pt,
    ymajorgrids=true,
    yminorgrids=true,
    minor y tick num=4,
    minor grid style={line width=.01pt,draw=black!10},
    major grid style={line width=.01pt,draw=black!30},
    ]

    \nextgroupplot[
	title={},
	xlabel={Average computing time (sec)},
	ylabel={Average \% gap},
    extra y ticks={1.08191, 1.062},
    extra y tick labels={1.081, 1.062},
    extra y tick style={grid=none},
    extra x ticks={88.97},
    extra x tick labels={89},
    extra x tick style={grid=none},
    yticklabel=\pgfkeys{/pgf/number format/.cd,fixed,precision=3,zerofill}\pgfmathprintnumber{\tick}
    ]

\addplot[] coordinates {

(14.7,1.07969)
(16.64,1.0709)
(24.1,1.07023)
(36.53,1.06606)
(88.97,1.06223)

};

\addplot[only marks, mark=square*, mark size=1pt, point meta=explicit symbolic, nodes near coords, text width=3cm, align=center, font=\tiny\linespread{0.8}\selectfont] coordinates {

(88.97,1.06223)[FILO2]
(53.91,1.06728)[2]
(36.53,1.06606)[3]
(28.91,1.07534)[4]
(24.1,1.07023)[5]
(20.96,1.07686)[6]
(15.36,1.08191)[9]
(14.7,1.07969)[10]

};

\addplot[only marks, mark=square*, mark size=1pt, point meta=explicit symbolic, nodes near coords, text width=3cm, align=center, every node near coord/.append style={yshift=-0.35cm}, font=\tiny\linespread{0.8}\selectfont] coordinates {

};

\addplot[only marks, mark=square*, mark size=1pt, point meta=explicit symbolic, nodes near coords, text width=3cm, align=center, every node near coord/.append style={xshift=-0.25cm, yshift=-0.16cm}, font=\tiny\linespread{0.8}\selectfont] coordinates {

(16.64,1.0709)[8]

};

\addplot[only marks, mark=square*, mark size=1pt, point meta=explicit symbolic, nodes near coords, text width=3cm, align=center, every node near coord/.append style={xshift=+0.20cm, yshift=-0.16cm}, font=\tiny\linespread{0.8}\selectfont] coordinates {

(18.51,1.07201)[7]

};

\nextgroupplot[
	title={},	
	xlabel={Average computing time (sec)},
	ylabel={Average \% gap},
        scaled y ticks=false,
    extra y ticks={0.37279},
    extra y tick labels={0.373},
    extra y tick style={grid=none},
yticklabel=\pgfkeys{/pgf/number format/.cd,fixed,precision=3,zerofill}\pgfmathprintnumber{\tick},
    ]

\addplot[] coordinates {

(168.91,0.36425)
(212.75,0.35542)

};

\addplot[only marks, mark=square*, mark size=1pt, point meta=explicit symbolic, nodes near coords, text width=3cm, align=center, font=\tiny\linespread{0.8}\selectfont] coordinates {

(1006.78,0.36322)[FILO2]
(598.21,0.36657)[2]
(419.8,0.35707)[3]
(332.15,0.36505)[4]
(276.97,0.357)[5]
(241.68,0.36495)[6]
(192.8,0.36482)[8]
(178.09,0.37279)[9]

};

\addplot[only marks, mark=square*, mark size=1pt, point meta=explicit symbolic, nodes near coords, text width=3cm, align=center, every node near coord/.append style={yshift=-0.35cm}, font=\tiny\linespread{0.8}\selectfont] coordinates {

};

\addplot[only marks, mark=square*, mark size=1pt, point meta=explicit symbolic, nodes near coords, text width=3cm, align=center, every node near coord/.append style={xshift=-0.25cm, yshift=-0.16cm}, font=\tiny\linespread{0.8}\selectfont] coordinates {

(168.91,0.36425)[10]

};

\addplot[only marks, mark=square*, mark size=1pt, point meta=explicit symbolic, nodes near coords, text width=3cm, align=center, every node near coord/.append style={xshift=+0.20cm, yshift=-0.16cm}, font=\tiny\linespread{0.8}\selectfont] coordinates {

(212.75,0.35542)[7]

};

\end{groupplot}

	\end{tikzpicture}
\label{fig:FILO2x.performance.b}
\end{figure}
The $\mathbb{B}$ dataset, introduced in \citet{arnold2019xxl}, contains ten very large-scale instances with up to thirty thousand customers.
These instances are modeled after real-world parcel distribution problems occurring in Belgium.
In some instances, the depot is located centrally with respect to the customers and relatively short routes are needed to serve them.
In other instances, the depot is eccentric with respect to the service zones, requiring much longer routes to visit the customers.

The performance charts shown in Figure \ref{fig:FILO2x.performance.b} highlight that when performing standard runs, the sequential FILO2 algorithm provides solutions that have, on average, a slightly better quality compared to those obtained by the parallel version.
On the other hand, when longer runs are performed, some parallel versions are able to provide better results. However, despite these minor differences, statistical analyses show that \FILOX and FILO2 both provide results of similar quality; see \ref{appendix:b}.

Finally, for a better comparison with state-of-the-art algorithms on this dataset, consider also the knowledge-guided local search algorithm for large-scale instances (KGLS$^\text{XXL}$) described in \citet{arnold2019xxl}, which is a deterministic algorithm originally proposed to solve the $\mathbb{B}$ instances. Short runs of $|V_c|/1000 \times 180$ seconds of KGLS$^\text{XXL}$ generate solutions having an average gap of $2.63\%$. Four times longer runs are capable of reducing the average gap to $1.77\%$. The percentage gaps are taken from \citet{accorsi2024}. Finally, \citet{ails2} report an average gap of $0.10\%$ obtained by their adaptive iterated local search heuristic, called AILS-II, in 600 minutes of computations.

\subsection{Testing on the \texorpdfstring{$\mathbb{I}$}{I} Instances}
The $\mathbb{I}$ dataset, originally introduced in \citet{accorsi2024}, contains twenty extremely large-scale instances with a number of customers ranging from $20,000$ to $1,000,000$.
These instances are based on projected coordinates of randomly sampled addresses in the various Italian regions with the depot located at the regional capital city position.
The customer demand and the vehicle capacity are defined to have half of the instances where we can expect relatively short routes and half where longer ones are required.

\begin{figure}
\centering
\caption{Performance charts of \FILOX and FILO2 on the $\mathbb{I}$ dataset when performing standard (left) and long (right) runs. The individual numbers identify how many solvers have been used to run the \FILOX algorithm.}
\begin{tikzpicture}[]
\scriptsize

\begin{groupplot}[
    group style={group size= 2 by 1, horizontal sep=2cm},
    height=180pt,
    width=220pt,
    ymajorgrids=true,
    yminorgrids=true,
    minor y tick num=4,
    minor grid style={line width=.01pt,draw=black!10},
    major grid style={line width=.01pt,draw=black!30},
    ]

    \nextgroupplot[
	title={},
	xlabel={Average computing time (sec)},
	ylabel={Average \% gap},
    yticklabel=\pgfkeys{/pgf/number format/.cd,fixed,precision=3,zerofill}\pgfmathprintnumber{\tick}
    ]

\addplot[] coordinates {

(64.965,0.69767)
(72.635,0.69411)
(84.735,0.69222)

};

\addplot[only marks, mark=square*, mark size=1pt, point meta=explicit symbolic, nodes near coords, text width=3cm, align=center, font=\tiny\linespread{0.8}\selectfont] coordinates {

(310.22,0.69832)[FILO2]
(186.365,0.69503)[2]
(132.5,0.6995)[3]
(107.195,0.69709)[4]
(92.725,0.69419)[5]

(76.06,0.69541)[7]
(67.075,0.6998)[9]
(64.965,0.69767)[10]

};

\addplot[only marks, mark=square*, mark size=1pt, point meta=explicit symbolic, nodes near coords, text width=3cm, align=center, every node near coord/.append style={yshift=-0.35cm}, font=\tiny\linespread{0.8}\selectfont] coordinates {

};

\addplot[only marks, mark=square*, mark size=1pt, point meta=explicit symbolic, nodes near coords, text width=3cm, align=center, every node near coord/.append style={xshift=-0.25cm, yshift=-0.16cm}, font=\tiny\linespread{0.8}\selectfont] coordinates {

(72.635,0.69411)[8]

};

\addplot[only marks, mark=square*, mark size=1pt, point meta=explicit symbolic, nodes near coords, text width=3cm, align=center, every node near coord/.append style={xshift=+0.20cm, yshift=-0.16cm}, font=\tiny\linespread{0.8}\selectfont] coordinates {

(84.735,0.69222)[6]

};

\nextgroupplot[
	title={},	
	xlabel={Average computing time (sec)},
	ylabel={Average \% gap},
    extra y ticks={0.29887},
    extra y tick labels={0.299},
    extra y tick style={grid=none},
yticklabel=\pgfkeys{/pgf/number format/.cd,fixed,precision=3,zerofill}\pgfmathprintnumber{\tick}
    ]

\addplot[] coordinates {

(399.55,0.29433)
(419.99,0.29364)
(789.035,0.29355)

};

\addplot[only marks, mark=square*, mark size=1pt, point meta=explicit symbolic, nodes near coords, text width=3cm, align=center, font=\tiny\linespread{0.8}\selectfont] coordinates {

(2493.38,0.29367)[FILO2]
(1433.13,0.29518)[2]
(1006.195,0.29887)[3]
(789.035,0.29355)[4]
(659.905,0.29405)[5]
(570.64,0.29697)[6]
(502.24,0.29395)[7]
(464.68,0.29513)[8]

(399.55,0.29433)[10]

};

\addplot[only marks, mark=square*, mark size=1pt, point meta=explicit symbolic, nodes near coords, text width=3cm, align=center, every node near coord/.append style={yshift=-0.35cm}, font=\tiny\linespread{0.8}\selectfont] coordinates {

};

\addplot[only marks, mark=square*, mark size=1pt, point meta=explicit symbolic, nodes near coords, text width=3cm, align=center, every node near coord/.append style={xshift=-0.25cm, yshift=-0.16cm}, font=\tiny\linespread{0.8}\selectfont] coordinates {

(419.99,0.29364)[9]

};

\addplot[only marks, mark=square*, mark size=1pt, point meta=explicit symbolic, nodes near coords, text width=3cm, align=center, every node near coord/.append style={xshift=+0.20cm, yshift=-0.16cm}, font=\tiny\linespread{0.8}\selectfont] coordinates {

};

\end{groupplot}

	\end{tikzpicture}
\label{fig:FILO2x.performance.i}
\end{figure}

Figure \ref{fig:FILO2x.performance.i} compares the computational results of \FILOX and FILO2.
In contrast to what is happening for the other datasets, several parallel versions are capable of finding slightly better solutions during standard runs. The results for long runs align with those obtained on other datasets.
However, as before, deeper statistical analyses show that the quality of the final solutions obtained by the parallel version does not differ from what the sequential algorithm can achieve; see \ref{appendix:i}. 

\section{Algorithmic Components Analysis} \label{sec:analysis}
In this section, we provide more detailed analyses and interpretations of the algorithm behavior, strengths, and weaknesses.
The analyses always consider computational results obtained by averaging ten distinct runs.

\subsection{Speedup}
The speedup, defined as the ratio between the computing times of the sequential and parallel versions, is a key property that justifies the design of a parallel approach.
As described in Section \ref{sec:filo2-and-filo}, a sequential execution of FILO2 consists of (i) a preprocessing instance that sets up fundamental data structures, (ii) a construction phase that builds an initial feasible solution, (iii) a greedy estimation of the routes required by a solution, (iv) a possible application of the route minimization procedure and, finally, (v) the application of the core optimization procedure.

Steps (i) and (v) are the two most time consuming steps performed by the algorithm, with (i) being negligible when solving moderate size instances, but becoming extremely costly when approaching extremely large-scale ones.
In particular, for standard FILO2 runs, (i) and (v) alone take more than $99\%$ of the total computing time when solving the $\mathbb{X}$ and $\mathbb{B}$ instances. This value decreases to $94\%$ when tackling the larger $\mathbb{I}$ dataset.
Clearly, when performing longer FILO2 runs, with ten times more core optimization procedure iterations, the time for (v) increases approximately tenfold, and (i) and (v) now take more than $99\%$ of the total computing time for all datasets.
Detailed statistics are available in the appendix in Tables \ref{tab:x-time-per-procedure} and \ref{tab:x-long-time-per-procedure} for the $\mathbb{X}$ instances, \ref{tab:b-time-per-procedure} and \ref{tab:b-long-time-per-procedure} for the $\mathbb{B}$ instances, and \ref{tab:i-time-per-procedure} and \ref{tab:i-long-time-per-procedure} for the $\mathbb{I}$ instances.

\FILOX takes advantage of multiple processors to parallelize these most time-consuming (i) and (v) procedures.
Table \ref{tab:speedup-by-procedure} shows the speedup for these procedures averaged over instances and run configurations.
\begin{table}
    \centering
    \footnotesize
    \caption{Speedup of the main algorithm procedures averaged over instances and run configurations for the different datasets. For reference, the Fraction column reports the percentage of time spent during the procedure, averaged across solvers. For the $\mathbb{X}$ instances we do not report results with 15 solvers as some of the smallest instances cannot be efficiently solved due to the high contention among solvers; see Section \ref{sec:analysis:infeasible-changes}.}
    \label{tab:speedup-by-procedure}
    \begin{tabular}{@{}c l rrrrrrrrrrr@{}}
        \toprule
        & Solvers & 2 & 3 & 4 & 5 & 6 & 7 & 8 & 9 & 10 & 15 & Fraction \\
        \midrule
        
        \multirow{2}{*}{$\mathbb{X}$} 
        & Instance preprocessing & 1.51 & 1.95 & 2.36 & 2.73 & 3.10 & 3.44 & 3.81 & 4.02 & 4.32 & \multicolumn{1}{c}{-} & 0.03$\pm$0.00 \\
        & Core optimization & 1.73 & 2.46 & 3.14 & 3.78 & 4.30 & 4.79 & 5.25 & 5.66 & 6.00 & \multicolumn{1}{c}{-} & 99.07$\pm$0.32 \\

        \midrule
        
        \multirow{2}{*}{$\mathbb{B}$} 
        & Instance preprocessing & 1.87 & 2.76 & 3.64 & 4.49 & 5.24 & 6.08 & 6.83 & 7.61 & 8.22 & 11.92 & 1.21$\pm$0.13\\
        & Core optimization & 1.68 & 2.40 & 3.03 & 3.64 & 4.17 & 4.74 & 5.23 & 5.67 & 5.97 & 7.34 & 96.90$\pm$0.64\\

        \midrule
        
        \multirow{2}{*}{$\mathbb{I}$} 
        & Instance preprocessing & 1.91 & 2.88 & 3.78 & 4.70 & 5.52 & 6.50 & 7.28 & 8.27 & 9.00 & 13.49 & 15.22$\pm$3.13\\
        & Core optimization & 1.74 & 2.48 & 3.18 & 3.81 & 4.42 & 5.04 & 5.47 & 6.06 & 6.38 & 8.27 & 70.41$\pm$2.37\\
        \bottomrule
    \end{tabular}
\end{table}
For the $\mathbb{B}$ and $\mathbb{I}$ instances, we see an almost linear speedup of the instance preprocessing procedure. This is much less pronounced for the $\mathbb{X}$ dataset, as this procedure already takes an incredibly short computing time of approximately 30 milliseconds when performed sequentially, and the time at this scale very often includes noise components not related to the algorithm (e.g., operating system routine operations).
On the other hand, the core optimization procedure exhibits a sub-linear, but steady, speedup increase that becomes increasingly marginal as we move to a higher number of solvers.

Ideally, the core optimization procedure should also scale linearly with the number of solvers, especially on the largest $\mathbb{I}$ instances.
However, there are several factors that could contribute to this sub-linear behavior. 
First, as described in Section \ref{sec:analysis:infeasible-changes}, the probability of generating a conflicting change grows steadily as the number of solvers increases, and whenever a change is discarded, an additional core optimization iteration must be performed.
Moreover, hardware characteristics such as CPU-level cache contention among threads begin to play an important role, especially for a complex procedure such as the core optimization, where every solver manages and uses considerably large data structures.
Finally, while instance preprocessing does only need very limited synchronization, core optimization requires acquiring locks when interacting with the dispatcher and solver queues, and despite this being minimal, it adds up to the other factors.

\begin{figure}[b!]
\centering
\caption{Speedup distribution obtained by \FILOX while varying the number of solvers. Each group refers from left-to-right to the $\mathbb{X}$, $\mathbb{B}$, and $\mathbb{I}$ datasets. In addition, within every group, each plot represents the speedup obtained by regular (left) and long (right) runs. Average values are denoted by the $\times$ symbol. Results for the $\mathbb{X}$ dataset with 15 solvers are not available due to the high contention among solvers; see \ref{sec:analysis:infeasible-changes}.}
\includegraphics{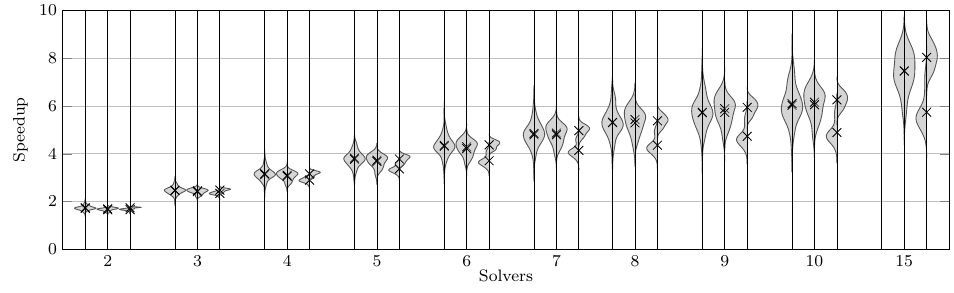}
\label{fig:filo2x.boxplots.speedup}
\end{figure}

Figure \ref{fig:filo2x.boxplots.speedup} shows the overall \FILOX speedup distribution obtained by standard and long runs. 
The graph shows that the algorithm speedup is similar across different datasets, except when solving $\mathbb{I}$ instances with standard runs.
In this case, the $6\%$ sequential processing still happening significantly reduces the overall algorithm speedup. 

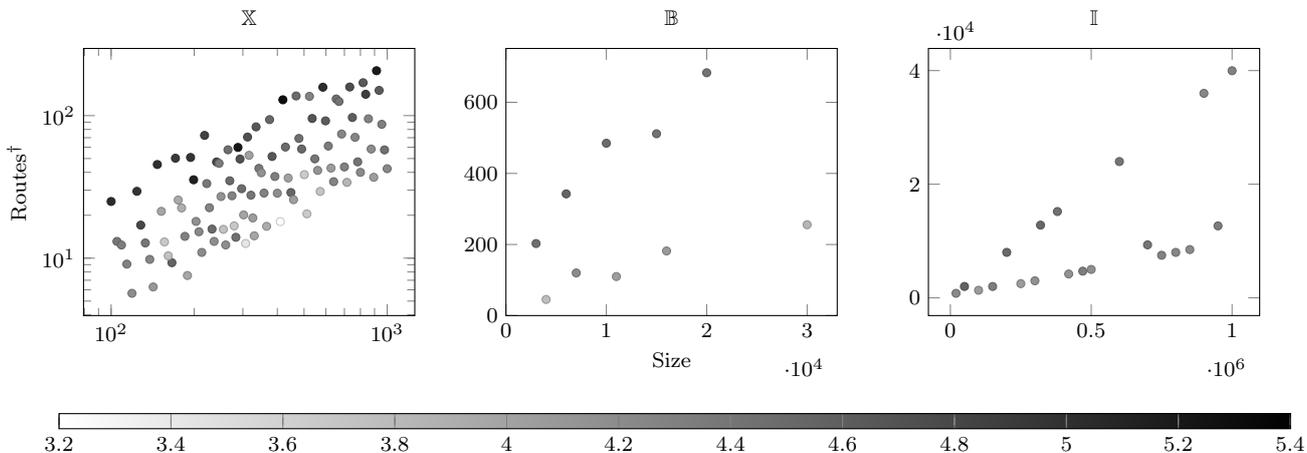
\begin{figure}[!b]
\centering
\caption{Median speedup averaged across parallel variants with 2 to 10 solvers. Dots represent instances. The $y$-axis denoted as Routes$^\dagger$ is defined as the sum of customer demands over the vehicle capacity, and provides a rough estimate on the number of routes required by a solution. Note that for better visualization, both axis for the $\mathbb{X}$ chart are logarithmic.}
\begin{tikzpicture}
\scriptsize
  \begin{groupplot}[
      group style={
          group size=3 by 1,
          horizontal sep=1.2cm,
      },
      width=0.33\textwidth,
      colormap={}{ gray(0cm)=(1); gray(1cm)=(0);},
      point meta min=3.2,
      point meta max=5.4,
    ]
    \nextgroupplot[title={$\mathbb{X}$}, xlabel={}, ylabel={Routes$^\dagger$}, xmode=log, ymode=log]
    \addplot[scatter, only marks, point meta=explicit, mark size=1.5pt,
    ] coordinates { 
(419,129.16667)[5.27448] 
(288,59.82772)[5.19252] 
(915,206.54545)[5.16452] 
(199,35.4801)[5.11331] 
(100,24.98544)[5.01354] 
(585,158.21429)[5.00054] 
(147,45.38889)[4.99032] 
(124,29.44681)[4.93916] 
(194,50.87293)[4.93837] 
(171,50.26087)[4.88597] 
(836,141.15909)[4.85349] 
(218,72.66667)[4.82618] 
(241,47.28571)[4.79615] 
(128,17.05128)[4.779] 
(732,158.76)[4.72789] 
(312,70.74597)[4.70046] 
(293,49.59649)[4.68651] 
(535,95.47978)[4.66736] 
(383,51.69681)[4.64517] 
(335,83.35468)[4.6397] 
(818,170.15084)[4.61444] 
(935,150.31159)[4.61406] 
(598,91.87064)[4.60693] 
(748,97.00505)[4.59099] 
(166,9.29323)[4.53451] 
(448,28.87902)[4.50943] 
(375,93.75)[4.49871] 
(490,58.31075)[4.49141] 
(269,34.90427)[4.47235] 
(468,137.38672)[4.46908] 
(232,15.99525)[4.4683] 
(297,30.70909)[4.45834] 
(428,60.18843)[4.45552] 
(265,57.57143)[4.44129] 
(654,130.8)[4.43514] 
(479,69.19231)[4.43117] 
(283,14.00917)[4.43108] 
(978,57.48397)[4.41632] 
(669,125.68992)[4.39677] 
(782,47.37139)[4.39618] 
(547,49.72727)[4.37709] 
(321,27.65323)[4.37679] 
(133,12.78383)[4.37199] 
(105,13.10667)[4.36561] 
(222,33.35135)[4.35951] 
(343,42.65574)[4.35913] 
(246,46.26866)[4.32019] 
(109,12.36364)[4.31719] 
(612,61.18929)[4.31353] 
(956,86.90909)[4.30219] 
(392,37.42308)[4.29622] 
(700,43.64368)[4.27861] 
(185,14.22279)[4.27845] 
(114,9.08284)[4.27402] 
(358,28.61765)[4.26915] 
(855,95.0)[4.26726] 
(523,136.536)[4.26646] 
(640,34.40333)[4.26202] 
(274,27.4)[4.2581] 
(227,22.58442)[4.25572] 
(138,9.80189)[4.24155] 
(765,70.45181)[4.24113] 
(684,74.26225)[4.24046] 
(1000,42.41985)[4.2199] 
(208,15.31683)[4.21793] 
(213,10.97034)[4.19377] 
(800,40.0)[4.19305] 
(119,5.66667)[4.18681] 
(875,58.29058)[4.1831] 
(400,28.55705)[4.1789] 
(260,12.37835)[4.16812] 
(236,13.11111)[4.1644] 
(203,18.10407)[4.15208] 
(350,39.71789)[4.14248] 
(250,27.05797)[4.13319] 
(560,41.28378)[4.11923] 
(142,6.28151)[4.08199] 
(626,42.84545)[4.04555] 
(326,19.15625)[4.04174] 
(894,36.92566)[4.0406] 
(458,25.70976)[4.03023] 
(316,52.66667)[4.0071] 
(438,36.5)[3.99588] 
(302,20.10957)[3.9838] 
(189,7.55797)[3.96482] 
(330,14.34783)[3.96477] 
(152,21.30556)[3.96203] 
(175,25.57746)[3.93021] 
(366,16.72936)[3.91187] 
(180,22.5)[3.88125] 
(715,34.02781)[3.82848] 
(161,10.38671)[3.72158] 
(255,15.9298)[3.71939] 
(572,29.39048)[3.70438] 
(279,16.83854)[3.67064] 
(156,13.0)[3.67035] 
(501,38.53846)[3.6562] 
(512,20.45775)[3.65381] 
(307,12.6748)[3.41632] 
(410,18.06019)[3.22734] 

};

    \nextgroupplot[title={$\mathbb{B}$}, xlabel={Size}, xmin=0, ymin=0]
    \addplot[scatter, only marks, point meta=explicit, mark size=1.5pt] coordinates { 
(6000,342.46667)[4.52199] 
(10000,484.91429)[4.46235] 
(3000,202.72)[4.44272] 
(15000,511.62)[4.4074] 
(20000,683.24)[4.40192] 
(7000,119.93)[4.19796] 
(16000,181.82667)[4.06682] 
(11000,109.85294)[4.0328] 
(30000,255.61)[3.82111] 
(4000,45.32)[3.75809] 

    };

    \nextgroupplot[title={$\mathbb{I}$}, xlabel={}]
    \addplot[scatter, only marks, point meta=explicit, mark size=1.5pt] coordinates { 
(50000,2006.38)[4.54207] 
(320000,12800.3)[4.49362] 
(200000,8005.68)[4.48387] 
(380000,15197.88)[4.42531] 
(700000,9334.58)[4.38542] 
(600000,23990.84)[4.38536] 
(950000,12668.02667)[4.32035] 
(150000,2001.73333)[4.31796] 
(470000,4699.96)[4.31016] 
(750000,7497.045)[4.28045] 
(1000000,39978.62)[4.27849] 
(800000,8002.89)[4.23657] 
(900000,35999.8)[4.22796] 
(20000,799.38)[4.2179] 
(850000,8497.5)[4.20347] 
(500000,5002.225)[4.15724] 
(300000,3003.335)[4.14158] 
(420000,4204.015)[4.13012] 
(250000,2499.07)[4.0605] 
(100000,1336.6)[4.05646] 

};

  \end{groupplot}

  \begin{axis}[
      hide axis,
      scale only axis,
      height=0pt,
      width=0pt,
      colormap={}{ gray(0cm)=(1); gray(1cm)=(0);},
      point meta min=3.2,
      point meta max=5.4,
      colorbar,
      colorbar horizontal,
      at={($(group c1r1.south)!-0.225!(group c3r1.south)$)},
      anchor=north,
      yshift=-1.0cm,
      colorbar style={
          width=0.9\textwidth,
          height=0.2cm,
      }
    ]
  \end{axis}
\end{tikzpicture}
\label{fig:filo2x.xyz.overall.speedup}
\end{figure}

Finally, Figure \ref{fig:filo2x.xyz.overall.speedup} provides some insight into the algorithm speedup with respect to some instance characteristics. 
In particular, the chart shows that the speedup does not depend on the instance size but rather on the estimated number of routes required by a solution. 
The more routes, the higher the expected speedup. This is especially visible for the $\mathbb{X}$ and $\mathbb{B}$ instances given the relatively small number of routes they require compared to the $\mathbb{I}$ instances, where the speedup differences are less pronounced. 
As will be illustrated in Section \ref{sec:analysis:infeasible-changes}, this behavior is closely linked to the likelihood of generating infeasible changes when few routes are available.

\subsection{Infeasible/Inapplicable Changes} \label{sec:analysis:infeasible-changes}
The main characteristic of \FILOX consists in allowing solvers to freely optimize localized solution portions without imposing any centralized guidance or rigid boundaries.
As a result, several solvers may ultimately optimize overlapping areas.
In general, the resulting changes can still produce CVRP-feasible solutions depending on when they were generated and the order in which they are sent to the dispatcher.

The asynchronous nature of \FILOX makes every solver have a reference solution that belongs to the same shared search trajectory but may be at a different point. As a result, it is very likely that all solvers will generate changes using slightly different reference solutions.
In general, incorporating a change that was generated on a slightly different solution can still produce a CVRP-feasible solution depending on the area affected by the change and on the entity of the difference.

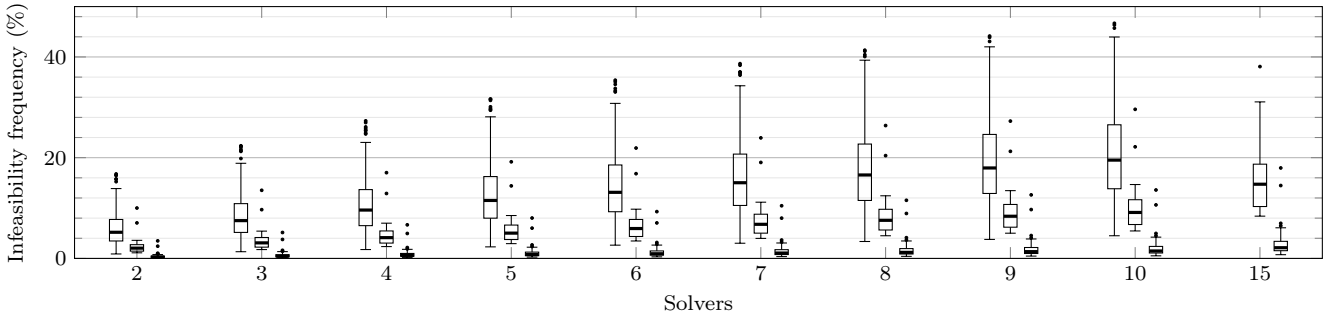
\begin{figure}
\centering
\caption{Percentage of candidate changes leading to infeasible solutions for different instances and run configurations. Each box plot group refers from left-to-right to the $\mathbb{X}$, $\mathbb{B}$, and $\mathbb{I}$ datasets. The results for 15 solvers only refer to the $\mathbb{B}$ and $\mathbb{I}$ instances due to the high contention among solvers when solving the $\mathbb{X}$ instances.}
\scriptsize
\makeatletter
\pgfplotsset{
    boxplot/draw/average/.code={ 
        \draw [/pgfplots/boxplot/every average/.try]
        \pgfextra
        \pgftransformshift{%
            \pgfplotsboxplotpointabbox
            {\pgfplotsboxplotvalue{average}}
            {0.5}%
        }%
        \endpgfextra
        ;
    }
}
\makeatother

\begin{tikzpicture}
\scriptsize

\begin{axis}[
ylabel style={align=center}, 
ylabel={Infeasibility frequency (\%)},
xlabel={Solvers},
ymajorgrids=true,
yminorgrids=true,
minor y tick num=4,
minor grid style={line width=.01pt,draw=black!10},
major grid style={line width=.01pt,draw=black!30},
clip=false,
boxplot/draw direction=y,
boxplot/variable width,
boxplot/every median/.style={black,very thick,solid},
width=\textwidth,
height=140pt,
ylabel style={align=center}, 
y tick label style={align=right},
x tick label style={align=center},
xtick={0, 1.5, 3.0, 4.5, 6.0, 7.5, 9, 10.5, 12, 13.5},
xticklabels={2, 3, 4, 5, 6, 7, 8, 9, 10, 15},
xmin=-0.75,
xmax=14.25,
ymin=0,
ymax=50,
scatter/classes={ a={mark=star}, b={mark=*}}]


\addplot[mark=*, mark size=0.5pt,boxplot, boxplot prepared={draw position=-0.25,
median=5.191,upper quartile=7.732,lower quartile=3.463,upper whisker=13.869,lower whisker=0.884,sample size=1}]
coordinates{
(-0.25,16.471)
(-0.25,16.547)
(-0.25,16.772)
(-0.25,15.785)
(-0.25,15.236)
(-0.25,15.701)
};
\addplot[mark=*, mark size=0.5pt,boxplot, boxplot prepared={draw position=1.25,
median=7.503,upper quartile=10.876,lower quartile=5.161,upper whisker=18.887,lower whisker=1.331,sample size=1}]
coordinates{
(1.25,22.264)
(1.25,22.363)
(1.25,22.098)
(1.25,21.66)
(1.25,19.837)
(1.25,21.369)
(1.25,21.25)
};
\addplot[mark=*, mark size=0.5pt,boxplot, boxplot prepared={draw position=2.75,
median=9.602,upper quartile=13.663,lower quartile=6.506,upper whisker=23.03,lower whisker=1.76,sample size=1}]
coordinates{
(2.75,26.972)
(2.75,27.314)
(2.75,25.493)
(2.75,26.106)
(2.75,24.884)
(2.75,24.711)
(2.75,25.34)
(2.75,25.816)
};
\addplot[mark=*, mark size=0.5pt,boxplot, boxplot prepared={draw position=4.25,
median=11.508,upper quartile=16.253,lower quartile=7.989,upper whisker=28.113,lower whisker=2.293,sample size=1}]
coordinates{
(4.25,31.382)
(4.25,31.627)
(4.25,31.63)
(4.25,29.773)
(4.25,29.423)
(4.25,29.555)
(4.25,30.104)
};
\addplot[mark=*, mark size=0.5pt,boxplot, boxplot prepared={draw position=5.75,
median=13.122,upper quartile=18.554,lower quartile=9.269,upper whisker=30.782,lower whisker=2.631,sample size=1}]
coordinates{
(5.75,35.075)
(5.75,35.389)
(5.75,34.559)
(5.75,33.295)
(5.75,33.052)
(5.75,33.742)
};
\addplot[mark=*, mark size=0.5pt,boxplot, boxplot prepared={draw position=7.25,
median=15.027,upper quartile=20.712,lower quartile=10.516,upper whisker=34.281,lower whisker=3.014,sample size=1}]
coordinates{
(7.25,38.355)
(7.25,38.668)
(7.25,37.011)
(7.25,36.996)
(7.25,36.388)
(7.25,36.68)
};
\addplot[mark=*, mark size=0.5pt,boxplot, boxplot prepared={draw position=8.75,
median=16.567,upper quartile=22.707,lower quartile=11.508,upper whisker=39.357,lower whisker=3.359,sample size=1}]
coordinates{
(8.75,41.14)
(8.75,41.344)
(8.75,40.443)
(8.75,40.083)
};
\addplot[mark=*, mark size=0.5pt,boxplot, boxplot prepared={draw position=10.25,
median=17.958,upper quartile=24.618,lower quartile=12.896,upper whisker=42.009,lower whisker=3.772,sample size=1}]
coordinates{
(10.25,43.895)
(10.25,44.149)
(10.25,43.083)
};
\addplot[mark=*, mark size=0.5pt,boxplot, boxplot prepared={draw position=11.75,
median=19.519,upper quartile=26.534,lower quartile=13.833,upper whisker=43.963,lower whisker=4.5,sample size=1}]
coordinates{
(11.75,46.352)
(11.75,46.684)
(11.75,45.724)
};


\addplot[mark=*, mark size=0.5pt,boxplot, boxplot prepared={draw position=0.0,
median=2.04,upper quartile=2.728,lower quartile=1.444,upper whisker=3.59,lower whisker=1.113,sample size=1}]
coordinates{
(0.0,7.052)
(0.0,10.004)
};
\addplot[mark=*, mark size=0.5pt,boxplot, boxplot prepared={draw position=1.5,
median=3.087,upper quartile=4.153,lower quartile=2.237,upper whisker=5.418,lower whisker=1.756,sample size=1}]
coordinates{
(1.5,9.68)
(1.5,13.527)
};
\addplot[mark=*, mark size=0.5pt,boxplot, boxplot prepared={draw position=3.0,
median=4.125,upper quartile=5.455,lower quartile=3.044,upper whisker=6.996,lower whisker=2.35,sample size=1}]
coordinates{
(3.0,12.892)
(3.0,17.023)
};
\addplot[mark=*, mark size=0.5pt,boxplot, boxplot prepared={draw position=4.5,
median=5.021,upper quartile=6.622,lower quartile=3.751,upper whisker=8.514,lower whisker=2.934,sample size=1}]
coordinates{
(4.5,14.402)
(4.5,19.181)
};
\addplot[mark=*, mark size=0.5pt,boxplot, boxplot prepared={draw position=6.0,
median=5.934,upper quartile=7.735,lower quartile=4.353,upper whisker=9.775,lower whisker=3.461,sample size=1}]
coordinates{
(6.0,16.825)
(6.0,21.914)
};
\addplot[mark=*, mark size=0.5pt,boxplot, boxplot prepared={draw position=7.5,
median=6.771,upper quartile=8.776,lower quartile=5.016,upper whisker=11.163,lower whisker=3.973,sample size=1}]
coordinates{
(7.5,19.056)
(7.5,23.933)
};
\addplot[mark=*, mark size=0.5pt,boxplot, boxplot prepared={draw position=9.0,
median=7.565,upper quartile=9.755,lower quartile=5.641,upper whisker=12.435,lower whisker=4.5,sample size=1}]
coordinates{
(9.0,20.407)
(9.0,26.379)
};
\addplot[mark=*, mark size=0.5pt,boxplot, boxplot prepared={draw position=10.5,
median=8.374,upper quartile=10.714,lower quartile=6.198,upper whisker=13.456,lower whisker=5.006,sample size=1}]
coordinates{
(10.5,21.259)
(10.5,27.263)
};
\addplot[mark=*, mark size=0.5pt,boxplot, boxplot prepared={draw position=12.0,
median=9.125,upper quartile=11.654,lower quartile=6.73,upper whisker=14.671,lower whisker=5.453,sample size=1}]
coordinates{
(12.0,22.159)
(12.0,29.601)
};
\addplot[mark=*, mark size=0.5pt,boxplot, boxplot prepared={draw position=13.5,
median=14.731,upper quartile=18.711,lower quartile=10.302,upper whisker=31.078,lower whisker=8.392,sample size=1}]
coordinates{
(13.5,38.089)
};


\addplot[mark=*, mark size=0.5pt,boxplot, boxplot prepared={draw position=0.25,
median=0.286,upper quartile=0.456,lower quartile=0.203,upper whisker=0.761,lower whisker=0.091,sample size=1}]
coordinates{
(0.25,2.427)
(0.25,3.486)
(0.25,0.874)
(0.25,1.027)
(0.25,0.918)
};
\addplot[mark=*, mark size=0.5pt,boxplot, boxplot prepared={draw position=1.75,
median=0.456,upper quartile=0.737,lower quartile=0.324,upper whisker=1.349,lower whisker=0.149,sample size=1}]
coordinates{
(1.75,3.766)
(1.75,5.147)
(1.75,1.611)
(1.75,1.416)
};
\addplot[mark=*, mark size=0.5pt,boxplot, boxplot prepared={draw position=3.25,
median=0.611,upper quartile=0.989,lower quartile=0.433,upper whisker=1.801,lower whisker=0.204,sample size=1}]
coordinates{
(3.25,4.952)
(3.25,6.669)
(3.25,2.156)
(3.25,1.887)
};
\addplot[mark=*, mark size=0.5pt,boxplot, boxplot prepared={draw position=4.75,
median=0.774,upper quartile=1.252,lower quartile=0.547,upper whisker=2.226,lower whisker=0.26,sample size=1}]
coordinates{
(4.75,6.012)
(4.75,8.02)
(4.75,2.682)
(4.75,2.353)
};
\addplot[mark=*, mark size=0.5pt,boxplot, boxplot prepared={draw position=6.25,
median=0.926,upper quartile=1.486,lower quartile=0.655,upper whisker=2.646,lower whisker=0.311,sample size=1}]
coordinates{
(6.25,7.049)
(6.25,9.291)
(6.25,3.159)
(6.25,2.812)
};
\addplot[mark=*, mark size=0.5pt,boxplot, boxplot prepared={draw position=7.75,
median=1.065,upper quartile=1.736,lower quartile=0.763,upper whisker=3.063,lower whisker=0.362,sample size=1}]
coordinates{
(7.75,7.997)
(7.75,10.453)
(7.75,3.65)
(7.75,3.238)
};
\addplot[mark=*, mark size=0.5pt,boxplot, boxplot prepared={draw position=9.25,
median=1.203,upper quartile=1.951,lower quartile=0.875,upper whisker=3.465,lower whisker=0.41,sample size=1}]
coordinates{
(9.25,8.923)
(9.25,11.543)
(9.25,4.115)
(9.25,3.649)
};
\addplot[mark=*, mark size=0.5pt,boxplot, boxplot prepared={draw position=10.75,
median=1.344,upper quartile=2.167,lower quartile=0.976,upper whisker=3.86,lower whisker=0.46,sample size=1}]
coordinates{
(10.75,9.699)
(10.75,12.616)
(10.75,4.538)
(10.75,4.072)
};
\addplot[mark=*, mark size=0.5pt,boxplot, boxplot prepared={draw position=12.25,
median=1.476,upper quartile=2.382,lower quartile=1.077,upper whisker=4.223,lower whisker=0.514,sample size=1}]
coordinates{
(12.25,10.627)
(12.25,13.588)
(12.25,4.959)
(12.25,4.495)
};
\addplot[mark=*, mark size=0.5pt,boxplot, boxplot prepared={draw position=13.75,
median=2.125,upper quartile=3.389,lower quartile=1.533,upper whisker=6.082,lower whisker=0.739,sample size=1}]
coordinates{
(13.75,14.471)
(13.75,17.962)
(13.75,6.967)
(13.75,6.478)
};

\end{axis}

\end{tikzpicture}
\label{fig:filo2x.coreopt.boxplots.conflicts}
\end{figure}

However, the combination of overlapping optimization on non-synchronized reference solutions has a high probability of generating changes producing infeasible solutions.
This issue is mitigated by re-validating the feasibility of candidate changes received during the synchronization stage. 
Figure \ref{fig:filo2x.coreopt.boxplots.conflicts} shows the fraction of candidate changes that when applied to a solver reference solution during the synchronization stage generate an infeasible solution.
Note that because every solver traverses the same search trajectory, they will all encounter the same feasible and infeasible solutions. 

For the $\mathbb{X}$, the median infeasibility frequency increases to up to $20\%$ when employing 10 solvers, with peaks well above twice that value for some runs of instances X-n157-k13, X-n162-k11, X-n411-k19, and X-n308-k13. 
For the larger $\mathbb{B}$ instances, the infeasibility frequency shows a more compact set of values around the median that reaches the value of $8\%$ when using 10 solvers, with Leuven2 as an outlier during both standard and long runs. The variability among infeasibility frequency values grows considerably when employing 15 solvers.
Finally, the $\mathbb{I}$ dataset shows a more consistent behavior across the considered number of solvers, with a median infeasibility frequency around $2\%$ when using 15 solvers. Instance Valle-D-Aosta shows a considerably higher number of infeasibility frequency during both standard and long runs.

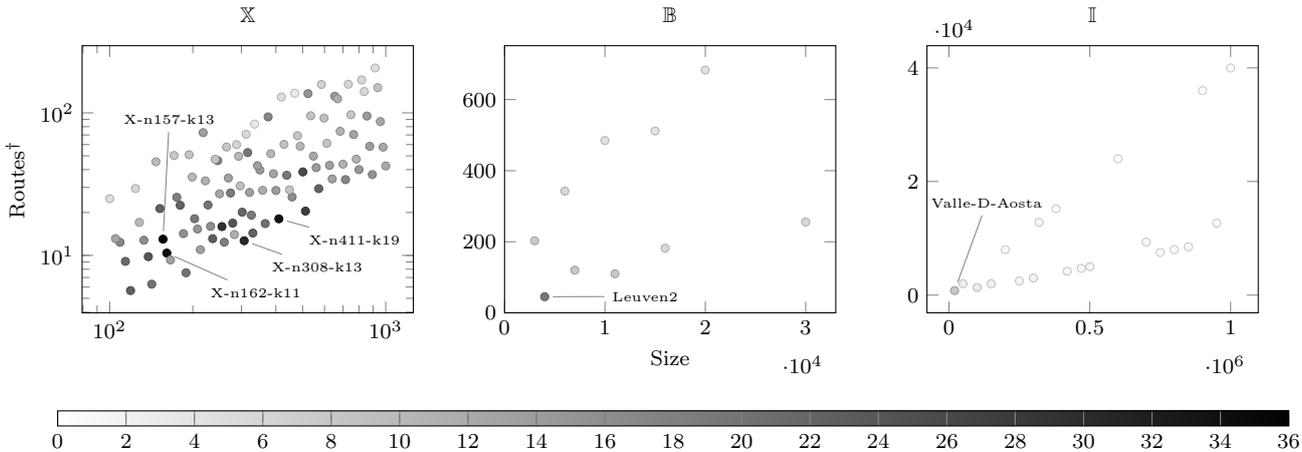
\begin{figure}[b!]
\centering
\caption{Median change infeasibility frequency averaged across parallel version with 2 to 10 solvers. Dots represent instances. The $y$-axis denoted as Routes$^\dagger$ is defined as the sum of customer demands over the vehicle capacity, and provides a rough estimate on the number of routes required by a solution. Note that for better visualization, both axis for the $\mathbb{X}$ chart are logarithmic.}
\begin{tikzpicture}
\scriptsize
  \begin{groupplot}[
      group style={
          group size=3 by 1,
          horizontal sep=1.2cm,
      },
      width=0.33\textwidth,
      colormap={}{ gray(0cm)=(1); gray(1cm)=(0);},
      point meta min=0.0,
      point meta max=36,
    ]
    \nextgroupplot[title={$\mathbb{X}$}, xlabel={}, ylabel={Routes$^\dagger$}, xmode=log, ymode=log]
    \addplot[scatter, only marks, point meta=explicit, mark size=1.5pt,
    ] coordinates { 
(156,13.0)[35.23211] 
(161,10.38671)[33.92705] 
(410,18.06019)[33.39704] 
(307,12.6748)[30.30662] 
(255,15.9298)[29.21452] 
(512,20.45775)[27.19838] 
(501,38.53846)[25.5326] 
(138,9.80189)[24.00943] 
(279,16.83854)[23.83262] 
(236,13.11111)[22.68587] 
(152,21.30556)[22.49183] 
(119,5.66667)[22.26313] 
(330,14.34783)[21.99485] 
(572,29.39048)[21.90047] 
(180,22.5)[21.719] 
(302,20.10957)[21.57792] 
(114,9.08284)[21.03852] 
(366,16.72936)[20.08799] 
(142,6.28151)[19.78536] 
(189,7.55797)[19.63009] 
(227,22.58442)[19.43145] 
(316,52.66667)[19.27629] 
(203,18.10407)[18.97505] 
(438,36.5)[18.86766] 
(326,19.15625)[18.85534] 
(274,27.4)[18.56] 
(246,46.26866)[18.30755] 
(260,12.37835)[18.05959] 
(175,25.57746)[17.89676] 
(715,34.02781)[17.12704] 
(375,93.75)[16.99626] 
(109,12.36364)[16.50198] 
(458,25.70976)[16.18669] 
(133,12.78383)[15.78321] 
(232,15.99525)[15.77434] 
(185,14.22279)[15.68268] 
(894,36.92566)[15.23026] 
(523,136.536)[15.00969] 
(208,15.31683)[14.91544] 
(800,40.0)[14.595] 
(654,130.8)[14.31285] 
(560,41.28378)[14.09419] 
(855,95.0)[14.02772] 
(400,28.55705)[14.02425] 
(350,39.71789)[13.95833] 
(875,58.29058)[13.59393] 
(213,10.97034)[13.37233] 
(626,42.84545)[13.3175] 
(218,72.66667)[13.2935] 
(321,27.65323)[13.27723] 
(343,42.65574)[13.10617] 
(269,34.90427)[13.07908] 
(166,9.29323)[13.04451] 
(640,34.40333)[12.86322] 
(105,13.10667)[12.7772] 
(765,70.45181)[12.55196] 
(392,37.42308)[12.50101] 
(250,27.05797)[12.33984] 
(956,86.90909)[12.3298] 
(547,49.72727)[12.27982] 
(1000,42.41985)[12.08322] 
(978,57.48397)[12.05069] 
(669,125.68992)[11.45924] 
(358,28.61765)[11.42936] 
(283,14.00917)[11.33366] 
(222,33.35135)[11.19687] 
(700,43.64368)[11.09113] 
(684,74.26225)[10.85847] 
(782,47.37139)[10.69787] 
(490,58.31075)[10.2267] 
(383,51.69681)[10.11358] 
(293,49.59649)[10.01761] 
(199,35.4801)[9.99828] 
(128,17.05128)[9.87931] 
(147,45.38889)[9.60738] 
(297,30.70909)[9.04092] 
(598,91.87064)[8.8801] 
(428,60.18843)[8.58324] 
(535,95.47978)[8.52731] 
(935,150.31159)[8.43933] 
(241,47.28571)[8.38294] 
(612,61.18929)[8.25308] 
(194,50.87293)[8.24434] 
(448,28.87902)[8.11314] 
(479,69.19231)[8.08793] 
(265,57.57143)[8.0508] 
(748,97.00505)[7.92066] 
(171,50.26087)[7.68548] 
(585,158.21429)[6.43493] 
(288,59.82772)[5.92255] 
(124,29.44681)[5.7327] 
(732,158.76)[5.60794] 
(836,141.15909)[5.53972] 
(818,170.15084)[5.53227] 
(312,70.74597)[5.30073] 
(100,24.98544)[4.96173] 
(419,129.16667)[4.64974] 
(915,206.54545)[4.53084] 
(335,83.35468)[3.03961] 
(468,137.38672)[2.96036] 
};

\node[
    pin={[font={\fontsize{3}{3}\selectfont}, 
        xshift=-28pt, 
        yshift=30pt
    ]45:X-n157-k13}
]
at (axis cs:156,13.0) {};

\node[
    pin={[font={\fontsize{3}{3}\selectfont}, 
        xshift=3pt, 
        yshift=-30pt
    ]45:X-n162-k11}
]
at (axis cs:161,10.38671) {};

\node[
    pin={[font={\fontsize{3}{3}\selectfont}, 
        xshift=-2pt, 
        yshift=-23pt
    ]45:X-n411-k19}
]
at (axis cs:410,18.06019) {};

\node[
    pin={[font={\fontsize{3}{3}\selectfont}, 
        xshift=-3pt, 
        yshift=-25pt
    ]45:X-n308-k13}
]
at (axis cs:307,12.6748) {};

    \nextgroupplot[title={$\mathbb{B}$}, xlabel={Size}, xmin=0, ymin=0]
    \addplot[scatter, only marks, point meta=explicit, mark size=1.5pt] coordinates {

(4000,45.32)[19.11866] 
(11000,109.85294)[7.78589] 
(7000,119.93)[7.73553] 
(3000,202.72)[7.54156] 
(30000,255.61)[5.53019] 
(16000,181.82667)[5.33018] 
(6000,342.46667)[5.17689] 
(10000,484.91429)[4.39658] 
(20000,683.24)[4.25474] 
(15000,511.62)[3.78304] 

    };

\node[
    pin={[font={\fontsize{3}{3}\selectfont}, 
        xshift=12pt, 
        yshift=-15pt
    ]45:Leuven2}
]
at (axis cs:4000,45.32) {};

    \nextgroupplot[title={$\mathbb{I}$}, xlabel={}]
    \addplot[scatter, only marks, point meta=explicit, mark size=1.5pt] coordinates { 

(20000,799.38)[8.0084] 
(100000,1336.6)[2.74673] 
(50000,2006.38)[2.34583] 
(150000,2001.73333)[1.92544] 
(300000,3003.335)[1.71072] 
(250000,2499.07)[1.38743] 
(500000,5002.225)[1.24482] 
(420000,4204.015)[1.17151] 
(320000,12800.3)[1.00164] 
(200000,8005.68)[0.91439] 
(700000,9334.58)[0.90779] 
(470000,4699.96)[0.87889] 
(950000,12668.02667)[0.83932] 
(850000,8497.5)[0.77377] 
(800000,8002.89)[0.64172] 
(750000,7497.045)[0.59126] 
(380000,15197.88)[0.55698] 
(600000,23990.84)[0.4773] 
(1000000,39978.62)[0.38163] 
(900000,35999.8)[0.33] 

};

\node[
    pin={[font={\fontsize{3}{3}\selectfont}, 
        xshift=12pt, 
        yshift=15pt
    ]90:Valle-D-Aosta}
]
at (axis cs:20000,799.38) {};

  \end{groupplot}

  \begin{axis}[
      hide axis,
      scale only axis,
      height=0pt,
      width=0pt,
      colormap={}{ gray(0cm)=(1); gray(1cm)=(0);},
      point meta min=0.0,
      point meta max=36,
      colorbar,
      colorbar horizontal,
      at={($(group c1r1.south)!-0.225!(group c3r1.south)$)},
      anchor=north,
      yshift=-1.0cm,
      colorbar style={
          width=0.9\textwidth,
          height=0.2cm,
      }
    ]
  \end{axis}
\end{tikzpicture}
\label{fig:filo2x.xyz.coreopt.conflicts}
\end{figure}
As better illustrated in Figure \ref{fig:filo2x.xyz.coreopt.conflicts}, all the instances mentioned above share the common characteristic of having a small number of routes. This chart is mostly a specular copy of the speedup chart of Figure \ref{fig:filo2x.xyz.overall.speedup}.
Indeed, for every discarded change, the algorithm requires an additional core optimization iteration to reach the required termination criterion, with a clear negative effect on overall speedup.
The likelihood of generating changes leading to infeasible solutions is closely related to the number of routes that make up a solution. The fewer the routes, the higher the chance that solvers optimize an overlapping set of routes, which makes generating a resulting CVRP-feasible solution from these independent optimizations extremely challenging.

\subsection{Route Minimization Procedure Parallelization}
\begin{figure}
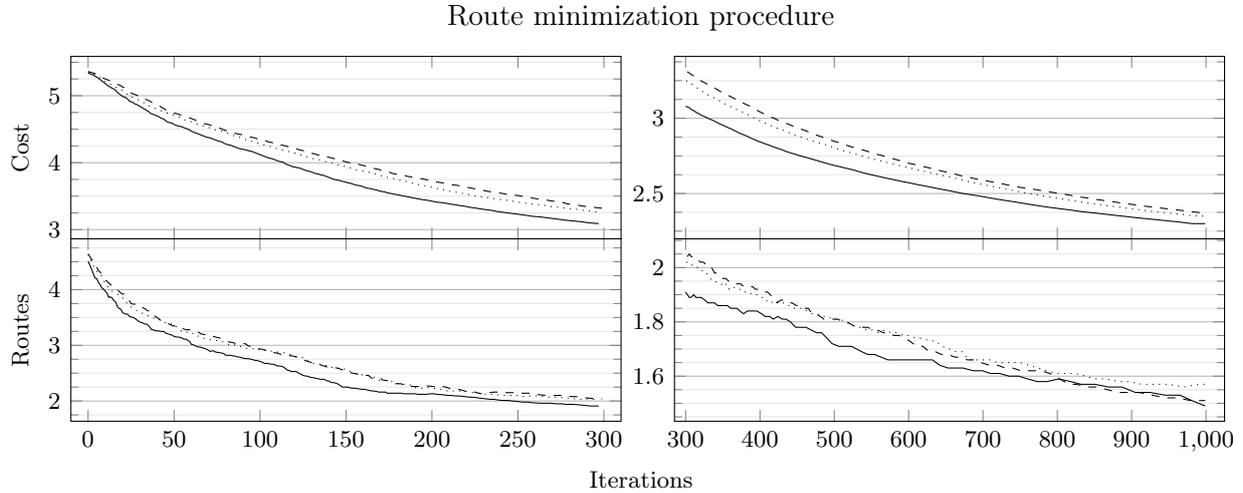

\centering
\caption{Search trajectory of the best-quality solution found during $\Delta_{RM} = 1000$ route minimization procedure iterations, averaged across datasets. 
Note that usually the route minimization procedure ends whenever the required number of routes is achieved. However, for the purpose of the analysis we performed exactly $\Delta_{RM}$ iterations, regardless of the previously mentioned termination criterion.
The Cost label denotes the average solution cost gap with respect to the best known solution.
The Routes label denotes the average route gap with respect to the greedy estimate employed to determine whether to run the route minimization procedure.
The continuous, dotted, and dashed lines refer to the sequential, and parallel versions with 5 and 10 solvers, respectively.
The graph is split in 0-300 and 300-999 iterations for better visibility.}
\include{imgs/filo2x.parallel.routemin}
\label{fig:filo2x.parallel.routemin}
\end{figure}

In principle, the route minimization procedure could be parallelized using the same schema applied for the core optimization procedure.
However, as shown in Figure \ref{fig:filo2x.parallel.routemin}, there is a significant degradation in the quality of the solutions obtained that increases with the number of solvers employed.
For the largest instances, this difference remains even after the application of the core optimization procedure, causing a statistically significant difference between the final results of the sequential algorithm compared to the parallel versions employing several solvers.

\begin{figure}
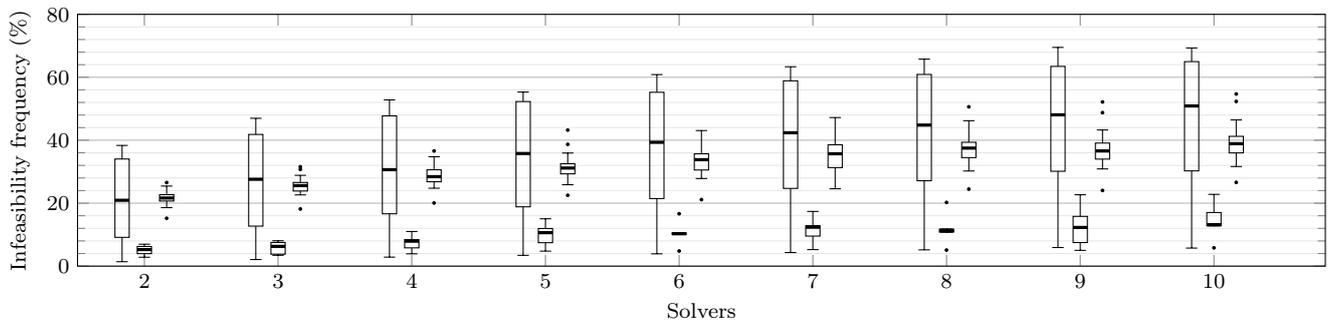

\centering
\caption{Percentage of candidate changes leading to infeasible solutions during the parallel route minimization procedure. Each box plot group refers from left-to-right to the $\mathbb{X}$, $\mathbb{B}$, and $\mathbb{I}$ datasets.}
\include{imgs/filo2x.routemin.boxplots.conflicts}
\label{fig:filo2x.routemin.boxplots.conflicts}
\end{figure}
The motivation behind such a reduction in solution quality is unclear and likely depends on the specifics of the route minimization procedure (see Algorithm \ref{algorithm:routemin}).
In particular, the key differences with the core optimization procedure are in the recreate step and in the neighbor acceptance criteria.
The recreate step can probabilistically leave some customers unserved for multiple iterations (lines \ref{algorithm:routemin:7} - \ref{algorithm:routemin:8}), while a generated neighbor solution is only accepted when improving and is reverted to the best solution found otherwise (lines \ref{algorithm:routemin:9} - \ref{algorithm:routemin:10}, \ref{algorithm:routemin:reverttobest}).
Most probably, these differences make the parallel versions less effective, even though we could not pinpoint the specific cause.

\begin{figure}
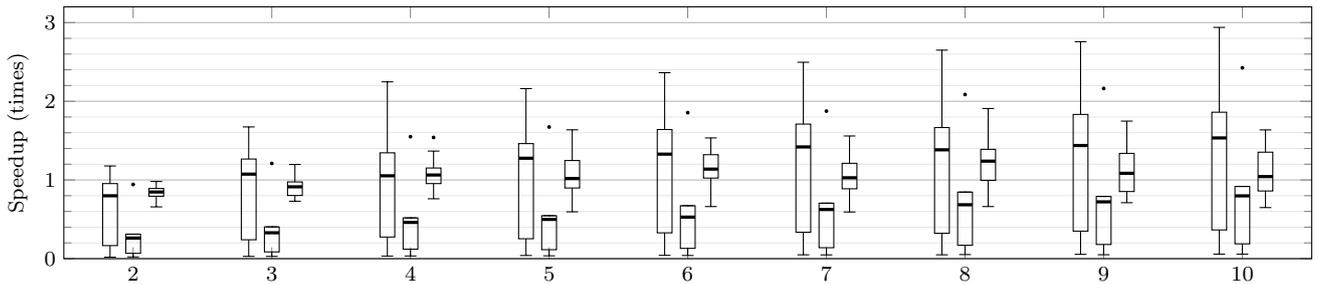

\centering
\caption{Speedup of the parallel route minimization procedure. Each box plot group refers from left-to-right to the $\mathbb{X}$, $\mathbb{B}$, and $\mathbb{I}$ datasets.}
\include{imgs/filo2x.boxplots.routemin.speedup}
\label{fig:filo2x.boxplots.routemin.speedup}
\end{figure}

Indeed, also Figure \ref{fig:filo2x.routemin.boxplots.conflicts} shows a surprisingly high chance of generating candidate changes leading to infeasible solutions, with a peak well above 60\% for some $\mathbb{X}$ instances.
This is most likely due to the strict reset-to-best rule (line \ref{algorithm:routemin:reverttobest}) which, when applied, can have a global effect on the reference solution, greatly enhancing the chance of invalidating any change produced on previous reference solutions. 
As shown in Figure \ref{fig:filo2x.boxplots.routemin.speedup}, this has a hugely negative effect on the procedure speedup, which even makes the parallel version slower than the sequential version for some configurations.

Because of these issues and the short computing time the sequential route minimization is anyway taking even for the largest instances (see Tables \ref{tab:i-time-per-procedure} and \ref{tab:i-long-time-per-procedure}), we decided not to invest more in its parallelization.

\subsection{Synchronization and Generation Time}
Synchronization is the key stage that allows solvers to incorporate changes generated by other solvers on potentially different reference solutions. 
It is the main additional step that the parallel version performs compared to the sequential algorithm, where synchronization and generation coexist into one single cohesive step.
For the parallel approach to be successful, it is crucial that the synchronization overhead, which includes re-validating a received change, is still negligible with respect to the execution of the generation stage.

\begin{table}[b!]
    \centering
    \footnotesize
    \caption{Fraction of core optimization procedure spent doing the synchronization and generation stages.}
    \label{tab:sync-vs-gen}
    \begin{tabular}{@{}c l rrrrrrrrr@{}}
        \toprule
        & Solvers & 2 & 3 & 4 & 5 & 6 & 7 & 8 & 9 & 10 \\
        \midrule
        
        \multirow{2}{*}{$\mathbb{X}$}
&Synchronization & 0.04 & 0.06 & 0.08 & 0.10 & 0.11 & 0.13 & 0.15 & 0.17 & 0.19\\
&Generation & 0.95 & 0.94 & 0.92 & 0.90 & 0.88 & 0.86 & 0.85 & 0.82 & 0.80\\

        \midrule        
        \multirow{2}{*}{$\mathbb{B}$} 
&Synchronization & 0.05 & 0.07 & 0.09 & 0.11 & 0.12 & 0.14 & 0.16 & 0.18 & 0.19\\
&Generation & 0.95 & 0.93 & 0.91 & 0.89 & 0.87 & 0.85 & 0.84 & 0.82 & 0.80\\

        \midrule
        
        \multirow{2}{*}{$\mathbb{I}$} 
&Synchronization & 0.09 & 0.13 & 0.16 & 0.19 & 0.22 & 0.25 & 0.27 & 0.30 & 0.32\\
&Generation & 0.90 & 0.86 & 0.82 & 0.78 & 0.75 & 0.72 & 0.69 & 0.67 & 0.64\\
        \bottomrule
    \end{tabular}

\end{table}

As can be seen in Table \ref{tab:sync-vs-gen}, the generation time takes most of the total time of the core optimization procedure, with the local search step taking more than 86\% of it. 
The synchronization time starts becoming relevant as the instances get larger and the number of solvers increase.

\section{Conclusions} \label{sec:conclusions}
In this paper, we introduced a schema for parallelizing iterations of optimization approaches that scope their actions on a localized solution portion, without the need for decomposition or rigid boundaries.
This schema is applied to FILO2, an already extremely efficient algorithm capable of scaling to instances with several hundreds of thousands of vertices.
The resulting \FILOX algorithm can be classified as an Independent Multi-Search (IMS/p-Control) approach with an indirect data exchange strategy, through the use of message queues, and an asynchronous cooperative optimization process.
The \FILOX algorithm maintains the effectiveness of FILO2, producing state-of-the-art results on datasets from the literature, but in a fraction of time.
Interestingly, the approach can be applied to small instances with a few hundred vertices still obtaining good speedups.
In general, the higher the number of routes used by a solution, the better the speedup, as this reduces the likelihood of generating changes leading to infeasible solutions.

\clearpage

\appendix

\section{Additional Computational Details}
In the following, \ref{appendix:x}, \ref{appendix:b}, and \ref{appendix:i} provide additional computational details for the $\mathbb{X}$, $\mathbb{B}$, and $\mathbb{I}$ datasets, respectively.
We also provide statistical validation of the algorithm results to better understand whether there are statistically significant differences between the behavior of \FILOX and FILO2.
In particular, as in \citet{accorsi2024}, we perform a one-tailed Wilcoxon signed rank test (see \citet{wilcoxon}) where the null and alternative hypotheses $H_0$ and $H_1$, respectively, are defined as follows.
\begin{equation*}
    H_0: \text{\textsc{AverageCost}}(\text{FILO2}^x) = \text{\textsc{AverageCost}}(\text{FILO2})
\end{equation*}
\begin{equation*}
    H_1: \text{\textsc{AverageCost}}(\text{FILO2}^x) > \text{\textsc{AverageCost}}(\text{FILO2})    
\end{equation*}
A hypothesis is rejected if its associated $p$-value is smaller than a significance level $\alpha$.
When the null hypothesis $H_0$ is not rejected, the average results of \FILOX are not statistically different from those obtained by FILO2.
In contrast, rejecting $H_0$ means that the average results obtained by the algorithms are statistically different.
The alternative hypothesis $H_1$ is thus tested to evaluate whether the average results obtained by \FILOX are statistically worse than those obtained by FILO2. If $H_1$ is rejected, \FILOX performs as well as or better than FILO2.

When multiple comparisons are performed with the same data, the probability of erroneously rejecting a null hypothesis increases. 
To better control these errors, the significance level $\alpha$ is typically adjusted to lower values. 
The Bonferroni correction (\cite{dunn}) is a simple method used to adjust $\alpha$ when performing multiple comparisons. In particular, given $n$ comparisons, the significance level is set to $\alpha / n$.
In particular, we set $\alpha = 0.025$ and $n = 9$, as we are performing one test consisting of two hypotheses for each configuration of the parallel version, using from 2 to 10 solvers.
The corrected significance level is therefore $\bar{\alpha} = 0.025 / 9 = 0.002778$.

\clearpage

\subsection{Computational Details for the \texorpdfstring{$\mathbb{X}$}{X} Instances} \label{appendix:x}
In the following, we report detailed results for the computations on the $\mathbb{X}$ dataset.
\begin{itemize}
    \item Figure \ref{fig:FILO2x.boxplots.quality.x} shows boxplots for the average gaps obtained by the approaches.
    \item Table \ref{table:p_values_x} shows the result of the one-tailed Wilcoxon signed rank test.
    \item Tables \ref{tab:x-time-per-procedure} and \ref{tab:x-long-time-per-procedure} report the computing time for the main algorithm procedures.
    \item Tables \ref{tab:x-small-gaps} -- \ref{tab:x-large-gaps} show the average gap for each instance in the dataset.
    \item Tables \ref{tab:x-small-speedups} -- \ref{tab:x-large-speedups} show the average speedup for each instance in the dataset.
\end{itemize}

\clearpage

\begin{figure}
\centering
\caption{Average gaps obtained by FILO2 and \FILOX (denoted by the number of solvers) on the $\mathbb{X}$ dataset when executing the standard (top) and long (bottom) versions. The median value is shown in the middle of each boxplot.}
\scriptsize
	\makeatletter
	\pgfplotsset{
		boxplot/draw/average/.code={ 
			\draw [/pgfplots/boxplot/every average/.try]
			\pgfextra
			\pgftransformshift{%
				\pgfplotsboxplotpointabbox
				{\pgfplotsboxplotvalue{average}}
				{0.5}%
			}%
			\endpgfextra
			;
		},
		boxplot/draw/median/.code={
			\draw [/pgfplots/boxplot/every median/.try]
			(boxplot box cs:\pgfplotsboxplotvalue{median},0)
			node[xshift=0.55cm, font=\tiny] {\pgfmathprintnumber{\pgfplotsboxplotvalue{median}}}
			--
			(boxplot box cs:\pgfplotsboxplotvalue{median},1);
		},
	}
	\makeatother
\begin{tikzpicture}
\scriptsize

\begin{axis}
[
xlabel={Algorithm},
ylabel={Average \% gap},
ymajorgrids=true,
yminorgrids=true,
minor y tick num=4,
minor grid style={line width=.01pt,draw=black!10},
major grid style={line width=.01pt,draw=black!30},
clip=false,
boxplot/draw direction=y,
boxplot/variable width,
boxplot/every median/.style={black,very thick,solid},
width=\textwidth,
height=150pt,
ylabel style={align=center}, 
y tick label style={align=right},
x tick label style={align=center},
xtick={0, 1, 2, 3, 4, 5, 6, 7, 8, 9, 10},
xticklabels={FILO2, 2, 3, 4, 5, 6, 7, 8, 9, 10},
xmin=-0.75,
xmax=10,
scatter/classes={ a={mark=star}, b={mark=*}}]

\addplot[mark=*, mark size=0.5pt,boxplot, boxplot prepared={draw position=0,
median=0.322,upper quartile=0.515,lower quartile=0.126,upper whisker=1.059,lower whisker=0.0,sample size=1}]
coordinates{
(0,1.101)
};
\addplot[mark=*, mark size=0.5pt,boxplot, boxplot prepared={draw position=1,
median=0.364,upper quartile=0.516,lower quartile=0.087,upper whisker=1.003,lower whisker=0.0,sample size=1}]
coordinates{
(1,1.215)
};
\addplot[mark=*, mark size=0.5pt,boxplot, boxplot prepared={draw position=2,
median=0.324,upper quartile=0.54,lower quartile=0.115,upper whisker=1.014,lower whisker=0.0,sample size=1}]
coordinates{
(2,1.227)
};
\addplot[mark=*, mark size=0.5pt,boxplot, boxplot prepared={draw position=3,
median=0.338,upper quartile=0.527,lower quartile=0.114,upper whisker=0.999,lower whisker=0.0,sample size=1}]
coordinates{
(3,1.169)
};
\addplot[mark=*, mark size=0.5pt,boxplot, boxplot prepared={draw position=4,
median=0.349,upper quartile=0.508,lower quartile=0.138,upper whisker=0.983,lower whisker=0.0,sample size=1}]
coordinates{
(4,1.081)
(4,1.184)
};
\addplot[mark=*, mark size=0.5pt,boxplot, boxplot prepared={draw position=5,
median=0.369,upper quartile=0.548,lower quartile=0.111,upper whisker=1.116,lower whisker=0.0,sample size=1}]
coordinates{
};
\addplot[mark=*, mark size=0.5pt,boxplot, boxplot prepared={draw position=6,
median=0.368,upper quartile=0.537,lower quartile=0.151,upper whisker=1.077,lower whisker=0.0,sample size=1}]
coordinates{
(6,1.249)
};
\addplot[mark=*, mark size=0.5pt,boxplot, boxplot prepared={draw position=7,
median=0.357,upper quartile=0.515,lower quartile=0.117,upper whisker=0.973,lower whisker=0.0,sample size=1}]
coordinates{
(7,1.118)
(7,1.226)
};
\addplot[mark=*, mark size=0.5pt,boxplot, boxplot prepared={draw position=8,
median=0.335,upper quartile=0.534,lower quartile=0.131,upper whisker=1.044,lower whisker=0.0,sample size=1}]
coordinates{
(8,1.238)
};
\addplot[mark=*, mark size=0.5pt,boxplot, boxplot prepared={draw position=9,
median=0.363,upper quartile=0.544,lower quartile=0.115,upper whisker=1.095,lower whisker=0.0,sample size=1}]
coordinates{
};

\end{axis}
\end{tikzpicture}
\vspace{-0.9cm}
\scriptsize
	\makeatletter
	\pgfplotsset{
		boxplot/draw/average/.code={ 
			\draw [/pgfplots/boxplot/every average/.try]
			\pgfextra
			\pgftransformshift{%
				\pgfplotsboxplotpointabbox
				{\pgfplotsboxplotvalue{average}}
				{0.5}%
			}%
			\endpgfextra
			;
		},
		boxplot/draw/median/.code={
			\draw [/pgfplots/boxplot/every median/.try]
			(boxplot box cs:\pgfplotsboxplotvalue{median},0)
			node[xshift=0.55cm, font=\tiny] {\pgfmathprintnumber{\pgfplotsboxplotvalue{median}}}
			--
			(boxplot box cs:\pgfplotsboxplotvalue{median},1);
		},
	}
	\makeatother
\begin{tikzpicture}
\scriptsize

\begin{axis}
[
xlabel={Algorithm},
ylabel={Average \% gap},
ymajorgrids=true,
yminorgrids=true,
minor y tick num=4,
minor grid style={line width=.01pt,draw=black!10},
major grid style={line width=.01pt,draw=black!30},
clip=false,
boxplot/draw direction=y,
boxplot/variable width,
boxplot/every median/.style={black,very thick,solid},
width=\textwidth,
height=150pt,
ylabel style={align=center}, 
y tick label style={align=right},
x tick label style={align=center},
xtick={0, 1, 2, 3, 4, 5, 6, 7, 8, 9, 10},
xticklabels={FILO2, 2, 3, 4, 5, 6, 7, 8, 9, 10},
xmin=-0.75,
xmax=10,
scatter/classes={ a={mark=star}, b={mark=*}}]

\addplot[mark=*, mark size=0.5pt,boxplot, boxplot prepared={draw position=0,
median=0.184,upper quartile=0.295,lower quartile=0.031,upper whisker=0.558,lower whisker=0.0,sample size=1}]
coordinates{
(0,0.75)
(0,0.954)
};
\addplot[mark=*, mark size=0.5pt,boxplot, boxplot prepared={draw position=1,
median=0.192,upper quartile=0.289,lower quartile=0.037,upper whisker=0.548,lower whisker=0.0,sample size=1}]
coordinates{
(1,0.741)
(1,0.977)
};
\addplot[mark=*, mark size=0.5pt,boxplot, boxplot prepared={draw position=2,
median=0.211,upper quartile=0.316,lower quartile=0.032,upper whisker=0.732,lower whisker=0.0,sample size=1}]
coordinates{
(2,0.865)
};
\addplot[mark=*, mark size=0.5pt,boxplot, boxplot prepared={draw position=3,
median=0.198,upper quartile=0.304,lower quartile=0.03,upper whisker=0.605,lower whisker=0.0,sample size=1}]
coordinates{
(3,0.744)
(3,0.941)
};
\addplot[mark=*, mark size=0.5pt,boxplot, boxplot prepared={draw position=4,
median=0.193,upper quartile=0.308,lower quartile=0.038,upper whisker=0.657,lower whisker=0.0,sample size=1}]
coordinates{
(4,0.734)
(4,0.843)
};
\addplot[mark=*, mark size=0.5pt,boxplot, boxplot prepared={draw position=5,
median=0.201,upper quartile=0.291,lower quartile=0.039,upper whisker=0.63,lower whisker=0.0,sample size=1}]
coordinates{
(5,0.714)
(5,0.886)
};
\addplot[mark=*, mark size=0.5pt,boxplot, boxplot prepared={draw position=6,
median=0.201,upper quartile=0.3,lower quartile=0.04,upper whisker=0.584,lower whisker=0.0,sample size=1}]
coordinates{
(6,0.727)
(6,0.919)
};
\addplot[mark=*, mark size=0.5pt,boxplot, boxplot prepared={draw position=7,
median=0.21,upper quartile=0.314,lower quartile=0.032,upper whisker=0.612,lower whisker=0.0,sample size=1}]
coordinates{
(7,0.745)
(7,1.017)
};
\addplot[mark=*, mark size=0.5pt,boxplot, boxplot prepared={draw position=8,
median=0.194,upper quartile=0.307,lower quartile=0.03,upper whisker=0.721,lower whisker=0.0,sample size=1}]
coordinates{
(8,0.874)
};
\addplot[mark=*, mark size=0.5pt,boxplot, boxplot prepared={draw position=9,
median=0.204,upper quartile=0.309,lower quartile=0.045,upper whisker=0.594,lower whisker=0.0,sample size=1}]
coordinates{
(9,0.707)
(9,0.938)
};

\end{axis}
\end{tikzpicture}
\label{fig:FILO2x.boxplots.quality.x}
\end{figure}

\begin{table}
	\caption{Computations on the $\mathbb{X}$ dataset: $p$-values for \FILOX vs FILO2 on the left and \FILOX (long) vs FILO2 (long) on the right.
	\\
	\footnotesize
	$p$-values in bold are associated with rejected hypothesis when $\bar{\alpha} = 0.002778$.\\
The last row of each group contains a $p$-value interpretation. 
In particular, \FILOX is not statistically different from FILO2 when $H_0$ cannot be rejected (Similar), \FILOX is statistically better when both $H_0$ and $H_1$ are rejected (Better), and, finally, \FILOX is statistically worse when $H_0$ is rejected and $H_1$ is not rejected (Worse).}
	\label{table:p_values_x}
	\centering
	\footnotesize
	\begin{tabular}{rrrrrrr}
	\toprule
        & & \multicolumn{2}{c}{\FILOX vs FILO2} & & \multicolumn{2}{c}{\FILOX (long) vs FILO2 (long)} \\
        \cmidrule{3-4}
        \cmidrule{6-7}
        
Solvers & & $H_0$ / $H_1$ & Outcome & & $H_0$ / $H_1$ & Outcome \\
	\midrule
2 &&   0.636031 /   0.318015 & Similar &&   0.686502 /   0.343251 & Similar \\
3 &&   0.861598 /   0.430799 & Similar &&   0.986721 /   0.493360 & Similar \\
4 &&   0.204705 /   0.897648 & Similar &&   0.918803 /   0.540598 & Similar \\
5 &&   0.400299 /   0.799850 & Similar &&   0.748675 /   0.374337 & Similar \\
6 &&   0.093292 /   0.953354 & Similar &&   0.702061 /   0.351031 & Similar \\
7 &&   0.019983 /   0.990009 & Similar &&   0.352392 /   0.823804 & Similar \\
8 &&   0.354791 /   0.822604 & Similar &&   0.184298 /   0.907851 & Similar \\
9 &&   0.146985 /   0.926507 & Similar &&   0.434540 /   0.782730 & Similar \\
10 &&   0.022747 /   0.988626 & Similar &&   0.112887 /   0.943557 & Similar \\
	\bottomrule
	\end{tabular}
\end{table}

\begin{table}
    \centering
    \footnotesize
    \caption{Computing time in seconds for the main algorithm FILO2 and \FILOX procedures when solving the $\mathbb{X}$ dataset.}
    \label{tab:x-time-per-procedure}
    \begin{tabular}{lrrrrrrrrrr}
        \toprule
        & & \multicolumn{8}{c}{\FILOX} \\
        \cmidrule{3-11}
        & FILO2 & 2 & 3 & 4 & 5 & 6 & 7 & 8 & 9 & 10\\
        \midrule
Instance preprocessing & 0.03 & 0.02 & 0.02 & 0.01 & 0.01 & 0.01 & 0.01 & 0.01 & 0.01 & 0.01\\
Construction phase & 0.00 & 0.00 & 0.00 & 0.00 & 0.00 & 0.00 & 0.00 & 0.00 & 0.00 & 0.00\\
Greedy route estimation & 0.00 & 0.00 & 0.00 & 0.00 & 0.00 & 0.00 & 0.00 & 0.00 & 0.00 & 0.00\\
Move generators initialization & 0.00 & 0.00 & 0.00 & 0.00 & 0.00 & 0.00 & 0.00 & 0.00 & 0.00 & 0.00\\
Route minimization procedure & 0.25 & 0.26 & 0.27 & 0.27 & 0.27 & 0.27 & 0.27 & 0.27 & 0.27 & 0.27\\
Core optimization procedure & 66.91 & 39.13 & 27.46 & 21.66 & 18.01 & 15.80 & 14.20 & 12.89 & 12.01 & 11.39\\
         \bottomrule
    \end{tabular}
\end{table}

\begin{table}
    \centering
    \footnotesize
    \caption{Computing time in seconds for the main algorithm FILO2 (long) and \FILOX (long) procedures when solving the $\mathbb{X}$ dataset.}
    \label{tab:x-long-time-per-procedure}
    \begin{tabular}{lrrrrrrrrrr}
        \toprule
        & & \multicolumn{8}{c}{\FILOX} \\
        \cmidrule{3-11}
        & FILO2 & 2 & 3 & 4 & 5 & 6 & 7 & 8 & 9 & 10\\
        \midrule
Instance preprocessing & 0.03 & 0.02 & 0.01 & 0.01 & 0.01 & 0.01 & 0.01 & 0.01 & 0.01 & 0.01\\
Construction phase & 0.00 & 0.00 & 0.00 & 0.00 & 0.00 & 0.00 & 0.00 & 0.00 & 0.00 & 0.00\\
Greedy route estimation & 0.00 & 0.00 & 0.00 & 0.00 & 0.00 & 0.00 & 0.00 & 0.00 & 0.00 & 0.00\\
Move generators initialization & 0.00 & 0.00 & 0.00 & 0.00 & 0.00 & 0.00 & 0.00 & 0.00 & 0.00 & 0.00\\
Route minimization procedure & 0.25 & 0.26 & 0.27 & 0.27 & 0.27 & 0.27 & 0.27 & 0.27 & 0.27 & 0.27\\
Core optimization procedure & 704.80 & 407.10 & 285.79 & 224.22 & 186.18 & 163.57 & 146.77 & 134.13 & 124.27 & 117.18\\

         \bottomrule
    \end{tabular}
\end{table}

\begin{sidewaystable}
\centering
\footnotesize
\caption{Average gap on small-sized $\mathbb{X}$ instances.}
\label{tab:x-small-gaps}

\begin{tabular}{lr c rrrrrrrrrr c rrrrrrrrrr}
\toprule
 &&& &\multicolumn{9}{c}{\FILOX} && FILO2& \multicolumn{9}{c}{\FILOX (long)} \\
\cmidrule{5-13}
\cmidrule{16-24}
Instance & BKS && FILO2 & 2 & 3 & 4 & 5 & 6 & 7 & 8 & 9 & 10 && (long) & 2 & 3 & 4 & 5 & 6 & 7 & 8 & 9 & 10 \\
\midrule
X-n101-k25 &  27591  && 0.00 & 0.00 & 0.00 & 0.00 & 0.00 & 0.00 & 0.02 & 0.00 & 0.03 & 0.00 && 0.00 & 0.00 & 0.00 & 0.00 & 0.00 & 0.00 & 0.00 & 0.00 & 0.00 & 0.00 \\
X-n106-k14 & 26362   && 0.06 & 0.06 & 0.05 & 0.03 & 0.06 & 0.05 & 0.06 & 0.06 & 0.06 & 0.05 && 0.00 & 0.00 & 0.01 & 0.01 & 0.01 & 0.00 & 0.00 & 0.01 & 0.00 & 0.01 \\
X-n110-k13 & 14971   && 0.00 & 0.00 & 0.00 & 0.00 & 0.00 & 0.00 & 0.00 & 0.00 & 0.00 & 0.00 && 0.00 & 0.00 & 0.00 & 0.00 & 0.00 & 0.00 & 0.00 & 0.00 & 0.00 & 0.00 \\
X-n115-k10 &  12747  && 0.00 & 0.00 & 0.00 & 0.00 & 0.00 & 0.00 & 0.00 & 0.00 & 0.00 & 0.00 && 0.00 & 0.00 & 0.00 & 0.00 & 0.00 & 0.00 & 0.00 & 0.00 & 0.00 & 0.00 \\
X-n120-k6  & 13332   && 0.00 & 0.00 & 0.00 & 0.00 & 0.00 & 0.00 & 0.00 & 0.00 & 0.00 & 0.00 && 0.00 & 0.00 & 0.00 & 0.00 & 0.00 & 0.00 & 0.00 & 0.00 & 0.00 & 0.00 \\
X-n125-k30 & 55539   && 0.51 & 0.61 & 0.65 & 0.69 & 0.78 & 0.55 & 0.49 & 0.51 & 0.53 & 0.57 && 0.22 & 0.13 & 0.32 & 0.16 & 0.11 & 0.05 & 0.32 & 0.22 & 0.19 & 0.06 \\
X-n129-k18 & 28940   && 0.13 & 0.05 & 0.09 & 0.05 & 0.04 & 0.06 & 0.16 & 0.07 & 0.05 & 0.05 && 0.03 & 0.04 & 0.03 & 0.03 & 0.04 & 0.03 & 0.02 & 0.02 & 0.03 & 0.04 \\
X-n134-k13 &  10916  && 0.12 & 0.13 & 0.16 & 0.13 & 0.15 & 0.15 & 0.14 & 0.16 & 0.15 & 0.16 && 0.02 & 0.06 & 0.03 & 0.02 & 0.05 & 0.07 & 0.07 & 0.05 & 0.04 & 0.05 \\
X-n139-k10 & 13590   && 0.00 & 0.00 & 0.00 & 0.00 & 0.00 & 0.00 & 0.00 & 0.00 & 0.00 & 0.00 && 0.00 & 0.00 & 0.00 & 0.00 & 0.00 & 0.00 & 0.00 & 0.00 & 0.00 & 0.00 \\
X-n143-k7  &  15700  && 0.16 & 0.13 & 0.17 & 0.17 & 0.16 & 0.13 & 0.17 & 0.12 & 0.17 & 0.16 && 0.11 & 0.10 & 0.08 & 0.11 & 0.13 & 0.08 & 0.11 & 0.13 & 0.10 & 0.09 \\
X-n148-k46 & 43448   && 0.23 & 0.15 & 0.15 & 0.27 & 0.16 & 0.20 & 0.18 & 0.11 & 0.22 & 0.12 && 0.00 & 0.00 & 0.00 & 0.00 & 0.02 & 0.04 & 0.01 & 0.03 & 0.01 & 0.07 \\
X-n153-k22 &  21220  && 0.18 & 0.07 & 0.24 & 0.29 & 0.22 & 0.18 & 0.26 & 0.18 & 0.26 & 0.28 && 0.02 & 0.02 & 0.04 & 0.03 & 0.02 & 0.03 & 0.02 & 0.04 & 0.05 & 0.02 \\
X-n157-k13 &  16876  && 0.00 & 0.00 & 0.00 & 0.00 & 0.00 & 0.00 & 0.00 & 0.00 & 0.00 & 0.00 && 0.00 & 0.00 & 0.00 & 0.00 & 0.00 & 0.00 & 0.00 & 0.00 & 0.00 & 0.00 \\
X-n162-k11 &  14138  && 0.17 & 0.12 & 0.13 & 0.11 & 0.17 & 0.14 & 0.16 & 0.17 & 0.19 & 0.14 && 0.03 & 0.03 & 0.04 & 0.06 & 0.04 & 0.02 & 0.07 & 0.05 & 0.03 & 0.04 \\
X-n167-k10 &  20557  && 0.08 & 0.07 & 0.08 & 0.12 & 0.07 & 0.10 & 0.11 & 0.10 & 0.13 & 0.10 && 0.00 & 0.01 & 0.01 & 0.02 & 0.01 & 0.01 & 0.01 & 0.00 & 0.03 & 0.01 \\
X-n172-k51 &  45607  && 0.01 & 0.00 & 0.03 & 0.01 & 0.04 & 0.01 & 0.03 & 0.00 & 0.01 & 0.01 && 0.00 & 0.00 & 0.00 & 0.00 & 0.00 & 0.00 & 0.00 & 0.00 & 0.00 & 0.00 \\
X-n176-k26 &  47812  && 0.52 & 0.49 & 0.53 & 0.77 & 0.27 & 0.39 & 0.58 & 0.47 & 0.57 & 0.73 && 0.08 & 0.17 & 0.23 & 0.22 & 0.13 & 0.15 & 0.18 & 0.33 & 0.26 & 0.24 \\
X-n181-k23 & 25569   && 0.01 & 0.02 & 0.02 & 0.02 & 0.03 & 0.02 & 0.00 & 0.01 & 0.03 & 0.02 && 0.00 & 0.00 & 0.00 & 0.00 & 0.00 & 0.00 & 0.00 & 0.00 & 0.00 & 0.00 \\
X-n186-k15 & 24145   && 0.12 & 0.08 & 0.11 & 0.10 & 0.05 & 0.10 & 0.04 & 0.09 & 0.08 & 0.07 && 0.03 & 0.04 & 0.02 & 0.01 & 0.03 & 0.02 & 0.04 & 0.03 & 0.05 & 0.04 \\
X-n190-k8 &  16980   && 0.04 & 0.07 & 0.07 & 0.09 & 0.04 & 0.06 & 0.13 & 0.09 & 0.13 & 0.06 && 0.02 & 0.02 & 0.01 & 0.02 & 0.01 & 0.02 & 0.01 & 0.02 & 0.02 & 0.02 \\
X-n195-k51 &  44225  && 0.20 & 0.16 & 0.19 & 0.16 & 0.25 & 0.19 & 0.22 & 0.15 & 0.18 & 0.26 && 0.06 & 0.05 & 0.04 & 0.08 & 0.05 & 0.09 & 0.08 & 0.16 & 0.09 & 0.09 \\
X-n200-k36 & 58578   && 0.53 & 0.80 & 0.70 & 0.64 & 0.73 & 0.77 & 1.08 & 0.65 & 0.76 & 0.57 && 0.44 & 0.35 & 0.33 & 0.31 & 0.39 & 0.29 & 0.35 & 0.42 & 0.37 & 0.33 \\
X-n204-k19 & 19565   && 0.02 & 0.02 & 0.12 & 0.11 & 0.07 & 0.03 & 0.15 & 0.15 & 0.12 & 0.10 && 0.00 & 0.00 & 0.00 & 0.00 & 0.00 & 0.00 & 0.00 & 0.01 & 0.00 & 0.01 \\
X-n209-k16 & 30656   && 0.05 & 0.07 & 0.09 & 0.10 & 0.14 & 0.13 & 0.18 & 0.12 & 0.07 & 0.07 && 0.06 & 0.05 & 0.05 & 0.05 & 0.06 & 0.07 & 0.05 & 0.05 & 0.03 & 0.05 \\
X-n214-k11 &  10856  && 0.28 & 0.27 & 0.30 & 0.25 & 0.30 & 0.35 & 0.21 & 0.32 & 0.28 & 0.27 && 0.16 & 0.19 & 0.16 & 0.16 & 0.15 & 0.21 & 0.16 & 0.17 & 0.15 & 0.16 \\
X-n219-k73 &  117595  && 0.00 & 0.00 & 0.00 & 0.00 & 0.00 & 0.00 & 0.00 & 0.00 & 0.00 & 0.00 && 0.00 & 0.00 & 0.00 & 0.00 & 0.00 & 0.00 & 0.00 & 0.00 & 0.00 & 0.00 \\
X-n223-k34 &  40437  && 0.28 & 0.37 & 0.28 & 0.26 & 0.33 & 0.35 & 0.29 & 0.31 & 0.32 & 0.30 && 0.17 & 0.20 & 0.21 & 0.20 & 0.18 & 0.22 & 0.19 & 0.19 & 0.19 & 0.19 \\
X-n228-k23 &  25742  && 0.18 & 0.13 & 0.16 & 0.22 & 0.15 & 0.19 & 0.17 & 0.18 & 0.15 & 0.20 && 0.11 & 0.16 & 0.17 & 0.14 & 0.11 & 0.12 & 0.17 & 0.11 & 0.18 & 0.12 \\
X-n233-k16 &  19230  && 0.50 & 0.40 & 0.50 & 0.56 & 0.43 & 0.49 & 0.43 & 0.43 & 0.44 & 0.44 && 0.24 & 0.31 & 0.30 & 0.31 & 0.30 & 0.23 & 0.24 & 0.34 & 0.28 & 0.31 \\
X-n237-k14 &  27042  && 0.00 & 0.03 & 0.01 & 0.00 & 0.03 & 0.02 & 0.01 & 0.01 & 0.00 & 0.01 && 0.01 & 0.01 & 0.01 & 0.01 & 0.01 & 0.00 & 0.01 & 0.00 & 0.01 & 0.01 \\
X-n242-k48 &  82751  && 0.32 & 0.25 & 0.31 & 0.33 & 0.32 & 0.29 & 0.28 & 0.27 & 0.31 & 0.39 && 0.11 & 0.16 & 0.14 & 0.14 & 0.13 & 0.16 & 0.13 & 0.14 & 0.14 & 0.16 \\
X-n247-k50 & 37274   && 0.79 & 0.78 & 0.68 & 0.75 & 0.79 & 0.79 & 0.77 & 0.71 & 0.73 & 0.72 && 0.50 & 0.54 & 0.44 & 0.50 & 0.50 & 0.41 & 0.32 & 0.45 & 0.59 & 0.50 \\
\midrule
Mean & && 0.17 & 0.17 & 0.18 & 0.20 & 0.18 & 0.18 & 0.20 & 0.17 & 0.19 & 0.18 && 0.08 & 0.08 & 0.08 & 0.08 & 0.08 & 0.07 & 0.08 & 0.09 & 0.09 & 0.08 \\
\bottomrule

\end{tabular}
\end{sidewaystable}

\begin{sidewaystable}
\centering
\footnotesize
\caption{Average gap on medium-sized $\mathbb{X}$ instances.}
\label{tab:x-medium-gaps}
\begin{tabular}{lr c rrrrrrrrrr c rrrrrrrrrr}
\toprule
 &&& &\multicolumn{9}{c}{\FILOX} && FILO2& \multicolumn{9}{c}{\FILOX (long)} \\
\cmidrule{5-13}
\cmidrule{16-24}
Instance & BKS && FILO2 & 2 & 3 & 4 & 5 & 6 & 7 & 8 & 9 & 10 && (long) & 2 & 3 & 4 & 5 & 6 & 7 & 8 & 9 & 10 \\
\midrule
X-n251-k28 &  38684    && 0.30 & 0.39 & 0.32 & 0.36 & 0.31 & 0.38 & 0.31 & 0.39 & 0.38 & 0.32 && 0.24 & 0.23 & 0.26 & 0.23 & 0.23 & 0.27 & 0.28 & 0.26 & 0.24 & 0.25 \\
X-n256-k16 &  18839    && 0.22 & 0.22 & 0.22 & 0.22 & 0.22 & 0.22 & 0.22 & 0.22 & 0.22 & 0.22 && 0.22 & 0.22 & 0.22 & 0.22 & 0.22 & 0.22 & 0.22 & 0.22 & 0.22 & 0.22 \\
X-n261-k13 &  26558    && 0.33 & 0.36 & 0.39 & 0.33 & 0.40 & 0.30 & 0.41 & 0.37 & 0.39 & 0.39 && 0.25 & 0.24 & 0.31 & 0.28 & 0.29 & 0.26 & 0.28 & 0.25 & 0.19 & 0.31 \\
X-n266-k58 &  75478    && 0.46 & 0.45 & 0.41 & 0.48 & 0.47 & 0.48 & 0.42 & 0.40 & 0.46 & 0.45 && 0.29 & 0.35 & 0.35 & 0.35 & 0.35 & 0.34 & 0.36 & 0.33 & 0.34 & 0.32 \\
X-n270-k35 &  35291    && 0.17 & 0.21 & 0.23 & 0.25 & 0.24 & 0.23 & 0.22 & 0.22 & 0.26 & 0.26 && 0.17 & 0.18 & 0.17 & 0.18 & 0.15 & 0.13 & 0.08 & 0.13 & 0.18 & 0.20 \\
X-n275-k28 &   21245    && 0.03 & 0.07 & 0.03 & 0.03 & 0.02 & 0.02 & 0.05 & 0.05 & 0.06 & 0.00 && 0.00 & 0.00 & 0.00 & 0.00 & 0.00 & 0.02 & 0.00 & 0.00 & 0.00 & 0.00 \\
X-n280-k17 &33503       && 0.35 & 0.40 & 0.30 & 0.35 & 0.44 & 0.39 & 0.40 & 0.34 & 0.29 & 0.42 && 0.22 & 0.28 & 0.31 & 0.29 & 0.31 & 0.24 & 0.25 & 0.22 & 0.25 & 0.22 \\
X-n284-k15 &  20215     && 0.44 & 0.52 & 0.58 & 0.39 & 0.39 & 0.38 & 0.49 & 0.32 & 0.43 & 0.63 && 0.24 & 0.23 & 0.24 & 0.20 & 0.22 & 0.26 & 0.24 & 0.27 & 0.25 & 0.24 \\
X-n289-k60 &  95151     && 0.59 & 0.55 & 0.60 & 0.52 & 0.53 & 0.56 & 0.61 & 0.60 & 0.57 & 0.62 && 0.43 & 0.41 & 0.42 & 0.39 & 0.36 & 0.39 & 0.36 & 0.38 & 0.40 & 0.37 \\
X-n294-k50 &  47161     && 0.32 & 0.31 & 0.32 & 0.28 & 0.30 & 0.36 & 0.30 & 0.32 & 0.32 & 0.29 && 0.20 & 0.21 & 0.21 & 0.24 & 0.19 & 0.21 & 0.20 & 0.23 & 0.26 & 0.20 \\
X-n298-k31 &   34231    && 0.24 & 0.25 & 0.18 & 0.27 & 0.18 & 0.29 & 0.30 & 0.34 & 0.25 & 0.33 && 0.19 & 0.17 & 0.15 & 0.14 & 0.17 & 0.17 & 0.18 & 0.17 & 0.20 & 0.13 \\
X-n303-k21 &  21736     && 0.44 & 0.47 & 0.43 & 0.51 & 0.46 & 0.48 & 0.47 & 0.51 & 0.51 & 0.41 && 0.31 & 0.30 & 0.33 & 0.31 & 0.31 & 0.34 & 0.30 & 0.29 & 0.31 & 0.31 \\
X-n308-k13 &   25859    && 0.73 & 0.48 & 0.75 & 1.00 & 1.08 & 0.74 & 0.91 & 1.12 & 0.91 & 0.79 && 0.29 & 0.25 & 0.53 & 0.37 & 0.66 & 0.36 & 0.34 & 0.23 & 0.34 & 0.41 \\
X-n313-k71 &  94043     && 0.51 & 0.52 & 0.53 & 0.57 & 0.52 & 0.55 & 0.59 & 0.46 & 0.48 & 0.58 && 0.28 & 0.28 & 0.35 & 0.27 & 0.26 & 0.29 & 0.27 & 0.27 & 0.35 & 0.33 \\
X-n317-k53 &  78355     && 0.01 & 0.00 & 0.00 & 0.01 & 0.01 & 0.01 & 0.01 & 0.01 & 0.00 & 0.00 && 0.00 & 0.00 & 0.00 & 0.00 & 0.00 & 0.00 & 0.00 & 0.00 & 0.00 & 0.00 \\
X-n322-k28 &  29834     && 0.49 & 0.43 & 0.43 & 0.41 & 0.37 & 0.42 & 0.50 & 0.40 & 0.49 & 0.43 && 0.29 & 0.21 & 0.38 & 0.31 & 0.26 & 0.32 & 0.34 & 0.35 & 0.31 & 0.31 \\
X-n327-k20 & 27532      && 0.50 & 0.40 & 0.54 & 0.40 & 0.43 & 0.53 & 0.42 & 0.47 & 0.43 & 0.46 && 0.28 & 0.31 & 0.25 & 0.30 & 0.26 & 0.29 & 0.29 & 0.30 & 0.28 & 0.29 \\
X-n331-k15 &  31102     && 0.03 & 0.06 & 0.01 & 0.03 & 0.01 & 0.04 & 0.05 & 0.03 & 0.08 & 0.08 && 0.01 & 0.01 & 0.01 & 0.00 & 0.00 & 0.01 & 0.02 & 0.01 & 0.00 & 0.01 \\
X-n336-k84 &  139111     && 0.56 & 0.51 & 0.54 & 0.56 & 0.52 & 0.58 & 0.58 & 0.58 & 0.61 & 0.53 && 0.35 & 0.30 & 0.35 & 0.34 & 0.34 & 0.31 & 0.32 & 0.31 & 0.30 & 0.27 \\
X-n344-k43 &  42050     && 0.52 & 0.56 & 0.55 & 0.48 & 0.55 & 0.55 & 0.54 & 0.54 & 0.50 & 0.53 && 0.34 & 0.27 & 0.30 & 0.37 & 0.40 & 0.29 & 0.32 & 0.35 & 0.34 & 0.34 \\
X-n351-k40 &  25896     && 0.62 & 0.60 & 0.62 & 0.63 & 0.69 & 0.51 & 0.60 & 0.61 & 0.59 & 0.67 && 0.42 & 0.42 & 0.39 & 0.43 & 0.39 & 0.45 & 0.46 & 0.44 & 0.47 & 0.42 \\
X-n359-k29 &  51505     && 0.41 & 0.44 & 0.41 & 0.43 & 0.36 & 0.44 & 0.45 & 0.41 & 0.44 & 0.46 && 0.19 & 0.20 & 0.16 & 0.18 & 0.20 & 0.20 & 0.20 & 0.22 & 0.17 & 0.20 \\
X-n367-k17 &  22814     && 0.19 & 0.06 & 0.10 & 0.14 & 0.15 & 0.08 & 0.18 & 0.16 & 0.12 & 0.23 && 0.03 & 0.04 & 0.03 & 0.04 & 0.02 & 0.04 & 0.02 & 0.02 & 0.02 & 0.04 \\
X-n376-k94 &   147713    && 0.01 & 0.01 & 0.01 & 0.01 & 0.01 & 0.01 & 0.01 & 0.01 & 0.01 & 0.01 && 0.01 & 0.00 & 0.00 & 0.00 & 0.00 & 0.00 & 0.00 & 0.01 & 0.00 & 0.00 \\
X-n384-k52 &  65928     && 0.33 & 0.41 & 0.39 & 0.45 & 0.36 & 0.38 & 0.39 & 0.36 & 0.41 & 0.41 && 0.31 & 0.28 & 0.27 & 0.26 & 0.26 & 0.27 & 0.26 & 0.28 & 0.27 & 0.25 \\
X-n393-k38 &   38260    && 0.17 & 0.19 & 0.23 & 0.32 & 0.26 & 0.27 & 0.22 & 0.19 & 0.20 & 0.20 && 0.09 & 0.11 & 0.09 & 0.12 & 0.09 & 0.09 & 0.10 & 0.08 & 0.13 & 0.12 \\
X-n401-k29 &   66154    && 0.24 & 0.26 & 0.23 & 0.22 & 0.27 & 0.24 & 0.23 & 0.24 & 0.23 & 0.22 && 0.17 & 0.14 & 0.13 & 0.13 & 0.12 & 0.14 & 0.15 & 0.15 & 0.13 & 0.13 \\
X-n411-k19 &  19712     && 0.45 & 0.36 & 0.55 & 0.44 & 0.40 & 0.40 & 0.45 & 0.43 & 0.36 & 0.47 && 0.33 & 0.33 & 0.34 & 0.34 & 0.32 & 0.38 & 0.32 & 0.32 & 0.32 & 0.32 \\
X-n420-k130 &  107798      && 0.26 & 0.27 & 0.22 & 0.25 & 0.27 & 0.31 & 0.26 & 0.26 & 0.26 & 0.23 && 0.17 & 0.15 & 0.13 & 0.17 & 0.16 & 0.12 & 0.17 & 0.14 & 0.13 & 0.13 \\
X-n429-k61 &   65449    && 0.40 & 0.38 & 0.36 & 0.34 & 0.39 & 0.40 & 0.44 & 0.40 & 0.35 & 0.43 && 0.19 & 0.25 & 0.18 & 0.22 & 0.23 & 0.21 & 0.24 & 0.22 & 0.21 & 0.22 \\
X-n439-k37 & 36391      && 0.05 & 0.09 & 0.06 & 0.08 & 0.05 & 0.10 & 0.10 & 0.06 & 0.11 & 0.06 && 0.03 & 0.03 & 0.01 & 0.01 & 0.04 & 0.01 & 0.04 & 0.03 & 0.02 & 0.01 \\
X-n449-k29 &   55233    && 0.62 & 0.60 & 0.64 & 0.69 & 0.68 & 0.70 & 0.64 & 0.59 & 0.72 & 0.64 && 0.36 & 0.32 & 0.31 & 0.30 & 0.32 & 0.31 & 0.31 & 0.28 & 0.32 & 0.35 \\
X-n459-k26 &  24139     && 0.29 & 0.33 & 0.29 & 0.39 & 0.29 & 0.38 & 0.41 & 0.45 & 0.30 & 0.36 && 0.21 & 0.25 & 0.23 & 0.17 & 0.25 & 0.29 & 0.21 & 0.23 & 0.27 & 0.24 \\
X-n469-k138 &  221824     && 1.06 & 1.00 & 1.01 & 0.99 & 0.98 & 0.96 & 0.97 & 0.97 & 1.04 & 0.94 && 0.54 & 0.55 & 0.59 & 0.60 & 0.62 & 0.60 & 0.58 & 0.55 & 0.63 & 0.57 \\
X-n480-k70 &   89449    && 0.35 & 0.38 & 0.39 & 0.35 & 0.36 & 0.30 & 0.38 & 0.36 & 0.35 & 0.36 && 0.18 & 0.17 & 0.17 & 0.16 & 0.22 & 0.15 & 0.19 & 0.17 & 0.22 & 0.17 \\
X-n491-k59 & 66483      && 0.61 & 0.49 & 0.48 & 0.50 & 0.48 & 0.43 & 0.54 & 0.51 & 0.53 & 0.53 && 0.30 & 0.30 & 0.29 & 0.32 & 0.33 & 0.28 & 0.30 & 0.37 & 0.32 & 0.30 \\
\midrule
Mean &       && 0.37 & 0.36 & 0.37 & 0.38 & 0.37 & 0.37 & 0.39 & 0.38 & 0.38 & 0.39 && 0.23 & 0.22 & 0.24 & 0.23 & 0.24 & 0.23 & 0.23 & 0.22 & 0.23 & 0.23 \\
\bottomrule
\end{tabular}
\end{sidewaystable}

\begin{sidewaystable}
\centering
\footnotesize
\caption{Average gap on large-sized $\mathbb{X}$ instances.}
\label{tab:x-large-gaps}
\begin{tabular}{lr c rrrrrrrrrr c rrrrrrrrrr}
\toprule
 &&& &\multicolumn{9}{c}{\FILOX} && FILO2& \multicolumn{9}{c}{\FILOX (long)} \\
\cmidrule{5-13}
\cmidrule{16-24}
Instance & BKS && FILO2 & 2 & 3 & 4 & 5 & 6 & 7 & 8 & 9 & 10 && (long) & 2 & 3 & 4 & 5 & 6 & 7 & 8 & 9 & 10 \\
\midrule
X-n502-k39 &   69226    && 0.04 & 0.04 & 0.03 & 0.06 & 0.05 & 0.06 & 0.06 & 0.05 & 0.06 & 0.04 && 0.03 & 0.03 & 0.03 & 0.03 & 0.04 & 0.04 & 0.04 & 0.03 & 0.03 & 0.05 \\
X-n513-k21 &  24201     && 0.32 & 0.25 & 0.35 & 0.26 & 0.35 & 0.28 & 0.27 & 0.24 & 0.31 & 0.27 && 0.08 & 0.08 & 0.13 & 0.21 & 0.14 & 0.09 & 0.19 & 0.09 & 0.12 & 0.13 \\
X-n524-k153 &  154593     && 0.41 & 0.48 & 0.41 & 0.51 & 0.50 & 0.56 & 0.53 & 0.52 & 0.57 & 0.53 && 0.15 & 0.19 & 0.27 & 0.30 & 0.23 & 0.27 & 0.28 & 0.38 & 0.31 & 0.25 \\
X-n536-k96 & 94846      && 0.86 & 0.82 & 0.83 & 0.87 & 0.87 & 0.84 & 0.83 & 0.87 & 0.84 & 0.88 && 0.75 & 0.74 & 0.73 & 0.74 & 0.73 & 0.71 & 0.73 & 0.74 & 0.72 & 0.71 \\
X-n548-k50 &  86700     && 0.12 & 0.09 & 0.10 & 0.08 & 0.10 & 0.05 & 0.11 & 0.07 & 0.09 & 0.09 && 0.04 & 0.02 & 0.04 & 0.02 & 0.05 & 0.04 & 0.04 & 0.03 & 0.03 & 0.05 \\
X-n561-k42 &  42717     && 0.37 & 0.38 & 0.35 & 0.43 & 0.37 & 0.38 & 0.38 & 0.42 & 0.36 & 0.43 && 0.21 & 0.24 & 0.22 & 0.26 & 0.18 & 0.18 & 0.21 & 0.21 & 0.27 & 0.24 \\
X-n573-k30 &  50673     && 0.35 & 0.45 & 0.51 & 0.37 & 0.40 & 0.47 & 0.45 & 0.39 & 0.40 & 0.46 && 0.24 & 0.19 & 0.21 & 0.21 & 0.20 & 0.26 & 0.20 & 0.19 & 0.22 & 0.25 \\
X-n586-k159 &  190316     && 0.60 & 0.68 & 0.63 & 0.60 & 0.66 & 0.63 & 0.65 & 0.64 & 0.62 & 0.62 && 0.41 & 0.42 & 0.44 & 0.38 & 0.34 & 0.38 & 0.36 & 0.39 & 0.36 & 0.38 \\
X-n599-k92 &   108451    && 0.45 & 0.39 & 0.48 & 0.41 & 0.42 & 0.50 & 0.41 & 0.36 & 0.34 & 0.47 && 0.27 & 0.30 & 0.25 & 0.28 & 0.27 & 0.23 & 0.29 & 0.25 & 0.28 & 0.28 \\
X-n613-k62 &  59535     && 0.71 & 0.68 & 0.64 & 0.61 & 0.64 & 0.66 & 0.75 & 0.69 & 0.68 & 0.64 && 0.32 & 0.36 & 0.36 & 0.35 & 0.32 & 0.34 & 0.41 & 0.36 & 0.30 & 0.34 \\
X-n627-k43 &  62164     && 0.34 & 0.38 & 0.32 & 0.32 & 0.35 & 0.36 & 0.32 & 0.36 & 0.34 & 0.39 && 0.22 & 0.21 & 0.21 & 0.23 & 0.20 & 0.20 & 0.24 & 0.25 & 0.22 & 0.23 \\
X-n641-k35 &  63682     && 0.32 & 0.41 & 0.35 & 0.40 & 0.39 & 0.45 & 0.37 & 0.41 & 0.34 & 0.36 && 0.18 & 0.21 & 0.26 & 0.21 & 0.25 & 0.21 & 0.28 & 0.24 & 0.18 & 0.22 \\
X-n655-k131 & 106780      && 0.04 & 0.03 & 0.04 & 0.03 & 0.03 & 0.04 & 0.04 & 0.04 & 0.04 & 0.04 && 0.02 & 0.02 & 0.02 & 0.03 & 0.02 & 0.02 & 0.02 & 0.02 & 0.02 & 0.02 \\
X-n670-k130 &  146332     && 1.10 & 1.21 & 1.23 & 1.17 & 1.18 & 1.12 & 1.25 & 1.23 & 1.24 & 1.10 && 0.95 & 0.98 & 0.86 & 0.94 & 0.84 & 0.89 & 0.92 & 1.02 & 0.87 & 0.94 \\
X-n685-k75 & 68205      && 0.64 & 0.60 & 0.55 & 0.64 & 0.61 & 0.64 & 0.60 & 0.58 & 0.55 & 0.58 && 0.36 & 0.43 & 0.42 & 0.46 & 0.43 & 0.37 & 0.38 & 0.44 & 0.43 & 0.43 \\
X-n701-k44 & 81923      && 0.46 & 0.47 & 0.45 & 0.46 & 0.45 & 0.47 & 0.43 & 0.48 & 0.47 & 0.41 && 0.13 & 0.17 & 0.21 & 0.20 & 0.20 & 0.19 & 0.17 & 0.21 & 0.17 & 0.20 \\
X-n716-k35 & 43373      && 0.64 & 0.61 & 0.73 & 0.73 & 0.64 & 0.72 & 0.62 & 0.72 & 0.64 & 0.69 && 0.33 & 0.28 & 0.31 & 0.24 & 0.29 & 0.28 & 0.30 & 0.34 & 0.28 & 0.28 \\
X-n733-k159 &  136187     && 0.31 & 0.34 & 0.31 & 0.33 & 0.36 & 0.39 & 0.36 & 0.36 & 0.33 & 0.35 && 0.20 & 0.20 & 0.21 & 0.19 & 0.21 & 0.21 & 0.20 & 0.21 & 0.22 & 0.22 \\
X-n749-k98 &  77269     && 0.72 & 0.71 & 0.71 & 0.72 & 0.73 & 0.69 & 0.74 & 0.73 & 0.74 & 0.77 && 0.46 & 0.43 & 0.43 & 0.48 & 0.41 & 0.46 & 0.42 & 0.52 & 0.48 & 0.48 \\
X-n766-k71 &  114417     && 0.69 & 0.67 & 0.63 & 0.65 & 0.60 & 0.61 & 0.60 & 0.61 & 0.57 & 0.66 && 0.44 & 0.45 & 0.36 & 0.33 & 0.47 & 0.38 & 0.46 & 0.44 & 0.42 & 0.42 \\
X-n783-k48 &  72386     && 0.50 & 0.52 & 0.52 & 0.50 & 0.50 & 0.56 & 0.52 & 0.57 & 0.63 & 0.51 && 0.23 & 0.26 & 0.33 & 0.27 & 0.30 & 0.22 & 0.26 & 0.30 & 0.24 & 0.24 \\
X-n801-k40 &  73305     && 0.24 & 0.22 & 0.21 & 0.21 & 0.22 & 0.28 & 0.24 & 0.24 & 0.22 & 0.24 && 0.13 & 0.12 & 0.11 & 0.12 & 0.10 & 0.11 & 0.15 & 0.16 & 0.15 & 0.14 \\
X-n819-k171 &  158121     && 0.78 & 0.78 & 0.78 & 0.76 & 0.74 & 0.79 & 0.83 & 0.75 & 0.81 & 0.80 && 0.56 & 0.54 & 0.54 & 0.56 & 0.53 & 0.54 & 0.56 & 0.53 & 0.53 & 0.53 \\
X-n837-k142 &   193737    && 0.50 & 0.52 & 0.48 & 0.48 & 0.47 & 0.47 & 0.48 & 0.49 & 0.47 & 0.49 && 0.28 & 0.27 & 0.30 & 0.27 & 0.31 & 0.29 & 0.28 & 0.27 & 0.31 & 0.29 \\
X-n856-k95 &  88965     && 0.16 & 0.17 & 0.16 & 0.18 & 0.17 & 0.16 & 0.15 & 0.17 & 0.15 & 0.14 && 0.11 & 0.10 & 0.08 & 0.07 & 0.09 & 0.08 & 0.08 & 0.08 & 0.09 & 0.08 \\
X-n876-k59 &   99299    && 0.46 & 0.42 & 0.39 & 0.46 & 0.44 & 0.40 & 0.47 & 0.46 & 0.45 & 0.43 && 0.26 & 0.25 & 0.25 & 0.25 & 0.27 & 0.22 & 0.25 & 0.26 & 0.25 & 0.24 \\
X-n895-k37 &  53860     && 0.63 & 0.59 & 0.62 & 0.72 & 0.64 & 0.61 & 0.67 & 0.67 & 0.66 & 0.71 && 0.38 & 0.38 & 0.33 & 0.34 & 0.36 & 0.41 & 0.31 & 0.32 & 0.31 & 0.36 \\
X-n916-k207 &  329179     && 0.68 & 0.68 & 0.67 & 0.66 & 0.67 & 0.66 & 0.64 & 0.68 & 0.67 & 0.68 && 0.41 & 0.38 & 0.38 & 0.40 & 0.38 & 0.40 & 0.37 & 0.43 & 0.35 & 0.36 \\
X-n936-k151 &   132715    && 0.90 & 0.91 & 0.90 & 0.95 & 0.93 & 0.97 & 0.96 & 0.93 & 1.00 & 0.96 && 0.52 & 0.50 & 0.47 & 0.56 & 0.48 & 0.63 & 0.54 & 0.61 & 0.59 & 0.59 \\
X-n957-k87 &  85465     && 0.14 & 0.14 & 0.13 & 0.13 & 0.13 & 0.11 & 0.14 & 0.14 & 0.16 & 0.13 && 0.09 & 0.07 & 0.10 & 0.08 & 0.09 & 0.08 & 0.08 & 0.09 & 0.07 & 0.10 \\
X-n979-k58 &  118976     && 0.70 & 0.67 & 0.77 & 0.77 & 0.76 & 0.70 & 0.67 & 0.65 & 0.78 & 0.57 && 0.15 & 0.15 & 0.15 & 0.14 & 0.15 & 0.18 & 0.16 & 0.16 & 0.18 & 0.16 \\
X-n1001-k43 & 72355      && 0.58 & 0.63 & 0.65 & 0.59 & 0.63 & 0.69 & 0.64 & 0.63 & 0.68 & 0.62 && 0.33 & 0.29 & 0.30 & 0.29 & 0.21 & 0.26 & 0.29 & 0.31 & 0.31 & 0.28 \\
\midrule
Mean &       && 0.49 & 0.50 & 0.50 & 0.50 & 0.50 & 0.51 & 0.51 & 0.50 & 0.51 & 0.50 && 0.29 & 0.29 & 0.29 & 0.30 & 0.28 & 0.29 & 0.30 & 0.31 & 0.29 & 0.30 \\
\bottomrule
\end{tabular}
\end{sidewaystable}

\begin{sidewaystable}
\centering
\footnotesize
\caption{Average speedup on small-sized $\mathbb{X}$ instances.}
\label{tab:x-small-speedups}

\begin{tabular}{l c rrrrrrrrr c rrrrrrrrr}
\toprule
 && \multicolumn{9}{c}{\FILOX} && \multicolumn{9}{c}{\FILOX (long)} \\
\cmidrule{3-11}
\cmidrule{13-21}
Instance && 2 & 3 & 4 & 5 & 6 & 7 & 8 & 9 & 10 && 2 & 3 & 4 & 5 & 6 & 7 & 8 & 9 & 10 \\
\midrule
X-n101-k25  && 1.75 & 2.71 & 3.32 & 4.30 & 5.02 & 5.45 & 6.46 & 6.75 & 7.06 && 1.86 & 2.73 & 3.55 & 4.35 & 5.06 & 5.67 & 6.38 & 6.96 & 7.31 \\
X-n106-k14  && 1.73 & 2.59 & 3.32 & 3.93 & 4.62 & 4.99 & 5.88 & 6.37 & 7.15 && 1.71 & 2.54 & 3.12 & 3.82 & 4.36 & 5.14 & 5.46 & 5.98 & 6.61 \\
X-n110-k13  && 1.74 & 2.51 & 3.21 & 3.96 & 4.46 & 4.88 & 5.51 & 6.16 & 6.34 && 1.72 & 2.47 & 3.15 & 3.79 & 4.33 & 4.82 & 5.32 & 5.78 & 6.11 \\
X-n115-k10  && 1.73 & 2.47 & 3.18 & 3.83 & 4.48 & 4.96 & 5.40 & 5.99 & 6.11 && 1.72 & 2.46 & 3.14 & 3.77 & 4.28 & 4.76 & 5.22 & 5.66 & 5.97 \\
X-n120-k6  && 1.73 & 2.45 & 3.12 & 3.75 & 4.25 & 4.68 & 5.24 & 5.57 & 5.85 && 1.71 & 2.45 & 3.09 & 3.69 & 4.20 & 4.63 & 5.11 & 5.50 & 5.83 \\
X-n125-k30  && 1.73 & 2.57 & 3.13 & 3.59 & 4.25 & 5.39 & 6.55 & 6.74 & 7.03 && 1.83 & 2.74 & 3.67 & 4.29 & 5.05 & 5.11 & 6.49 & 6.90 & 7.70 \\
X-n129-k18  && 1.83 & 2.65 & 3.24 & 4.31 & 4.94 & 6.06 & 6.83 & 7.21 & 7.30 && 1.81 & 2.61 & 3.36 & 4.03 & 4.79 & 5.42 & 6.18 & 6.63 & 6.78 \\
X-n134-k13  && 1.73 & 2.50 & 3.21 & 3.88 & 4.46 & 5.08 & 5.48 & 6.25 & 6.31 && 1.74 & 2.46 & 3.09 & 3.85 & 4.38 & 4.71 & 5.41 & 5.86 & 6.06 \\
X-n139-k10  && 1.73 & 2.48 & 3.14 & 3.78 & 4.28 & 4.83 & 5.28 & 5.82 & 5.99 && 1.72 & 2.44 & 3.08 & 3.71 & 4.25 & 4.68 & 5.17 & 5.63 & 5.99 \\
X-n143-k7  && 1.82 & 2.63 & 3.35 & 3.78 & 4.34 & 4.93 & 5.34 & 5.82 & 5.77 && 1.77 & 2.72 & 3.08 & 3.62 & 4.08 & 4.52 & 5.18 & 5.53 & 5.90 \\
X-n148-k46  && 1.72 & 2.57 & 3.07 & 4.19 & 4.85 & 4.89 & 5.88 & 6.08 & 6.91 && 1.89 & 2.73 & 3.59 & 4.43 & 5.04 & 5.77 & 6.46 & 6.88 & 7.57 \\
X-n153-k22  && 1.93 & 2.33 & 2.86 & 3.34 & 4.23 & 4.49 & 4.63 & 4.81 & 5.24 && 1.68 & 2.30 & 2.76 & 3.36 & 3.96 & 4.25 & 4.34 & 4.76 & 5.41 \\
X-n157-k13  && 1.55 & 2.21 & 2.79 & 3.11 & 3.50 & 3.93 & 4.27 & 4.45 & 4.68 && 1.63 & 2.30 & 2.82 & 3.27 & 3.70 & 4.00 & 4.40 & 4.61 & 4.90 \\
X-n162-k11  && 1.59 & 2.23 & 3.11 & 3.26 & 3.60 & 4.17 & 4.20 & 4.59 & 4.71 && 1.66 & 2.29 & 2.74 & 3.36 & 3.75 & 3.87 & 4.19 & 4.68 & 4.81 \\
X-n167-k10  && 1.67 & 2.54 & 3.35 & 4.03 & 4.71 & 4.86 & 5.65 & 6.12 & 6.99 && 1.74 & 2.52 & 3.40 & 3.99 & 4.54 & 5.13 & 5.49 & 5.92 & 6.22 \\
X-n172-k51  && 1.92 & 2.75 & 3.68 & 4.41 & 4.75 & 5.37 & 5.75 & 6.52 & 7.30 && 1.86 & 2.68 & 3.34 & 4.24 & 4.93 & 5.54 & 6.11 & 6.56 & 7.09 \\
X-n176-k26  && 1.59 & 2.33 & 2.78 & 3.94 & 4.10 & 4.39 & 5.10 & 5.21 & 5.43 && 1.66 & 2.45 & 3.11 & 3.50 & 3.94 & 4.94 & 4.76 & 5.13 & 5.49 \\
X-n181-k23  && 1.57 & 2.30 & 2.83 & 3.40 & 3.77 & 4.32 & 4.71 & 5.00 & 5.36 && 1.65 & 2.33 & 2.85 & 3.47 & 3.90 & 4.37 & 4.67 & 5.14 & 5.45 \\
X-n186-k15  && 1.89 & 2.50 & 3.23 & 4.14 & 4.53 & 5.13 & 5.65 & 6.15 & 6.82 && 1.75 & 2.39 & 3.09 & 3.81 & 4.28 & 4.62 & 5.07 & 5.46 & 5.71 \\
X-n190-k8  && 1.69 & 2.09 & 2.72 & 3.63 & 3.77 & 4.51 & 4.91 & 5.18 & 5.71 && 1.73 & 2.43 & 2.96 & 3.72 & 4.00 & 4.63 & 4.91 & 5.31 & 5.69 \\
X-n195-k51  && 1.63 & 2.32 & 3.12 & 3.92 & 4.61 & 4.87 & 5.52 & 6.05 & 6.28 && 1.93 & 2.77 & 3.59 & 4.48 & 5.01 & 5.48 & 5.86 & 6.28 & 7.20 \\
X-n200-k36  && 1.69 & 2.44 & 3.21 & 3.77 & 4.20 & 4.48 & 5.69 & 5.52 & 6.58 && 1.92 & 2.66 & 3.71 & 4.26 & 5.25 & 5.54 & 5.77 & 6.70 & 7.26 \\
X-n204-k19  && 1.63 & 2.23 & 3.02 & 3.55 & 3.98 & 4.38 & 4.42 & 4.96 & 5.31 && 1.72 & 2.42 & 3.07 & 3.68 & 4.19 & 4.56 & 5.11 & 5.42 & 5.70 \\
X-n209-k16  && 1.75 & 2.30 & 2.99 & 3.60 & 4.10 & 4.13 & 5.05 & 5.56 & 5.81 && 1.70 & 2.58 & 3.25 & 3.70 & 4.25 & 4.97 & 5.32 & 5.78 & 5.83 \\
X-n214-k11  && 1.75 & 2.40 & 3.22 & 3.79 & 4.07 & 4.48 & 4.88 & 5.54 & 5.67 && 1.81 & 2.45 & 3.12 & 3.94 & 4.23 & 4.75 & 5.28 & 5.65 & 5.99 \\
X-n219-k73  && 1.77 & 2.62 & 3.38 & 4.12 & 4.79 & 5.74 & 6.06 & 6.59 & 7.29 && 1.79 & 2.64 & 3.46 & 4.15 & 4.84 & 5.64 & 6.17 & 6.75 & 7.42 \\
X-n223-k34  && 1.69 & 2.53 & 3.29 & 3.89 & 4.57 & 4.84 & 5.68 & 6.10 & 6.96 && 1.69 & 2.52 & 3.17 & 3.84 & 4.38 & 4.85 & 5.41 & 6.00 & 6.24 \\
X-n228-k23  && 1.71 & 2.37 & 3.04 & 3.52 & 4.19 & 4.41 & 4.83 & 5.34 & 5.39 && 1.73 & 2.36 & 3.17 & 3.73 & 4.28 & 4.43 & 5.12 & 5.21 & 5.76 \\
X-n233-k16  && 1.54 & 2.18 & 2.99 & 3.57 & 4.13 & 4.22 & 5.25 & 5.16 & 5.51 && 1.81 & 2.49 & 3.37 & 3.95 & 4.53 & 4.89 & 5.56 & 5.68 & 6.44 \\
X-n237-k14  && 1.61 & 2.32 & 2.95 & 3.57 & 4.29 & 4.63 & 5.07 & 5.60 & 5.53 && 1.68 & 2.31 & 2.96 & 3.61 & 4.17 & 4.41 & 5.02 & 5.30 & 5.38 \\
X-n242-k48  && 2.01 & 2.76 & 3.68 & 4.57 & 5.00 & 5.66 & 6.60 & 7.36 & 7.19 && 1.82 & 2.53 & 3.17 & 4.13 & 4.81 & 5.10 & 5.67 & 6.63 & 6.91 \\
X-n247-k50  && 1.69 & 2.46 & 3.08 & 3.72 & 4.25 & 4.67 & 5.29 & 5.53 & 5.69 && 1.78 & 2.46 & 3.29 & 3.92 & 4.35 & 5.21 & 5.36 & 5.61 & 6.36 \\
\midrule
Mean  && 1.71 & 2.44 & 3.14 & 3.79 & 4.32 & 4.80 & 5.34 & 5.75 & 6.09 && 1.75 & 2.50 & 3.18 & 3.83 & 4.37 & 4.85 & 5.33 & 5.75 & 6.15 \\
\bottomrule
\end{tabular}
\end{sidewaystable}

\begin{sidewaystable}
\centering
\footnotesize
\caption{Average speedup on medium-sized $\mathbb{X}$ instances.}
\label{tab:x-medium-speedups}
\begin{tabular}{l c rrrrrrrrr c rrrrrrrrr}
\toprule
 && \multicolumn{9}{c}{\FILOX} && \multicolumn{9}{c}{\FILOX (long)} \\
\cmidrule{3-11}
\cmidrule{13-21}
Instance && 2 & 3 & 4 & 5 & 6 & 7 & 8 & 9 & 10 && 2 & 3 & 4 & 5 & 6 & 7 & 8 & 9 & 10 \\
\midrule
X-n251-k28  && 1.75 & 2.64 & 3.19 & 3.84 & 4.59 & 5.20 & 5.29 & 6.06 & 6.30 && 1.67 & 2.24 & 3.21 & 3.91 & 4.11 & 4.60 & 5.46 & 5.93 & 5.98 \\
X-n256-k16  && 1.60 & 2.32 & 2.79 & 3.40 & 3.84 & 4.11 & 4.53 & 4.97 & 5.11 && 1.61 & 2.18 & 2.79 & 3.31 & 3.72 & 4.04 & 4.35 & 4.72 & 4.99 \\
X-n261-k13  && 1.57 & 2.36 & 2.99 & 3.50 & 4.25 & 4.35 & 4.98 & 5.17 & 5.59 && 1.91 & 2.54 & 3.15 & 3.68 & 4.18 & 4.44 & 5.18 & 5.56 & 5.81 \\
X-n266-k58  && 1.76 & 2.77 & 3.32 & 4.04 & 4.43 & 5.54 & 5.85 & 5.90 & 6.83 && 1.75 & 2.26 & 3.07 & 3.72 & 4.46 & 4.94 & 5.23 & 5.79 & 6.57 \\
X-n270-k35  && 1.71 & 2.39 & 2.88 & 3.63 & 4.22 & 4.51 & 5.24 & 5.35 & 5.83 && 1.77 & 2.59 & 3.24 & 4.02 & 4.52 & 5.25 & 5.48 & 5.99 & 6.29 \\
X-n275-k28  && 1.66 & 2.48 & 3.09 & 3.63 & 4.20 & 4.75 & 5.32 & 5.66 & 6.00 && 1.71 & 2.53 & 3.13 & 3.86 & 4.28 & 4.81 & 5.19 & 5.75 & 6.08 \\
X-n280-k17  && 1.64 & 2.26 & 2.74 & 3.16 & 3.70 & 3.95 & 4.53 & 4.87 & 4.74 && 1.55 & 2.19 & 2.52 & 3.15 & 3.69 & 4.01 & 4.40 & 4.61 & 4.95 \\
X-n284-k15  && 1.67 & 2.47 & 3.48 & 3.86 & 4.74 & 5.23 & 5.96 & 6.14 & 6.40 && 1.70 & 2.47 & 3.24 & 3.79 & 4.42 & 4.74 & 5.29 & 5.70 & 5.89 \\
X-n289-k60  && 1.91 & 2.84 & 3.50 & 4.30 & 4.62 & 5.42 & 5.84 & 6.38 & 7.38 && 1.83 & 2.67 & 3.71 & 4.51 & 5.28 & 6.27 & 6.27 & 6.37 & 8.08 \\
X-n294-k50  && 1.89 & 2.66 & 3.41 & 4.02 & 4.59 & 5.00 & 5.42 & 5.84 & 6.34 && 1.72 & 2.50 & 3.14 & 3.99 & 4.75 & 5.10 & 5.48 & 5.90 & 6.38 \\
X-n298-k31  && 1.71 & 2.59 & 3.22 & 3.98 & 4.42 & 5.03 & 5.54 & 6.00 & 6.08 && 1.84 & 2.53 & 3.30 & 3.87 & 4.49 & 4.85 & 5.48 & 5.88 & 6.36 \\
X-n303-k21  && 1.67 & 2.41 & 2.92 & 3.59 & 3.96 & 4.28 & 4.62 & 4.92 & 5.19 && 1.73 & 2.48 & 3.02 & 3.58 & 4.02 & 4.29 & 4.67 & 4.92 & 5.07 \\
X-n308-k13  && 1.53 & 2.22 & 2.44 & 2.90 & 3.32 & 3.70 & 3.77 & 4.27 & 4.12 && 1.60 & 2.16 & 2.68 & 3.18 & 3.44 & 3.84 & 4.13 & 4.48 & 4.52 \\
X-n313-k71  && 1.67 & 2.50 & 3.13 & 3.75 & 4.59 & 4.79 & 5.62 & 6.12 & 6.19 && 1.88 & 2.47 & 3.37 & 4.12 & 4.75 & 5.44 & 6.23 & 5.88 & 6.52 \\
X-n317-k53  && 1.66 & 2.33 & 2.88 & 3.53 & 3.93 & 4.44 & 4.85 & 5.17 & 5.50 && 1.64 & 2.34 & 2.97 & 3.50 & 4.02 & 4.43 & 4.88 & 5.24 & 5.59 \\
X-n322-k28  && 1.78 & 2.53 & 3.30 & 4.06 & 4.38 & 4.92 & 5.69 & 5.82 & 6.17 && 1.83 & 2.39 & 3.24 & 3.85 & 4.41 & 4.88 & 5.13 & 5.82 & 5.70 \\
X-n327-k20  && 1.75 & 2.33 & 3.17 & 3.74 & 4.10 & 4.54 & 5.00 & 5.62 & 5.72 && 1.71 & 2.48 & 3.18 & 3.85 & 4.05 & 4.44 & 5.02 & 5.53 & 5.69 \\
X-n331-k15  && 1.70 & 2.43 & 2.95 & 3.51 & 4.15 & 4.59 & 5.08 & 5.19 & 5.65 && 1.70 & 2.38 & 3.06 & 3.57 & 3.96 & 4.31 & 4.92 & 5.15 & 5.64 \\
X-n336-k84  && 1.91 & 2.55 & 3.26 & 4.16 & 4.67 & 5.13 & 6.14 & 6.29 & 6.86 && 1.84 & 2.58 & 3.53 & 4.01 & 4.67 & 5.27 & 6.28 & 6.21 & 7.22 \\
X-n344-k43  && 1.65 & 2.33 & 3.04 & 3.46 & 3.93 & 4.24 & 4.83 & 5.21 & 5.37 && 1.77 & 2.49 & 3.09 & 3.75 & 4.43 & 4.93 & 4.95 & 5.38 & 6.06 \\
X-n351-k40  && 1.67 & 2.28 & 2.91 & 3.29 & 4.17 & 4.39 & 4.59 & 5.15 & 5.26 && 1.74 & 2.54 & 3.16 & 3.91 & 4.17 & 4.55 & 5.28 & 5.49 & 5.78 \\
X-n359-k29  && 1.73 & 2.41 & 3.25 & 3.86 & 4.21 & 4.79 & 5.15 & 5.91 & 6.11 && 1.74 & 2.49 & 3.12 & 3.82 & 4.30 & 4.84 & 5.10 & 5.70 & 6.18 \\
X-n367-k17  && 1.75 & 2.40 & 3.06 & 3.49 & 4.04 & 4.38 & 4.93 & 5.11 & 5.02 && 1.68 & 2.30 & 2.95 & 3.49 & 3.92 & 4.36 & 4.79 & 5.04 & 5.37 \\
X-n376-k94  && 1.70 & 2.37 & 3.02 & 3.83 & 4.18 & 4.83 & 5.40 & 5.78 & 6.16 && 1.79 & 2.60 & 3.30 & 3.91 & 4.54 & 5.18 & 5.64 & 6.16 & 6.59 \\
X-n384-k52  && 1.64 & 2.48 & 3.10 & 3.79 & 4.31 & 4.96 & 5.41 & 6.06 & 6.49 && 1.74 & 2.57 & 3.36 & 4.10 & 4.70 & 5.07 & 5.70 & 6.54 & 6.50 \\
X-n393-k38  && 1.74 & 2.51 & 2.98 & 3.70 & 4.25 & 4.82 & 5.19 & 5.51 & 5.74 && 1.71 & 2.49 & 3.08 & 3.81 & 4.33 & 4.82 & 5.44 & 5.67 & 6.01 \\
X-n401-k29  && 1.58 & 2.39 & 2.96 & 3.57 & 4.20 & 4.60 & 5.07 & 5.61 & 5.78 && 1.71 & 2.46 & 3.07 & 3.81 & 4.20 & 4.50 & 5.06 & 5.51 & 5.98 \\
X-n411-k19  && 1.50 & 1.99 & 2.50 & 2.97 & 3.22 & 3.43 & 3.72 & 4.03 & 4.16 && 1.48 & 2.14 & 2.60 & 2.94 & 3.24 & 3.55 & 3.81 & 3.97 & 4.19 \\
X-n420-k130  && 1.84 & 2.58 & 3.57 & 3.98 & 4.53 & 5.22 & 5.77 & 6.55 & 6.93 && 1.98 & 2.82 & 3.86 & 4.37 & 5.40 & 5.74 & 6.51 & 7.31 & 7.74 \\
X-n429-k61  && 1.69 & 2.43 & 3.35 & 3.82 & 4.28 & 4.68 & 5.43 & 5.90 & 6.24 && 1.70 & 2.53 & 3.29 & 3.88 & 4.50 & 4.85 & 5.42 & 6.08 & 6.34 \\
X-n439-k37  && 1.71 & 2.34 & 3.03 & 3.76 & 4.22 & 4.70 & 5.10 & 5.44 & 5.77 && 1.69 & 2.35 & 2.91 & 3.54 & 4.00 & 4.53 & 4.98 & 5.20 & 5.44 \\
X-n449-k29  && 1.77 & 2.47 & 3.11 & 3.87 & 4.39 & 4.85 & 5.37 & 5.70 & 5.88 && 1.84 & 2.50 & 3.29 & 4.06 & 4.55 & 5.23 & 5.67 & 6.02 & 6.50 \\
X-n459-k26  && 1.71 & 2.39 & 2.98 & 3.50 & 3.97 & 4.45 & 4.86 & 5.25 & 5.60 && 1.74 & 2.50 & 3.11 & 3.69 & 4.05 & 4.64 & 5.03 & 5.31 & 5.64 \\
X-n469-k138  && 1.98 & 2.88 & 3.74 & 4.38 & 5.70 & 6.14 & 6.84 & 7.64 & 7.64 && 1.84 & 2.50 & 3.46 & 4.34 & 4.39 & 5.30 & 6.10 & 6.70 & 7.33 \\
X-n480-k70  && 1.71 & 2.45 & 3.23 & 3.91 & 4.47 & 5.06 & 5.56 & 6.05 & 6.49 && 1.66 & 2.44 & 3.28 & 3.88 & 4.45 & 5.10 & 5.61 & 6.17 & 6.37 \\
X-n491-k59  && 1.73 & 2.46 & 3.16 & 3.68 & 4.32 & 4.61 & 5.04 & 5.38 & 5.55 && 1.74 & 2.51 & 3.19 & 3.84 & 4.54 & 4.93 & 5.30 & 5.71 & 6.18 \\
\midrule
Mean  && 1.70 & 2.44 & 3.07 & 3.68 & 4.21 & 4.67 & 5.15 & 5.54 & 5.82 && 1.73 & 2.44 & 3.13 & 3.76 & 4.25 & 4.73 & 5.19 & 5.58 & 5.95 \\
\bottomrule
\end{tabular}
\end{sidewaystable}

\begin{sidewaystable}
\centering
\footnotesize
\caption{Average speedup on large-sized $\mathbb{X}$ instances.}
\label{tab:x-large-speedups}
\begin{tabular}{l c rrrrrrrrr c rrrrrrrrr}
\toprule
 && \multicolumn{9}{c}{\FILOX} && \multicolumn{9}{c}{\FILOX (long)} \\
\cmidrule{3-11}
\cmidrule{13-21}
Instance && 2 & 3 & 4 & 5 & 6 & 7 & 8 & 9 & 10 && 2 & 3 & 4 & 5 & 6 & 7 & 8 & 9 & 10 \\
\midrule
X-n502-k39  && 1.63 & 2.24 & 2.81 & 3.37 & 3.69 & 4.02 & 4.41 & 4.65 & 4.93 && 1.59 & 2.26 & 2.78 & 3.23 & 3.67 & 4.02 & 4.38 & 4.66 & 4.84 \\
X-n513-k21  && 1.57 & 2.29 & 2.78 & 3.23 & 3.64 & 4.03 & 4.49 & 4.62 & 4.95 && 1.61 & 2.26 & 2.69 & 3.32 & 3.67 & 3.89 & 4.35 & 4.77 & 4.95 \\
X-n524-k153  && 1.72 & 2.51 & 3.14 & 3.76 & 4.26 & 4.82 & 5.23 & 5.55 & 6.03 && 1.75 & 2.48 & 3.14 & 3.86 & 4.30 & 4.87 & 5.31 & 5.70 & 6.11 \\
X-n536-k96  && 1.76 & 2.53 & 3.23 & 3.92 & 4.26 & 4.80 & 5.37 & 5.86 & 6.03 && 1.79 & 2.56 & 3.28 & 3.95 & 4.73 & 5.31 & 5.71 & 6.47 & 6.80 \\
X-n548-k50  && 1.76 & 2.52 & 3.23 & 3.91 & 4.55 & 5.10 & 5.68 & 6.02 & 6.36 && 1.72 & 2.46 & 3.20 & 3.73 & 4.37 & 4.86 & 5.22 & 5.73 & 6.03 \\
X-n561-k42  && 1.69 & 2.50 & 2.94 & 3.68 & 4.28 & 4.64 & 5.07 & 5.40 & 5.58 && 1.67 & 2.43 & 3.01 & 3.71 & 4.13 & 4.55 & 5.02 & 5.14 & 5.53 \\
X-n573-k30  && 1.63 & 2.35 & 2.84 & 3.47 & 3.81 & 4.47 & 5.01 & 5.09 & 5.36 && 1.67 & 2.35 & 2.92 & 3.48 & 3.71 & 4.34 & 4.71 & 4.94 & 4.85 \\
X-n586-k159  && 1.76 & 2.54 & 3.16 & 4.22 & 4.75 & 5.18 & 5.52 & 6.17 & 6.79 && 1.74 & 2.75 & 3.46 & 4.40 & 5.05 & 5.80 & 6.32 & 6.88 & 7.40 \\
X-n599-k92  && 1.84 & 2.59 & 3.30 & 4.17 & 4.88 & 5.36 & 5.89 & 6.54 & 6.60 && 1.76 & 2.60 & 3.23 & 4.07 & 4.61 & 5.06 & 5.78 & 6.01 & 6.50 \\
X-n613-k62  && 1.75 & 2.51 & 3.26 & 3.78 & 4.37 & 4.79 & 5.29 & 5.56 & 5.84 && 1.71 & 2.42 & 3.15 & 3.84 & 4.34 & 4.88 & 5.25 & 5.57 & 5.92 \\
X-n627-k43  && 1.70 & 2.44 & 3.09 & 3.71 & 4.26 & 4.61 & 5.15 & 5.58 & 5.63 && 1.71 & 2.34 & 3.04 & 3.57 & 4.05 & 4.66 & 5.07 & 5.40 & 5.61 \\
X-n641-k35  && 1.67 & 2.51 & 3.13 & 3.80 & 4.31 & 4.64 & 5.33 & 5.71 & 6.15 && 1.69 & 2.50 & 3.16 & 3.61 & 4.27 & 4.80 & 5.06 & 5.86 & 5.86 \\
X-n655-k131  && 1.76 & 2.51 & 3.17 & 3.87 & 4.54 & 5.17 & 5.58 & 6.00 & 6.50 && 1.75 & 2.46 & 3.18 & 3.88 & 4.44 & 5.05 & 5.56 & 6.11 & 6.34 \\
X-n670-k130  && 1.76 & 2.51 & 3.28 & 3.77 & 4.37 & 4.93 & 5.39 & 5.89 & 6.20 && 1.70 & 2.38 & 3.17 & 3.84 & 4.42 & 5.11 & 5.35 & 6.05 & 6.13 \\
X-n685-k75  && 1.69 & 2.40 & 3.01 & 3.49 & 4.19 & 4.54 & 4.91 & 5.24 & 5.56 && 1.66 & 2.41 & 3.03 & 3.71 & 4.27 & 4.78 & 5.07 & 5.53 & 5.77 \\
X-n701-k44  && 1.76 & 2.55 & 3.19 & 3.79 & 4.44 & 4.99 & 5.45 & 5.93 & 6.18 && 1.72 & 2.42 & 3.11 & 3.65 & 4.29 & 4.85 & 5.17 & 5.69 & 5.93 \\
X-n716-k35  && 1.73 & 2.37 & 3.03 & 3.69 & 4.10 & 4.54 & 4.89 & 5.26 & 5.54 && 1.62 & 2.29 & 2.86 & 3.40 & 3.82 & 4.29 & 4.60 & 4.84 & 5.16 \\
X-n733-k159  && 1.78 & 2.53 & 3.31 & 3.92 & 4.58 & 5.02 & 5.68 & 6.08 & 6.54 && 1.77 & 2.60 & 3.17 & 4.15 & 4.77 & 5.23 & 5.93 & 6.39 & 6.71 \\
X-n749-k98  && 1.67 & 2.44 & 3.20 & 3.83 & 4.31 & 4.92 & 5.30 & 5.67 & 5.79 && 1.79 & 2.67 & 3.35 & 4.00 & 4.65 & 5.18 & 5.47 & 6.15 & 6.40 \\
X-n766-k71  && 1.69 & 2.39 & 3.02 & 3.61 & 4.14 & 4.69 & 5.00 & 5.41 & 5.71 && 1.69 & 2.42 & 3.10 & 3.63 & 4.28 & 4.77 & 5.11 & 5.50 & 5.64 \\
X-n783-k48  && 1.72 & 2.46 & 3.25 & 3.91 & 4.36 & 5.03 & 5.30 & 5.94 & 6.21 && 1.76 & 2.52 & 3.16 & 3.83 & 4.42 & 5.00 & 5.48 & 5.78 & 6.09 \\
X-n801-k40  && 1.65 & 2.42 & 3.07 & 3.64 & 4.10 & 4.71 & 5.08 & 5.52 & 5.88 && 1.71 & 2.43 & 3.07 & 3.70 & 4.21 & 4.66 & 5.02 & 5.45 & 5.78 \\
X-n819-k171  && 1.80 & 2.54 & 3.38 & 4.09 & 4.61 & 5.45 & 5.96 & 6.15 & 6.89 && 1.77 & 2.45 & 3.26 & 4.15 & 4.65 & 5.33 & 5.98 & 6.46 & 6.53 \\
X-n837-k142  && 1.78 & 2.63 & 3.50 & 4.24 & 4.94 & 5.69 & 6.18 & 6.83 & 7.17 && 1.77 & 2.57 & 3.53 & 4.26 & 4.87 & 5.38 & 6.35 & 6.52 & 6.83 \\
X-n856-k95  && 1.73 & 2.46 & 3.14 & 3.79 & 4.38 & 4.96 & 5.40 & 6.02 & 6.12 && 1.71 & 2.47 & 3.10 & 3.74 & 4.27 & 4.85 & 5.22 & 5.72 & 6.01 \\
X-n876-k59  && 1.72 & 2.44 & 3.13 & 3.76 & 4.25 & 4.79 & 5.20 & 5.59 & 5.93 && 1.70 & 2.38 & 3.03 & 3.65 & 4.20 & 4.64 & 5.04 & 5.39 & 5.65 \\
X-n895-k37  && 1.66 & 2.36 & 3.00 & 3.64 & 4.11 & 4.59 & 4.93 & 5.49 & 5.78 && 1.66 & 2.37 & 2.98 & 3.60 & 4.05 & 4.59 & 4.94 & 5.26 & 5.60 \\
X-n916-k207  && 1.81 & 2.69 & 3.50 & 4.09 & 5.17 & 5.84 & 6.35 & 6.77 & 7.17 && 1.87 & 2.79 & 3.55 & 4.48 & 5.19 & 5.96 & 6.39 & 7.16 & 7.76 \\
X-n936-k151  && 1.78 & 2.51 & 3.19 & 3.91 & 4.49 & 4.94 & 5.39 & 5.92 & 5.99 && 1.80 & 2.64 & 3.31 & 4.06 & 4.67 & 5.33 & 5.76 & 6.27 & 6.55 \\
X-n957-k87  && 1.73 & 2.51 & 3.18 & 3.86 & 4.43 & 4.99 & 5.47 & 6.02 & 6.46 && 1.69 & 2.44 & 3.10 & 3.71 & 4.31 & 4.81 & 5.23 & 5.72 & 5.98 \\
X-n979-k58  && 1.71 & 2.45 & 3.04 & 3.71 & 4.37 & 4.67 & 5.14 & 5.67 & 5.90 && 1.77 & 2.53 & 3.27 & 3.86 & 4.44 & 5.05 & 5.35 & 5.94 & 6.38 \\
X-n1001-k43  && 1.75 & 2.45 & 3.13 & 3.85 & 4.27 & 5.01 & 5.39 & 5.78 & 6.17 && 1.72 & 2.42 & 3.01 & 3.77 & 4.24 & 4.76 & 5.15 & 5.62 & 5.84 \\
\midrule
Mean  && 1.72 & 2.47 & 3.13 & 3.78 & 4.33 & 4.84 & 5.29 & 5.71 & 6.02 && 1.72 & 2.46 & 3.13 & 3.78 & 4.32 & 4.85 & 5.27 & 5.71 & 5.98 \\
\bottomrule
\end{tabular}
\end{sidewaystable}

\clearpage

\subsection{Computational Details for the \texorpdfstring{$\mathbb{B}$}{B} Instances} \label{appendix:b}
In the following, we report detailed results for the computations on the $\mathbb{B}$ dataset.
\begin{itemize}
    \item Figure \ref{fig:FILO2x.boxplots.quality.b} shows boxplots for the average gaps obtained by the approaches.
    \item Table \ref{table:p_values_b} shows the result of the one-tailed Wilcoxon signed rank test.
    \item Tables \ref{tab:b-time-per-procedure} and \ref{tab:b-long-time-per-procedure} report the computing time for the main algorithm procedures.
    \item Table \ref{tab:b-gaps} shows the average gap for each instance in the dataset.
    \item Table \ref{tab:b-speedups} shows the average speedup for each instance in the dataset.
\end{itemize}

\clearpage

\begin{figure}
\centering
\caption{Average gaps obtained by FILO2 and \FILOX (denoted by the number of solvers) on the $\mathbb{B}$ dataset when executing the standard (top) and long (bottom) versions. The median value is shown on the right of each boxplot.}
\scriptsize
	\makeatletter
	\pgfplotsset{
		boxplot/draw/average/.code={ 
			\draw [/pgfplots/boxplot/every average/.try]
			\pgfextra
			\pgftransformshift{%
				\pgfplotsboxplotpointabbox
				{\pgfplotsboxplotvalue{average}}
				{0.5}%
			}%
			\endpgfextra
			;
		},
		boxplot/draw/median/.code={
			\draw [/pgfplots/boxplot/every median/.try]
			(boxplot box cs:\pgfplotsboxplotvalue{median},0)
			node[xshift=0.55cm, font=\tiny] {\pgfmathprintnumber{\pgfplotsboxplotvalue{median}}}
			--
			(boxplot box cs:\pgfplotsboxplotvalue{median},1);
		},
	}
	\makeatother
\begin{tikzpicture}
\scriptsize

\begin{axis}
[
xlabel={Algorithm},
ylabel={Average \% gap},
ymajorgrids=true,
yminorgrids=true,
minor y tick num=4,
minor grid style={line width=.01pt,draw=black!10},
major grid style={line width=.01pt,draw=black!30},
clip=false,
boxplot/draw direction=y,
boxplot/variable width,
boxplot/every median/.style={black,very thick,solid},
width=\textwidth,
height=150pt,
ylabel style={align=center}, 
y tick label style={align=right},
x tick label style={align=center},
xtick={0, 1, 2, 3, 4, 5, 6, 7, 8, 9, 10},
xticklabels={FILO2, 2, 3, 4, 5, 6, 7, 8, 9, 10},
xmin=-0.75,
xmax=10,
scatter/classes={ a={mark=star}, b={mark=*}}]

\addplot[mark=*, mark size=0.5pt,boxplot, boxplot prepared={draw position=0,
median=0.97,upper quartile=1.326,lower quartile=0.651,upper whisker=2.052,lower whisker=0.412,sample size=1}]
coordinates{
};
\addplot[mark=*, mark size=0.5pt,boxplot, boxplot prepared={draw position=1,
median=0.935,upper quartile=1.38,lower quartile=0.667,upper whisker=2.073,lower whisker=0.416,sample size=1}]
coordinates{
};
\addplot[mark=*, mark size=0.5pt,boxplot, boxplot prepared={draw position=2,
median=0.943,upper quartile=1.381,lower quartile=0.664,upper whisker=2.061,lower whisker=0.413,sample size=1}]
coordinates{
};
\addplot[mark=*, mark size=0.5pt,boxplot, boxplot prepared={draw position=3,
median=0.962,upper quartile=1.357,lower quartile=0.657,upper whisker=2.104,lower whisker=0.413,sample size=1}]
coordinates{
};
\addplot[mark=*, mark size=0.5pt,boxplot, boxplot prepared={draw position=4,
median=0.954,upper quartile=1.372,lower quartile=0.673,upper whisker=2.044,lower whisker=0.42,sample size=1}]
coordinates{
};
\addplot[mark=*, mark size=0.5pt,boxplot, boxplot prepared={draw position=5,
median=0.983,upper quartile=1.354,lower quartile=0.66,upper whisker=2.066,lower whisker=0.424,sample size=1}]
coordinates{
};
\addplot[mark=*, mark size=0.5pt,boxplot, boxplot prepared={draw position=6,
median=0.953,upper quartile=1.39,lower quartile=0.651,upper whisker=2.073,lower whisker=0.42,sample size=1}]
coordinates{
};
\addplot[mark=*, mark size=0.5pt,boxplot, boxplot prepared={draw position=7,
median=0.951,upper quartile=1.394,lower quartile=0.658,upper whisker=2.062,lower whisker=0.394,sample size=1}]
coordinates{
};
\addplot[mark=*, mark size=0.5pt,boxplot, boxplot prepared={draw position=8,
median=0.969,upper quartile=1.39,lower quartile=0.672,upper whisker=2.055,lower whisker=0.431,sample size=1}]
coordinates{
};
\addplot[mark=*, mark size=0.5pt,boxplot, boxplot prepared={draw position=9,
median=0.964,upper quartile=1.378,lower quartile=0.658,upper whisker=2.095,lower whisker=0.451,sample size=1}]
coordinates{
};

\end{axis}
\end{tikzpicture}
\vspace{-0.9cm}
\scriptsize
	\makeatletter
	\pgfplotsset{
		boxplot/draw/average/.code={ 
			\draw [/pgfplots/boxplot/every average/.try]
			\pgfextra
			\pgftransformshift{%
				\pgfplotsboxplotpointabbox
				{\pgfplotsboxplotvalue{average}}
				{0.5}%
			}%
			\endpgfextra
			;
		},
		boxplot/draw/median/.code={
			\draw [/pgfplots/boxplot/every median/.try]
			(boxplot box cs:\pgfplotsboxplotvalue{median},0)
			node[xshift=0.55cm, font=\tiny] {\pgfmathprintnumber{\pgfplotsboxplotvalue{median}}}
			--
			(boxplot box cs:\pgfplotsboxplotvalue{median},1);
		},
	}
	\makeatother
\begin{tikzpicture}
\scriptsize

\begin{axis}
[
xlabel={Algorithm},
ylabel={Average \% gap},
ymajorgrids=true,
yminorgrids=true,
minor y tick num=4,
minor grid style={line width=.01pt,draw=black!10},
major grid style={line width=.01pt,draw=black!30},
clip=false,
boxplot/draw direction=y,
boxplot/variable width,
boxplot/every median/.style={black,very thick,solid},
width=\textwidth,
height=150pt,
ylabel style={align=center}, 
y tick label style={align=right},
x tick label style={align=center},
xtick={0, 1, 2, 3, 4, 5, 6, 7, 8, 9, 10},
xticklabels={FILO2, 2, 3, 4, 5, 6, 7, 8, 9, 10},
xmin=-0.75,
xmax=10,
scatter/classes={ a={mark=star}, b={mark=*}}]

\addplot[mark=*, mark size=0.5pt,boxplot, boxplot prepared={draw position=0,
median=0.343,upper quartile=0.393,lower quartile=0.277,upper whisker=0.538,lower whisker=0.195,sample size=1}]
coordinates{
(0,0.636)
};
\addplot[mark=*, mark size=0.5pt,boxplot, boxplot prepared={draw position=1,
median=0.374,upper quartile=0.396,lower quartile=0.273,upper whisker=0.52,lower whisker=0.193,sample size=1}]
coordinates{
(1,0.602)
};
\addplot[mark=*, mark size=0.5pt,boxplot, boxplot prepared={draw position=2,
median=0.343,upper quartile=0.391,lower quartile=0.282,upper whisker=0.478,lower whisker=0.211,sample size=1}]
coordinates{
(2,0.585)
};
\addplot[mark=*, mark size=0.5pt,boxplot, boxplot prepared={draw position=3,
median=0.349,upper quartile=0.386,lower quartile=0.282,upper whisker=0.387,lower whisker=0.19,sample size=1}]
coordinates{
(3,0.548)
(3,0.608)
};
\addplot[mark=*, mark size=0.5pt,boxplot, boxplot prepared={draw position=4,
median=0.343,upper quartile=0.388,lower quartile=0.268,upper whisker=0.525,lower whisker=0.194,sample size=1}]
coordinates{
(4,0.593)
};
\addplot[mark=*, mark size=0.5pt,boxplot, boxplot prepared={draw position=5,
median=0.366,upper quartile=0.404,lower quartile=0.266,upper whisker=0.583,lower whisker=0.2,sample size=1}]
coordinates{
};
\addplot[mark=*, mark size=0.5pt,boxplot, boxplot prepared={draw position=6,
median=0.339,upper quartile=0.395,lower quartile=0.258,upper whisker=0.565,lower whisker=0.203,sample size=1}]
coordinates{
};
\addplot[mark=*, mark size=0.5pt,boxplot, boxplot prepared={draw position=7,
median=0.348,upper quartile=0.394,lower quartile=0.272,upper whisker=0.536,lower whisker=0.219,sample size=1}]
coordinates{
(7,0.614)
};
\addplot[mark=*, mark size=0.5pt,boxplot, boxplot prepared={draw position=8,
median=0.38,upper quartile=0.437,lower quartile=0.27,upper whisker=0.592,lower whisker=0.214,sample size=1}]
coordinates{
};
\addplot[mark=*, mark size=0.5pt,boxplot, boxplot prepared={draw position=9,
median=0.374,upper quartile=0.388,lower quartile=0.275,upper whisker=0.538,lower whisker=0.213,sample size=1}]
coordinates{
(9,0.56)
};

\end{axis}
\end{tikzpicture}
\label{fig:FILO2x.boxplots.quality.b}
\end{figure}

\begin{table}
	\caption{Computations on the $\mathbb{B}$ dataset: $p$-values for \FILOX vs FILO2 on the left and \FILOX (long) vs FILO2 (long) on the right.
	\\
	\footnotesize
		$p$-values in bold are associated with rejected hypothesis when $\bar{\alpha} = 0.002778$.\\
The last row of each group contains a $p$-value interpretation. 
In particular, \FILOX is not statistically different from FILO2 when $H_0$ cannot be rejected (Similar), \FILOX is statistically better when both $H_0$ and $H_1$ are rejected (Better), and, finally, \FILOX is statistically worse when $H_0$ is rejected and $H_1$ is not rejected (Worse).}
	\label{table:p_values_b}
	\centering
	\footnotesize
	\begin{tabular}{rrrrrrr}
	\toprule
        & & \multicolumn{2}{c}{\FILOX vs FILO2} & & \multicolumn{2}{c}{\FILOX (long) vs FILO2 (long)} \\
        \cmidrule{3-4}
        \cmidrule{6-7}
Solvers & & $H_0$ / $H_1$ & Outcome & & $H_0$ / $H_1$ & Outcome \\
	\midrule
2 &&   0.375000 /   0.838867 & Similar &&   0.695312 /   0.687500 & Similar \\
3 &&   0.556641 /   0.753906 & Similar &&   0.695312 /   0.687500 & Similar \\
4 &&   0.492188 /   0.784180 & Similar &&   0.556641 /   0.753906 & Similar \\
5 &&   0.845703 /   0.615234 & Similar &&   0.625000 /   0.312500 & Similar \\
6 &&   0.232422 /   0.903320 & Similar &&   0.921875 /   0.577148 & Similar \\
7 &&   0.695312 /   0.687500 & Similar &&   0.322266 /   0.161133 & Similar \\
8 &&   0.695312 /   0.687500 & Similar &&   1.000000 /   0.539062 & Similar \\
9 &&   0.083984 /   0.967773 & Similar &&   0.625000 /   0.312500 & Similar \\
10 &&   0.375000 /   0.838867 & Similar &&   0.769531 /   0.652344 & Similar \\
	\bottomrule
	\end{tabular}
\end{table}

\begin{table}
    \centering
    \footnotesize
    \caption{Computing time in seconds for the main algorithm FILO2 and \FILOX procedures when solving the $\mathbb{B}$ dataset.}
    \label{tab:b-time-per-procedure}
    \begin{tabular}{lrrrrrrrrrr}
        \toprule
        & & \multicolumn{8}{c}{\FILOX} \\
        \cmidrule{3-11}
        & FILO2 & 2 & 3 & 4 & 5 & 6 & 7 & 8 & 9 & 10\\
        \midrule
Instance preprocessing & 2.55 & 1.37 & 0.92 & 0.70 & 0.57 & 0.49 & 0.42 & 0.37 & 0.33 & 0.31\\
Construction phase & 0.10 & 0.10 & 0.10 & 0.10 & 0.10 & 0.10 & 0.10 & 0.10 & 0.10 & 0.10\\
Greedy route estimation & 0.00 & 0.00 & 0.00 & 0.00 & 0.00 & 0.00 & 0.00 & 0.00 & 0.00 & 0.00\\
Move generators initialization & 0.03 & 0.05 & 0.05 & 0.05 & 0.05 & 0.05 & 0.06 & 0.06 & 0.06 & 0.06\\
Route minimization procedure & 0.60 & 0.54 & 0.54 & 0.55 & 0.55 & 0.55 & 0.55 & 0.55 & 0.56 & 0.57\\
Core optimization procedure & 86.24 & 52.35 & 35.41 & 28.02 & 23.32 & 20.27 & 17.86 & 16.11 & 14.87 & 14.16\\

         \bottomrule
    \end{tabular}
\end{table}

\begin{table}
    \centering
    \footnotesize
    \caption{Computing time in seconds for the main algorithm FILO2 (long) and \FILOX (long) procedures when solving the $\mathbb{B}$ dataset.}
    \label{tab:b-long-time-per-procedure}
    \begin{tabular}{lrrrrrrrrrr}
        \toprule
        & & \multicolumn{8}{c}{\FILOX} \\
        \cmidrule{3-11}
        & FILO2 & 2 & 3 & 4 & 5 & 6 & 7 & 8 & 9 & 10\\
        \midrule
Instance preprocessing & 2.55 & 1.35 & 0.92 & 0.70 & 0.57 & 0.49 & 0.42 & 0.37 & 0.34 & 0.31\\
Construction phase & 0.10 & 0.10 & 0.10 & 0.10 & 0.10 & 0.10 & 0.10 & 0.10 & 0.10 & 0.10\\
Greedy route estimation & 0.00 & 0.00 & 0.00 & 0.00 & 0.00 & 0.00 & 0.00 & 0.00 & 0.00 & 0.00\\
Move generators initialization & 0.03 & 0.05 & 0.05 & 0.05 & 0.05 & 0.06 & 0.06 & 0.06 & 0.06 & 0.06\\
Route minimization procedure & 0.60 & 0.53 & 0.55 & 0.55 & 0.55 & 0.56 & 0.56 & 0.56 & 0.56 & 0.57\\
Core optimization procedure & 1003.96 & 596.74 & 418.68 & 331.28 & 276.20 & 240.92 & 212.14 & 192.24 & 177.54 & 168.38\\
         \bottomrule
    \end{tabular}
\end{table}

\begin{sidewaystable}
\centering
\footnotesize
\caption{Average gap on the $\mathbb{B}$ instances.}
\label{tab:b-gaps}
\begin{tabular}{lr c rrrrrrrrrr c rrrrrrrrrr}
\toprule
 &&& &\multicolumn{9}{c}{\FILOX} && FILO2& \multicolumn{9}{c}{\FILOX (long)} \\
\cmidrule{5-13}
\cmidrule{16-24}
Instance & BKS && FILO2 & 2 & 3 & 4 & 5 & 6 & 7 & 8 & 9 & 10 && (long) & 2 & 3 & 4 & 5 & 6 & 7 & 8 & 9 & 10 \\
\midrule
Leuven1 &  192848   && 0.41 & 0.42 & 0.41 & 0.41 & 0.42 & 0.42 & 0.42 & 0.39 & 0.43 & 0.45 && 0.20 & 0.19 & 0.21 & 0.19 & 0.19 & 0.20 & 0.20 & 0.22 & 0.21 & 0.21 \\
Leuven2 &  111391    && 0.93 & 0.85 & 0.87 & 0.91 & 0.91 & 0.97 & 0.92 & 0.89 & 0.95 & 0.92 && 0.30 & 0.37 & 0.32 & 0.32 & 0.32 & 0.34 & 0.30 & 0.33 & 0.45 & 0.38 \\
Antwerp1 &  477277     && 0.52 & 0.52 & 0.54 & 0.52 & 0.51 & 0.53 & 0.51 & 0.53 & 0.53 & 0.51 && 0.22 & 0.24 & 0.24 & 0.25 & 0.24 & 0.23 & 0.22 & 0.24 & 0.25 & 0.23 \\
Antwerp2 &  291350    && 1.01 & 1.08 & 1.03 & 1.06 & 1.06 & 1.09 & 1.05 & 1.08 & 1.08 & 1.07 && 0.39 & 0.40 & 0.39 & 0.38 & 0.37 & 0.39 & 0.39 & 0.37 & 0.37 & 0.39 \\
Ghent1 &  469531    && 0.63 & 0.65 & 0.65 & 0.64 & 0.66 & 0.64 & 0.64 & 0.64 & 0.65 & 0.64 && 0.27 & 0.26 & 0.27 & 0.27 & 0.25 & 0.25 & 0.24 & 0.26 & 0.26 & 0.26 \\
Ghent2 &  257748    && 1.42 & 1.48 & 1.50 & 1.46 & 1.48 & 1.44 & 1.50 & 1.50 & 1.49 & 1.48 && 0.40 & 0.38 & 0.37 & 0.37 & 0.39 & 0.41 & 0.40 & 0.40 & 0.39 & 0.37 \\
Brussels1 &  501719    && 1.04 & 1.02 & 1.02 & 1.01 & 1.00 & 1.00 & 0.99 & 1.01 & 0.99 & 1.01 && 0.38 & 0.39 & 0.39 & 0.39 & 0.38 & 0.40 & 0.37 & 0.38 & 0.39 & 0.39 \\
Brussels2 &  345468    && 1.89 & 1.86 & 1.87 & 1.93 & 1.90 & 1.89 & 1.93 & 1.89 & 1.91 & 1.91 && 0.54 & 0.52 & 0.48 & 0.55 & 0.52 & 0.55 & 0.55 & 0.54 & 0.51 & 0.54 \\
Flanders1 &   7240118   && 0.72 & 0.72 & 0.71 & 0.72 & 0.73 & 0.72 & 0.70 & 0.71 & 0.73 & 0.71 && 0.31 & 0.31 & 0.32 & 0.31 & 0.31 & 0.30 & 0.30 & 0.31 & 0.30 & 0.32 \\
Flanders2 &   4373244   && 2.05 & 2.07 & 2.06 & 2.10 & 2.04 & 2.07 & 2.07 & 2.06 & 2.06 & 2.09 && 0.64 & 0.60 & 0.59 & 0.61 & 0.59 & 0.58 & 0.57 & 0.61 & 0.59 & 0.56 \\
\midrule
Mean &      && 1.06 & 1.07 & 1.07 & 1.08 & 1.07 & 1.08 & 1.07 & 1.07 & 1.08 & 1.08 && 0.36 & 0.37 & 0.36 & 0.37 & 0.36 & 0.36 & 0.36 & 0.36 & 0.37 & 0.36 \\
\bottomrule
\end{tabular}
\end{sidewaystable}

\begin{sidewaystable}
\centering
\footnotesize
\caption{Average speedup on the $\mathbb{B}$ instances.}
\label{tab:b-speedups}
\begin{tabular}{l c rrrrrrrrr c rrrrrrrrr}
\toprule
 && \multicolumn{9}{c}{\FILOX} && \multicolumn{9}{c}{\FILOX (long)} \\
\cmidrule{3-11}
\cmidrule{13-21}
Instance && 2 & 3 & 4 & 5 & 6 & 7 & 8 & 9 & 10 && 2 & 3 & 4 & 5 & 6 & 7 & 8 & 9 & 10 \\
\midrule
Leuven1  && 1.70 & 2.51 & 3.26 & 3.93 & 4.47 & 5.14 & 5.76 & 6.15 & 6.43 && 1.73 & 2.48 & 3.20 & 3.85 & 4.45 & 5.03 & 5.53 & 6.01 & 6.35 \\
Leuven2  && 1.59 & 2.40 & 2.97 & 3.60 & 4.09 & 4.50 & 4.93 & 5.46 & 5.77 && 1.56 & 2.26 & 2.78 & 3.34 & 3.74 & 4.22 & 4.51 & 4.97 & 5.18 \\
Antwerp1  && 1.71 & 2.57 & 3.31 & 3.98 & 4.68 & 5.30 & 6.02 & 6.57 & 6.62 && 1.74 & 2.52 & 3.24 & 3.94 & 4.53 & 5.17 & 5.73 & 6.22 & 6.58 \\
Antwerp2  && 1.65 & 2.44 & 3.10 & 3.72 & 4.27 & 4.82 & 5.31 & 5.74 & 6.04 && 1.71 & 2.43 & 3.09 & 3.70 & 4.21 & 4.78 & 5.25 & 5.67 & 6.00 \\
Ghent1  && 1.69 & 2.53 & 3.22 & 3.97 & 4.60 & 5.23 & 5.94 & 6.39 & 6.86 && 1.74 & 2.51 & 3.21 & 3.89 & 4.47 & 5.11 & 5.68 & 6.14 & 6.56 \\
Ghent2  && 1.66 & 2.43 & 3.05 & 3.64 & 4.18 & 4.73 & 5.31 & 5.64 & 6.02 && 1.69 & 2.37 & 2.98 & 3.55 & 4.03 & 4.56 & 5.04 & 5.47 & 5.75 \\
Brussels1  && 1.69 & 2.50 & 3.22 & 3.91 & 4.47 & 5.14 & 5.87 & 6.32 & 6.74 && 1.73 & 2.49 & 3.18 & 3.84 & 4.42 & 5.08 & 5.65 & 6.12 & 6.53 \\
Brussels2  && 1.63 & 2.38 & 3.00 & 3.55 & 4.06 & 4.66 & 5.15 & 5.53 & 5.71 && 1.68 & 2.36 & 2.97 & 3.54 & 4.08 & 4.63 & 5.16 & 5.53 & 5.77 \\
Flanders1  && 1.68 & 2.49 & 3.17 & 3.84 & 4.46 & 5.11 & 5.65 & 6.32 & 6.59 && 1.73 & 2.47 & 3.17 & 3.82 & 4.41 & 5.05 & 5.59 & 6.09 & 6.43 \\
Flanders2  && 1.55 & 2.22 & 2.73 & 3.18 & 3.65 & 4.07 & 4.43 & 4.75 & 4.85 && 1.61 & 2.25 & 2.80 & 3.33 & 3.84 & 4.34 & 4.82 & 5.15 & 5.42 \\
\midrule
Mean  && 1.65 & 2.44 & 3.08 & 3.69 & 4.24 & 4.81 & 5.35 & 5.79 & 6.05 && 1.68 & 2.40 & 3.03 & 3.63 & 4.17 & 4.73 & 5.22 & 5.65 & 5.96 \\
\bottomrule
\end{tabular}
\end{sidewaystable}

\clearpage

\subsection{Computational Details for the \texorpdfstring{$\mathbb{I}$}{I} Instances} \label{appendix:i}

In the following, we report detailed results for the computations on the $\mathbb{I}$ dataset.
\begin{itemize}
    \item Figure \ref{fig:FILO2x.boxplots.quality.i} shows boxplots for the average gaps obtained by the approaches.
    \item Table \ref{table:p_values_i} shows the result of the one-tailed Wilcoxon signed rank test.
    \item Tables \ref{tab:i-time-per-procedure} and \ref{tab:i-long-time-per-procedure} report the computing time for the main algorithm procedures.
    \item Table \ref{tab:i-gaps} shows the average gap for each instance in the dataset.
    \item Table \ref{tab:i-speedups} shows the average speedup for each instance in the dataset.
\end{itemize}

\clearpage

\begin{figure}
\centering
\caption{Average gaps obtained by FILO2 and \FILOX (denoted by the number of solvers) on the $\mathbb{I}$ dataset when executing the standard (top) and long (bottom) versions. The median value is shown on the right of each boxplot.}
\scriptsize
	\makeatletter
	\pgfplotsset{
		boxplot/draw/average/.code={ 
			\draw [/pgfplots/boxplot/every average/.try]
			\pgfextra
			\pgftransformshift{%
				\pgfplotsboxplotpointabbox
				{\pgfplotsboxplotvalue{average}}
				{0.5}%
			}%
			\endpgfextra
			;
		},
		boxplot/draw/median/.code={
			\draw [/pgfplots/boxplot/every median/.try]
			(boxplot box cs:\pgfplotsboxplotvalue{median},0)
			node[xshift=0.55cm, font=\tiny] {\pgfmathprintnumber{\pgfplotsboxplotvalue{median}}}
			--
			(boxplot box cs:\pgfplotsboxplotvalue{median},1);
		},
	}
	\makeatother
\begin{tikzpicture}
\scriptsize

\begin{axis}
[
xlabel={Algorithm},
ylabel={Average \% gap},
ymajorgrids=true,
yminorgrids=true,
minor y tick num=4,
minor grid style={line width=.01pt,draw=black!10},
major grid style={line width=.01pt,draw=black!30},
clip=false,
boxplot/draw direction=y,
boxplot/variable width,
boxplot/every median/.style={black,very thick,solid},
width=\textwidth,
height=150pt,
ylabel style={align=center}, 
y tick label style={align=right},
x tick label style={align=center},
xtick={0, 1, 2, 3, 4, 5, 6, 7, 8, 9, 10},
xticklabels={FILO2, 2, 3, 4, 5, 6, 7, 8, 9, 10},
xmin=-0.75,
xmax=10,
scatter/classes={ a={mark=star}, b={mark=*}}]

\addplot[mark=*, mark size=0.5pt,boxplot, boxplot prepared={draw position=0,
median=0.69,upper quartile=0.811,lower quartile=0.366,upper whisker=1.323,lower whisker=0.205,sample size=1}]
coordinates{
(0,1.677)
(0,1.536)
};
\addplot[mark=*, mark size=0.5pt,boxplot, boxplot prepared={draw position=1,
median=0.672,upper quartile=0.799,lower quartile=0.367,upper whisker=1.374,lower whisker=0.207,sample size=1}]
coordinates{
(1,1.66)
(1,1.514)
};
\addplot[mark=*, mark size=0.5pt,boxplot, boxplot prepared={draw position=2,
median=0.684,upper quartile=0.812,lower quartile=0.366,upper whisker=1.398,lower whisker=0.203,sample size=1}]
coordinates{
(2,1.613)
(2,1.525)
};
\addplot[mark=*, mark size=0.5pt,boxplot, boxplot prepared={draw position=3,
median=0.681,upper quartile=0.813,lower quartile=0.372,upper whisker=1.341,lower whisker=0.207,sample size=1}]
coordinates{
(3,1.63)
(3,1.526)
};
\addplot[mark=*, mark size=0.5pt,boxplot, boxplot prepared={draw position=4,
median=0.686,upper quartile=0.806,lower quartile=0.372,upper whisker=1.364,lower whisker=0.205,sample size=1}]
coordinates{
(4,1.612)
(4,1.499)
};
\addplot[mark=*, mark size=0.5pt,boxplot, boxplot prepared={draw position=5,
median=0.67,upper quartile=0.802,lower quartile=0.375,upper whisker=1.357,lower whisker=0.209,sample size=1}]
coordinates{
(5,1.616)
(5,1.495)
};
\addplot[mark=*, mark size=0.5pt,boxplot, boxplot prepared={draw position=6,
median=0.682,upper quartile=0.807,lower quartile=0.37,upper whisker=1.364,lower whisker=0.209,sample size=1}]
coordinates{
(6,1.62)
(6,1.511)
};
\addplot[mark=*, mark size=0.5pt,boxplot, boxplot prepared={draw position=7,
median=0.684,upper quartile=0.796,lower quartile=0.366,upper whisker=1.38,lower whisker=0.203,sample size=1}]
coordinates{
(7,1.613)
(7,1.472)
};
\addplot[mark=*, mark size=0.5pt,boxplot, boxplot prepared={draw position=8,
median=0.683,upper quartile=0.817,lower quartile=0.372,upper whisker=1.379,lower whisker=0.209,sample size=1}]
coordinates{
(8,1.65)
(8,1.514)
};
\addplot[mark=*, mark size=0.5pt,boxplot, boxplot prepared={draw position=9,
median=0.687,upper quartile=0.808,lower quartile=0.377,upper whisker=1.359,lower whisker=0.214,sample size=1}]
coordinates{
(9,1.659)
(9,1.488)
};

\end{axis}
\end{tikzpicture}
\vspace{-0.9cm}
\scriptsize
	\makeatletter
	\pgfplotsset{
		boxplot/draw/average/.code={ 
			\draw [/pgfplots/boxplot/every average/.try]
			\pgfextra
			\pgftransformshift{%
				\pgfplotsboxplotpointabbox
				{\pgfplotsboxplotvalue{average}}
				{0.5}%
			}%
			\endpgfextra
			;
		},
		boxplot/draw/median/.code={
			\draw [/pgfplots/boxplot/every median/.try]
			(boxplot box cs:\pgfplotsboxplotvalue{median},0)
			node[xshift=0.55cm, font=\tiny] {\pgfmathprintnumber{\pgfplotsboxplotvalue{median}}}
			--
			(boxplot box cs:\pgfplotsboxplotvalue{median},1);
		},
	}
	\makeatother
\begin{tikzpicture}
\scriptsize

\begin{axis}
[
xlabel={Algorithm},
ylabel={Average \% gap},
ymajorgrids=true,
yminorgrids=true,
minor y tick num=4,
minor grid style={line width=.01pt,draw=black!10},
major grid style={line width=.01pt,draw=black!30},
clip=false,
boxplot/draw direction=y,
boxplot/variable width,
boxplot/every median/.style={black,very thick,solid},
width=\textwidth,
height=150pt,
ylabel style={align=center}, 
y tick label style={align=right},
x tick label style={align=center},
xtick={0, 1, 2, 3, 4, 5, 6, 7, 8, 9, 10},
xticklabels={FILO2, 2, 3, 4, 5, 6, 7, 8, 9, 10},
xmin=-0.75,
xmax=10,
scatter/classes={ a={mark=star}, b={mark=*}}]

\addplot[mark=*, mark size=0.5pt,boxplot, boxplot prepared={draw position=0,
median=0.268,upper quartile=0.343,lower quartile=0.14,upper whisker=0.629,lower whisker=0.068,sample size=1}]
coordinates{
(0,0.738)
(0,0.688)
};
\addplot[mark=*, mark size=0.5pt,boxplot, boxplot prepared={draw position=1,
median=0.269,upper quartile=0.349,lower quartile=0.144,upper whisker=0.642,lower whisker=0.067,sample size=1}]
coordinates{
(1,0.737)
(1,0.716)
};
\addplot[mark=*, mark size=0.5pt,boxplot, boxplot prepared={draw position=2,
median=0.279,upper quartile=0.353,lower quartile=0.14,upper whisker=0.652,lower whisker=0.066,sample size=1}]
coordinates{
(2,0.78)
(2,0.723)
};
\addplot[mark=*, mark size=0.5pt,boxplot, boxplot prepared={draw position=3,
median=0.274,upper quartile=0.354,lower quartile=0.144,upper whisker=0.639,lower whisker=0.067,sample size=1}]
coordinates{
(3,0.722)
(3,0.715)
};
\addplot[mark=*, mark size=0.5pt,boxplot, boxplot prepared={draw position=4,
median=0.271,upper quartile=0.351,lower quartile=0.143,upper whisker=0.636,lower whisker=0.068,sample size=1}]
coordinates{
(4,0.724)
(4,0.736)
};
\addplot[mark=*, mark size=0.5pt,boxplot, boxplot prepared={draw position=5,
median=0.274,upper quartile=0.349,lower quartile=0.143,upper whisker=0.375,lower whisker=0.068,sample size=1}]
coordinates{
(5,0.78)
(5,0.685)
(5,0.668)
};
\addplot[mark=*, mark size=0.5pt,boxplot, boxplot prepared={draw position=6,
median=0.277,upper quartile=0.353,lower quartile=0.143,upper whisker=0.654,lower whisker=0.067,sample size=1}]
coordinates{
(6,0.702)
(6,0.707)
};
\addplot[mark=*, mark size=0.5pt,boxplot, boxplot prepared={draw position=7,
median=0.274,upper quartile=0.361,lower quartile=0.139,upper whisker=0.628,lower whisker=0.068,sample size=1}]
coordinates{
(7,0.724)
(7,0.703)
};
\addplot[mark=*, mark size=0.5pt,boxplot, boxplot prepared={draw position=8,
median=0.275,upper quartile=0.353,lower quartile=0.147,upper whisker=0.628,lower whisker=0.068,sample size=1}]
coordinates{
(8,0.704)
(8,0.696)
};
\addplot[mark=*, mark size=0.5pt,boxplot, boxplot prepared={draw position=9,
median=0.282,upper quartile=0.352,lower quartile=0.14,upper whisker=0.613,lower whisker=0.068,sample size=1}]
coordinates{
(9,0.737)
(9,0.721)
};

\end{axis}
\end{tikzpicture}
\label{fig:FILO2x.boxplots.quality.i}
\end{figure}

\begin{table}
	\caption{Computations on the $\mathbb{I}$ dataset: $p$-values for \FILOX vs FILO2 on the left and \FILOX (long) vs FILO2 (long) on the right.
	\\
	\footnotesize
	$p$-values in bold are associated with rejected hypothesis when $\bar{\alpha} = 0.002778$.\\
The last row of each group contains a $p$-value interpretation. 
In particular, \FILOX is not statistically different from FILO2 when $H_0$ cannot be rejected (Similar), \FILOX is statistically better when both $H_0$ and $H_1$ are rejected (Better), and, finally, \FILOX is statistically worse when $H_0$ is rejected and $H_1$ is not rejected (Worse).}
	\label{table:p_values_i}
	\centering
	\footnotesize
	\begin{tabular}{rrrrrrr}
	\toprule

        & & \multicolumn{2}{c}{\FILOX vs FILO2} & & \multicolumn{2}{c}{\FILOX (long) vs FILO2 (long)} \\
        \cmidrule{3-4}
        \cmidrule{6-7}
        
Solvers & & $H_0$ / $H_1$ & Outcome & & $H_0$ / $H_1$ & Outcome \\
	\midrule
2 &&   0.674223 /   0.337111 & Similar &&   0.277355 /   0.869451 & Similar \\
3 &&   0.985435 /   0.492718 & Similar &&   0.082550 /   0.962074 & Similar \\
4 &&   0.570597 /   0.727062 & Similar &&   0.164957 /   0.923177 & Similar \\
5 &&   0.956329 /   0.478165 & Similar &&   0.701181 /   0.662889 & Similar \\
6 &&   0.898317 /   0.565256 & Similar &&   0.142906 /   0.933637 & Similar \\
7 &&   0.985435 /   0.492718 & Similar &&   0.311794 /   0.852874 & Similar \\
8 &&   0.927279 /   0.463639 & Similar &&   0.348810 /   0.835009 & Similar \\
9 &&   0.570597 /   0.727062 & Similar &&   0.294252 /   0.861322 & Similar \\
10 &&   0.498009 /   0.762547 & Similar &&   0.521673 /   0.750996 & Similar \\
	\bottomrule
	\end{tabular}
\end{table}

\begin{table}
    \centering
    \footnotesize
    \caption{Computing time in seconds for the main algorithm FILO2 and \FILOX procedures when solving the $\mathbb{I}$ dataset.}
    \label{tab:i-time-per-procedure}
    \begin{tabular}{lrrrrrrrrrr}
        \toprule
        & & \multicolumn{8}{c}{\FILOX} \\
        \cmidrule{3-11}
        & FILO2 & 2 & 3 & 4 & 5 & 6 & 7 & 8 & 9 & 10\\
        \midrule
Instance preprocessing & 126.49 & 67.00 & 43.86 & 33.43 & 26.91 & 22.93 & 19.46 & 17.37 & 15.29 & 14.03\\
Construction phase & 5.23 & 5.44 & 5.77 & 5.13 & 5.21 & 5.94 & 5.68 & 5.99 & 5.66 & 5.72\\
Greedy route estimation & 3.37 & 3.35 & 3.34 & 3.34 & 3.34 & 3.35 & 3.34 & 3.35 & 3.34 & 3.35\\
Move generators initialization & 3.60 & 4.09 & 4.03 & 4.08 & 4.24 & 4.88 & 4.83 & 5.33 & 5.14 & 5.31\\
Route minimization procedure & 5.48 & 6.31 & 6.30 & 6.37 & 6.52 & 6.77 & 6.76 & 7.01 & 6.96 & 7.20\\
Core optimization procedure & 166.58 & 100.68 & 69.74 & 55.33 & 46.99 & 41.36 & 36.50 & 34.10 & 31.18 & 29.86\\
         \bottomrule
    \end{tabular}

\end{table}

\begin{table}
    \centering
    \footnotesize
    \caption{Computing time in seconds for the main algorithm FILO2 (long) and \FILOX (long) procedures when solving the $\mathbb{I}$ dataset.}
    \label{tab:i-long-time-per-procedure}
    
    \begin{tabular}{lrrrrrrrrrr}
        \toprule
        & & \multicolumn{8}{c}{\FILOX} \\
        \cmidrule{3-11}
        & FILO2 & 2 & 3 & 4 & 5 & 6 & 7 & 8 & 9 & 10\\
        \midrule
Instance preprocessing & 126.53 & 65.51 & 44.04 & 33.46 & 26.93 & 22.87 & 19.46 & 17.39 & 15.29 & 14.08\\
Construction phase & 5.23 & 5.18 & 5.65 & 5.16 & 5.22 & 5.84 & 5.63 & 5.96 & 5.68 & 5.76\\
Greedy route estimation & 3.37 & 3.34 & 3.34 & 3.34 & 3.34 & 3.35 & 3.34 & 3.36 & 3.34 & 3.35\\
Move generators initialization & 3.58 & 3.87 & 4.01 & 4.09 & 4.25 & 4.81 & 4.82 & 5.24 & 5.19 & 5.36\\
Route minimization procedure & 5.48 & 6.11 & 6.28 & 6.39 & 6.51 & 6.70 & 6.75 & 7.02 & 6.98 & 7.24\\
Core optimization procedure & 2349.73 & 1349.64 & 943.40 & 737.08 & 614.16 & 527.56 & 462.72 & 426.20 & 384.02 & 364.27\\
         \bottomrule
    \end{tabular}

\end{table}

\begin{sidewaystable}
\centering
\footnotesize
\caption{Average gap on the $\mathbb{I}$ instances.}
\label{tab:i-gaps}
\begin{tabular}{lr c rrrrrrrrrr c rrrrrrrrrr}
\toprule
 &&& &\multicolumn{9}{c}{\FILOX} && FILO2& \multicolumn{9}{c}{\FILOX (long)} \\
\cmidrule{5-13}
\cmidrule{16-24}
Instance & BKS && FILO2 & 2 & 3 & 4 & 5 & 6 & 7 & 8 & 9 & 10 && (long) & 2 & 3 & 4 & 5 & 6 & 7 & 8 & 9 & 10 \\
\midrule
Valle-D-Aosta &  21679514    && 0.27 & 0.26 & 0.26 & 0.26 & 0.24 & 0.25 & 0.25 & 0.27 & 0.26 & 0.25 && 0.13 & 0.14 & 0.14 & 0.13 & 0.14 & 0.14 & 0.13 & 0.13 & 0.13 & 0.13 \\
Molise &   111184982   && 0.41 & 0.41 & 0.42 & 0.42 & 0.41 & 0.42 & 0.42 & 0.41 & 0.42 & 0.42 && 0.16 & 0.15 & 0.14 & 0.15 & 0.15 & 0.16 & 0.16 & 0.15 & 0.15 & 0.15 \\
Trentino-Alto-Adige &   102063181   && 0.89 & 0.89 & 0.91 & 0.92 & 0.90 & 0.89 & 0.90 & 0.90 & 0.92 & 0.91 && 0.36 & 0.36 & 0.38 & 0.37 & 0.37 & 0.35 & 0.37 & 0.39 & 0.36 & 0.37 \\
Basilicata &  175623919    && 0.74 & 0.69 & 0.73 & 0.70 & 0.71 & 0.68 & 0.70 & 0.71 & 0.70 & 0.73 && 0.31 & 0.30 & 0.30 & 0.29 & 0.30 & 0.29 & 0.30 & 0.30 & 0.33 & 0.29 \\
Umbria &   545507981   && 0.28 & 0.28 & 0.28 & 0.29 & 0.29 & 0.28 & 0.29 & 0.28 & 0.28 & 0.29 && 0.13 & 0.13 & 0.13 & 0.13 & 0.13 & 0.13 & 0.13 & 0.13 & 0.12 & 0.13 \\
Abruzzo &   311712556   && 0.70 & 0.73 & 0.74 & 0.73 & 0.74 & 0.74 & 0.72 & 0.74 & 0.74 & 0.72 && 0.26 & 0.26 & 0.27 & 0.27 & 0.26 & 0.27 & 0.27 & 0.26 & 0.27 & 0.27 \\
Friuli-Venezia-Giulia &  415805616    && 0.62 & 0.63 & 0.65 & 0.64 & 0.65 & 0.64 & 0.64 & 0.64 & 0.66 & 0.64 && 0.28 & 0.28 & 0.29 & 0.28 & 0.29 & 0.28 & 0.28 & 0.28 & 0.28 & 0.29 \\
Liguria &  1426389867    && 0.21 & 0.21 & 0.20 & 0.21 & 0.21 & 0.21 & 0.21 & 0.20 & 0.21 & 0.21 && 0.07 & 0.07 & 0.07 & 0.07 & 0.07 & 0.07 & 0.07 & 0.07 & 0.07 & 0.07 \\
Calabria & 1964651530     && 0.38 & 0.38 & 0.38 & 0.39 & 0.39 & 0.39 & 0.38 & 0.38 & 0.39 & 0.39 && 0.13 & 0.14 & 0.14 & 0.13 & 0.13 & 0.14 & 0.13 & 0.13 & 0.14 & 0.14 \\
Marche &   420484426   && 0.68 & 0.65 & 0.67 & 0.66 & 0.67 & 0.66 & 0.66 & 0.67 & 0.67 & 0.67 && 0.25 & 0.25 & 0.24 & 0.24 & 0.24 & 0.24 & 0.24 & 0.24 & 0.24 & 0.24 \\
Sardegna &  827934149    && 1.68 & 1.66 & 1.61 & 1.63 & 1.61 & 1.62 & 1.62 & 1.61 & 1.65 & 1.66 && 0.74 & 0.74 & 0.78 & 0.72 & 0.72 & 0.78 & 0.70 & 0.72 & 0.70 & 0.74 \\
Campania &  391859276    && 1.04 & 1.02 & 1.01 & 1.01 & 1.02 & 1.01 & 1.03 & 1.02 & 1.01 & 1.01 && 0.40 & 0.37 & 0.38 & 0.38 & 0.37 & 0.38 & 0.38 & 0.39 & 0.38 & 0.38 \\
Piemonte &  2627446164    && 0.32 & 0.33 & 0.33 & 0.33 & 0.32 & 0.33 & 0.33 & 0.33 & 0.33 & 0.33 && 0.15 & 0.15 & 0.15 & 0.15 & 0.15 & 0.15 & 0.15 & 0.15 & 0.15 & 0.15 \\
Toscana &   1084417188   && 0.78 & 0.77 & 0.78 & 0.78 & 0.77 & 0.77 & 0.78 & 0.76 & 0.78 & 0.77 && 0.33 & 0.32 & 0.33 & 0.33 & 0.33 & 0.35 & 0.33 & 0.34 & 0.35 & 0.33 \\
Puglia &  1464797603    && 1.54 & 1.51 & 1.53 & 1.53 & 1.50 & 1.49 & 1.51 & 1.47 & 1.51 & 1.49 && 0.69 & 0.72 & 0.72 & 0.72 & 0.74 & 0.68 & 0.71 & 0.70 & 0.70 & 0.72 \\
Sicilia & 1774262462     && 1.32 & 1.37 & 1.40 & 1.34 & 1.36 & 1.36 & 1.36 & 1.38 & 1.38 & 1.36 && 0.63 & 0.64 & 0.65 & 0.64 & 0.64 & 0.67 & 0.65 & 0.63 & 0.63 & 0.61 \\
Veneto & 1050488613     && 0.72 & 0.70 & 0.70 & 0.71 & 0.70 & 0.72 & 0.71 & 0.70 & 0.70 & 0.71 && 0.30 & 0.29 & 0.30 & 0.29 & 0.29 & 0.29 & 0.29 & 0.29 & 0.28 & 0.29 \\
Emilia-Romagna & 5405446715     && 0.24 & 0.24 & 0.24 & 0.24 & 0.24 & 0.24 & 0.23 & 0.24 & 0.24 & 0.24 && 0.09 & 0.09 & 0.09 & 0.09 & 0.09 & 0.09 & 0.09 & 0.09 & 0.09 & 0.09 \\
Lombardia &  1339900081    && 0.75 & 0.77 & 0.76 & 0.75 & 0.75 & 0.76 & 0.77 & 0.76 & 0.75 & 0.76 && 0.34 & 0.35 & 0.35 & 0.35 & 0.34 & 0.34 & 0.35 & 0.35 & 0.35 & 0.35 \\
Lazio &  3145381332    && 0.40 & 0.39 & 0.40 & 0.40 & 0.40 & 0.40 & 0.40 & 0.40 & 0.40 & 0.40 && 0.14 & 0.14 & 0.15 & 0.15 & 0.15 & 0.14 & 0.15 & 0.14 & 0.15 & 0.14 \\
\midrule
Mean &      && 0.70 & 0.70 & 0.70 & 0.70 & 0.69 & 0.69 & 0.70 & 0.69 & 0.70 & 0.70 && 0.29 & 0.30 & 0.30 & 0.29 & 0.29 & 0.30 & 0.29 & 0.30 & 0.29 & 0.29 \\
\bottomrule
\end{tabular}
\end{sidewaystable}

\begin{sidewaystable}
\centering
\footnotesize
\caption{Average speedup on the $\mathbb{I}$ instances.}
\label{tab:i-speedups}
\begin{tabular}{l c rrrrrrrrr c rrrrrrrrr}
\toprule
 && \multicolumn{9}{c}{\FILOX} && \multicolumn{9}{c}{\FILOX (long)} \\
\cmidrule{3-11}
\cmidrule{13-21}
Instance && 2 & 3 & 4 & 5 & 6 & 7 & 8 & 9 & 10 && 2 & 3 & 4 & 5 & 6 & 7 & 8 & 9 & 10 \\
\midrule
Valle-D-Aosta  && 1.69 & 2.37 & 3.00 & 3.59 & 4.09 & 4.65 & 5.00 & 5.57 & 5.84 && 1.69 & 2.42 & 3.07 & 3.70 & 4.23 & 4.85 & 5.29 & 5.82 & 6.16 \\
Molise  && 1.66 & 2.43 & 3.06 & 3.62 & 4.15 & 4.64 & 5.01 & 5.53 & 5.77 && 1.75 & 2.53 & 3.25 & 3.95 & 4.58 & 5.22 & 5.75 & 6.35 & 6.73 \\
Trentino-Alto-Adige  && 1.54 & 2.19 & 2.68 & 3.12 & 3.49 & 3.93 & 4.13 & 4.48 & 4.62 && 1.68 & 2.36 & 2.98 & 3.55 & 4.12 & 4.68 & 5.07 & 5.57 & 5.85 \\
Basilicata  && 1.67 & 2.29 & 2.82 & 3.29 & 3.65 & 4.10 & 4.31 & 4.65 & 4.76 && 1.72 & 2.46 & 3.13 & 3.76 & 4.38 & 4.96 & 5.37 & 5.96 & 6.25 \\
Umbria  && 1.70 & 2.43 & 3.05 & 3.60 & 4.04 & 4.56 & 4.90 & 5.36 & 5.57 && 1.75 & 2.52 & 3.24 & 3.92 & 4.55 & 5.21 & 5.67 & 6.27 & 6.63 \\
Abruzzo  && 1.65 & 2.25 & 2.78 & 3.20 & 3.54 & 3.99 & 4.18 & 4.51 & 4.70 && 1.69 & 2.36 & 2.99 & 3.53 & 4.13 & 4.59 & 4.96 & 5.48 & 5.71 \\
Friuli-Venezia-Giulia  && 1.61 & 2.29 & 2.85 & 3.28 & 3.67 & 4.07 & 4.31 & 4.62 & 4.84 && 1.71 & 2.38 & 3.01 & 3.56 & 4.20 & 4.71 & 5.07 & 5.54 & 5.91 \\
Liguria  && 1.72 & 2.44 & 3.04 & 3.60 & 3.98 & 4.53 & 4.81 & 5.20 & 5.34 && 1.77 & 2.53 & 3.28 & 3.96 & 4.57 & 5.24 & 5.71 & 6.34 & 6.67 \\
Calabria  && 1.67 & 2.38 & 2.95 & 3.44 & 3.80 & 4.22 & 4.47 & 4.86 & 4.99 && 1.76 & 2.53 & 3.25 & 3.91 & 4.53 & 5.13 & 5.64 & 6.23 & 6.57 \\
Marche  && 1.62 & 2.30 & 2.88 & 3.34 & 3.70 & 4.11 & 4.29 & 4.75 & 4.80 && 1.70 & 2.41 & 3.03 & 3.61 & 4.19 & 4.74 & 5.09 & 5.56 & 5.85 \\
Sardegna  && 1.64 & 2.32 & 2.87 & 3.30 & 3.64 & 4.01 & 4.14 & 4.56 & 4.68 && 1.75 & 2.51 & 3.18 & 3.83 & 4.42 & 5.01 & 5.45 & 5.98 & 6.31 \\
Campania  && 1.62 & 2.33 & 2.87 & 3.34 & 3.68 & 4.11 & 4.35 & 4.65 & 4.87 && 1.72 & 2.46 & 3.10 & 3.68 & 4.22 & 4.83 & 5.17 & 5.82 & 6.05 \\
Piemonte  && 1.67 & 2.36 & 2.92 & 3.37 & 3.65 & 4.10 & 4.31 & 4.62 & 4.77 && 1.76 & 2.53 & 3.25 & 3.89 & 4.53 & 5.11 & 5.57 & 6.13 & 6.48 \\
Toscana  && 1.69 & 2.37 & 2.94 & 3.37 & 3.68 & 4.11 & 4.31 & 4.68 & 4.81 && 1.76 & 2.55 & 3.24 & 3.94 & 4.49 & 5.13 & 5.51 & 6.13 & 6.51 \\
Puglia  && 1.67 & 2.34 & 2.88 & 3.32 & 3.59 & 4.00 & 4.12 & 4.45 & 4.59 && 1.76 & 2.51 & 3.22 & 3.84 & 4.42 & 5.00 & 5.39 & 5.97 & 6.26 \\
Sicilia  && 1.65 & 2.33 & 2.88 & 3.31 & 3.66 & 3.98 & 4.12 & 4.47 & 4.61 && 1.75 & 2.49 & 3.17 & 3.74 & 4.35 & 4.97 & 5.29 & 5.81 & 6.11 \\
Veneto  && 1.71 & 2.40 & 2.96 & 3.42 & 3.71 & 4.13 & 4.29 & 4.69 & 4.85 && 1.76 & 2.46 & 3.16 & 3.71 & 4.29 & 4.86 & 5.28 & 5.78 & 6.02 \\
Emilia-Romagna  && 1.68 & 2.32 & 2.82 & 3.23 & 3.44 & 3.82 & 4.01 & 4.25 & 4.42 && 1.76 & 2.52 & 3.22 & 3.87 & 4.41 & 5.04 & 5.42 & 6.05 & 6.34 \\
Lombardia  && 1.68 & 2.37 & 2.92 & 3.37 & 3.61 & 4.07 & 4.24 & 4.60 & 4.72 && 1.76 & 2.50 & 3.21 & 3.84 & 4.42 & 5.10 & 5.48 & 6.20 & 6.45 \\
Lazio  && 1.67 & 2.30 & 2.80 & 3.21 & 3.43 & 3.76 & 3.90 & 4.21 & 4.31 && 1.77 & 2.54 & 3.23 & 3.87 & 4.47 & 5.04 & 5.45 & 6.01 & 6.34 \\
\midrule
Mean  && 1.66 & 2.34 & 2.89 & 3.35 & 3.66 & 4.08 & 4.27 & 4.62 & 4.78 && 1.74 & 2.48 & 3.16 & 3.78 & 4.37 & 4.96 & 5.37 & 5.94 & 6.24 \\
\bottomrule
\end{tabular}
\end{sidewaystable}

\clearpage

\bibliographystyle{plainnat}
\bibliography{references.bib}

@article{dantzig1959,
  title={The truck dispatching problem},
  author={Dantzig, G. B. and Ramser, J. H.},
  journal={Management Science},
  volume={6},
  number={1},
  pages={80–-91},
  year={1959}
}

@article{clarke1964,
  title={Scheduling of vehicles from a central depot to a number of delivery points},
  author={Clarke, G. and Wright, J. W.},
  journal={Operations Research},
  volume={12},
  number={4},
  pages={568–-581},
  year={1964}
}

@inbook{Toth2014,
    author = {Semet, F. and Toth, P. and Vigo, D.},
    title = {Chapter 2: Classical Exact Algorithms for the Capacitated Vehicle Routing Problem},
    booktitle = {Vehicle Routing},
    year = {2014},
    publisher = {Society for Industrial and Applied Mathematics},
    pages = {37-57}
}

@book{Toth2002,
    editor = {Paolo Toth and Daniele Vigo},
    title = {The Vehicle Routing Problem},
    publisher = {Society for Industrial and Applied Mathematics},
    year = {2002},
    chapter= {2}
}

@article{accorsi2021,
author = {Accorsi, Luca and Vigo, Daniele},
title = {A Fast and Scalable Heuristic for the Solution of Large-Scale Capacitated Vehicle Routing Problems},
journal = {Transportation Science},
volume = {55},
number = {4},
pages = {832-856},
year = {2021},
doi = {10.1287/trsc.2021.1059},
URL = {https://doi.org/10.1287/trsc.2021.1059}
}

@article{accorsi2024,
    author = {Accorsi, L. and Vigo, D.},
    title = {Routing one million customers in a handful of minutes},
    journal = {Computers \& Operations Research},
    volume = {164},
    pages = {106562},
    year = {2024}
}

@Inbook{crainic2010,
    author = { Crainic, T. G. and Toulouse, M.},
    title = {Parallel Meta-heuristics},
    bookTitle = {Handbook of Metaheuristics},
    year = {2010},
    publisher = {Springer US},
    address = {Boston, MA},
    pages = {497--541}
}

@inbook{crainic2005,
    author = {Crainic, T. G. and Gendreau, M. and Potvin, J.-Y.},
    publisher = {John Wiley \& Sons, Ltd},
    title = {Parallel Tabu Search},
    booktitle = {Parallel Metaheuristics},
    chapter = {13},
    pages = {289-313},
    year = {2005}
}

@article{toth2003,
  author    = {Paolo Toth and
               Daniele Vigo},
  title     = {The Granular Tabu Search and Its Application to the Vehicle-Routing
               Problem},
  journal   = {{INFORMS} Journal on Computing},
  volume    = {15},
  number    = {4},
  pages     = {333--346},
  year      = {2003},
  url       = {https://doi.org/10.1287/ijoc.15.4.333.24890},
  doi       = {10.1287/ijoc.15.4.333.24890},
  bibsource = {dblp computer science bibliography, https://dblp.org}
}

@InProceedings{doerner2004,
    author = {Doerner, K. F. and Hartl, R. F. and Kiechle, G. and Lucka, M. and Reimann, M.},
    editor = {Gottlieb, J. and Raidl, G. R.},
    title = {Parallel Ant Systems for the Capacitated Vehicle Routing Problem},
    booktitle = {Evolutionary Computation in Combinatorial Optimization},
    year = {2004},
    publisher = {Springer Berlin Heidelberg},
    address = {Berlin, Heidelberg},
    pages = {72--83}
}

@article{gendron1994,
    author = {Gendron, B. and Crainic, T. G.},
    title = {Parallel Branch-and-Branch Algorithms: Survey and Synthesis},
    journal = {Operations Research},
    volume = {42},
    number = {6},
    pages = {1042--1066},
    year = {1994}
}

@article{kindervater1986,
    title = {An introduction to parallelism in combinatorial optimization},
    journal = {Discrete Applied Mathematics},
    volume = {14},
    number = {2},
    pages = {135--156},
    year = {1986},
    author = {Kindervater, G. A. P. and Lenstra, J. K.}
}

@book{talbi2006,
    author = {Talbi, El-Ghazali},
    title = {Parallel Combinatorial Optimization (Wiley Series on Parallel and Distributed Computing)},
    year = {2006},
    publisher = {Wiley-Interscience},
    address = {USA}
}

@inbook{crainic2006,
    author = {Crainic, T. G. and Bertrand, Le C. and Roucairol, C.},
    title = {Parallel Branch-and-Bound Algorithms},
    booktitle = {Parallel Combinatorial Optimization},
    publisher = {John Wiley \& Sons, Ltd},
    address = {USA},
    chapter = {1},
    pages = {1--28},
    year = {2006}
}

@book{alba2005,
    author = {Alba, Enrique},
    title = {Parallel Metaheuristics: A New Class of Algorithms},
    year = {2005},
    publisher = {Wiley-Interscience},
    address = {USA}
}

@Inbook{crainic2005b,
    author = {Crainic, T. G.},
    title = {Parallel Computation, Co-operation, Tabu Search},
    bookTitle = {Metaheuristic Optimization via Memory and Evolution: Tabu Search and Scatter Search},
    year = {2005},
    publisher = {Springer US},
    address = {Boston, MA},
    pages = {283--302}
}

@InProceedings{crainic2008,
    author = {Crainic, T. G. and Toulouse, M. },
    editor = {Maniezzo, V. and Battiti, R. and Watson, J.-P.},
    title = {Explicit and Emergent Cooperation Schemes for Search Algorithms},
    booktitle = {Learning and Intelligent Optimization},
    year = {2008},
    publisher = {Springer Berlin Heidelberg},
    address = {Berlin, Heidelberg},
    pages = {95--109}
}

@Inbook{crainic2008b,
    author = {Crainic, T. G.},
    title = {Parallel Solution Methods for Vehicle Routing Problems},
    bookTitle = {The Vehicle Routing Problem: Latest Advances and New Challenges},
    year = {2008},
    publisher = {Springer US},
    address = {Boston, MA},
    pages = {171--198}
}

@inbook{crainic2005c,
    author = {Crainic, T. G. and Hail, N.},
    publisher = {John Wiley \& Sons, Ltd},
    title = {Parallel Metaheuristics Applications},
    booktitle = {Parallel Metaheuristics},
    chapter = {19},
    pages = {447--494},
    year = {2005}
}

@Inbook{Barbucha2014,
    author = {Barbucha, D.},
    title = {A Cooperative Agent-Based Multiple Neighborhood Search for the Capacitated Vehicle Routing Problem},
    bookTitle = {Recent Advances in Knowledge-based Paradigms and Applications: Enhanced Applications Using Hybrid Artificial Intelligence Techniques},
    year = {2014},
    publisher = {Springer International Publishing},
    address = {Cham},
    pages = {129--143}
}

@inproceedings{Abdelatti2020,
    title = { An improved GPU-accelerated heuristic technique applied to the capacitated vehicle routing problem},
    author = {Abdelatti, M. F. and Sodhi, M. S.},
    booktitle = {GECCO '20: Proceedings of the 2020 Genetic and Evolutionary Computation Conference},
    organization = {ACM},
    year={2020},
    pages = {663–-671}
}

@article{Kalatzantonakis2005,
    author = { Kalatzantonakis, P. and Sifaleras, A. and Samaras, N.},
    title = { Cooperative versus non-cooperative parallel variable neighborhood search strategies: a case study on the capacitated vehicle routing problem},
    journal = {Journal of Global Optimization},
    volume = {78},
    number = {2},
    pages = {327--348},
    year = {2005}
}

@inproceedings{ Muniasamy2023,
    title = {Effective Parallelization of the Vehicle Routing Problem},
    author = {Rajesh Pandian, M. and Somesh, S. and Rupesh, N. Narayanaswamy, N. S.},
    booktitle = {GECCO '23: Proceedings of the Genetic and Evolutionary Computation Conference},
    organization = {ACM},
    year={2023},
    pages = {1036–-1044}
}

@article{Yelmewad2021,
    author = {Yelmewad, P. and Talawar, B.},
    title = {Parallel Version of Local Search Heuristic Algorithm to Solve Capacitated Vehicle Routing Problem},
    journal = {Cluster Computing},
    volume = {24},
    number = {4},
    pages = {3671--3692},
    year = {2021}
}

@InProceedings{Barbucha2011,
    author = {Barbucha, D.},
    editor = {J{\k{e}}drzejowicz, P. and Nguyen, N. T. and Hoang, K.},
    title = {Solving the Capacitated Vehicle Routing Problem by a Team of Parallel Heterogeneous Cooperating Agents},
    booktitle = {Computational Collective Intelligence. Technologies and Applications},
    year = {2011},
    publisher = {Springer Berlin Heidelberg},
    address = {Berlin, Heidelberg},
    pages = {332--341}
}

@INPROCEEDINGS{Borcinova2018,
  author = {Borčinová, Z.},
  booktitle = {2018 IEEE Workshop on Complexity in Engineering (COMPENG)}, 
  title = {Solving the Capacitated Vehicle Routing Problem Using a Parallel Micro Genetic Algorithm}, 
  year = {2018},
  pages = {1--6}
}

@article{uchoa2017,
    title = {New benchmark instances for the Capacitated Vehicle Routing Problem},
    journal = {European Journal of Operational Research},
    volume = {257},
    number = {3},
    pages = {845--858},
    year = {2017},
    author = {Uchoa, E. and Pecin, D. and Pessoa, A. and Poggi, M. and Vidal, T. and Subramanian, A.}
}

@article{arnold2019xxl,
    title = {Efficiently solving very large-scale routing problems},
    journal = {Computers \& Operations Research},
    volume = {107},
    pages = {32 - 42},
    year = {2019},
    author = {Arnold, F. and Gendreau, M. and Sörensen, K.},
}

@article{BEEK20181,
    title = {An Efficient Implementation of a Static Move Descriptor-based Local Search Heuristic},
    journal = {Computers \& Operations Research},
    volume = {94},
    pages = {1 - 10},
    year = {2018},
    author = {Beek, O. and Raa, B. and Dullaert, W. and Vigo, D.},
}

@book{christofides1979,
  title =  {Combinatorial Optimization},
  author = {Christofides, N. and Mingozzi, A. and Toth, P. and Sandi, C.},
  year = {1979},
  publisher = {John Wiley},
  address = {Chichester}
}

@article{laporte2000,
    title = {Classical and modern heuristics for the vehicle routing problem},
    journal = {International Transactions in Operational Research},
    volume = {7},
    number = {4},
    pages = {285--300},
    year = {2000},
    author = {Laporte, G. and Gendreau, M. and Potvin, J.-Y. and Semet, F.}
}

@article{augerat1995,
    author = {Augerat, P. and Belenguer, J. M. and Benavent, E. and Corber{\'a}n, {\'A}. and Naddef, D. and Rinaldi, G.},
    year = {1995},
    volume = {34},
    title = {Computational results with a branch and cut code for the capacitated vehicle routing problem},
    journal = {IMAG}
}

@article{christofides1969,
    title = {An Algorithm for the Vehicle-dispatching Problem},
    author = {Christofides, N. and and Eilon, S.},
    journal = {Journal of the Operational Research Society},
    year = {1969},
    volume = {20},
    pages = {309--318}
}

@article{christofides1981,
    title = {Exact algorithms for the vehicle routing problem, based on spanning tree and shortest path relaxations},
    journal = {Mathematical Programming},
    volume = {20},
    pages = {255–-282},
    year = {1981},
    author = {Christofides, N. and Mingozzi, A. and Toth, P.}
}

@article{fisher1994,
    author = {Fisher, M. L.},
    title = {Optimal Solution of Vehicle Routing Problems Using Minimum K-Trees},
    journal = {Operations Research},
    volume = {42},
    number = {4},
    pages = {626--642},
    year = {1994}
}

@article{gillet1976,
    title = {Multi-terminal vehicle-dispatch algorithm},
    journal = {Omega},
    volume = {4},
    number = {6},
    pages = {711--718},
    year = {1976},
    author = {Gillett, B. E. and Johnson, J. G.}
}

@article{feiyue2005,
    title = {Very large-scale vehicle routing: new test problems, algorithms, and results},
    journal = {Computers \& Operations Research},
    volume = {32},
    number = {5},
    pages = {1165--1179},
    year = {2005},
    author = {Li, F. and Golden, B. and Wasil, E.}
}

@Inbook{golden1998,
    author = {Golden, B. L. and Wasil, E. A. and Kelly, J. P. and Chao, I-M.},
    editor = {Crainic, T. G. and Laporte, G.},
    title = {The Impact of Metaheuristics on Solving the Vehicle Routing Problem: Algorithms, Problem Sets, and Computational Results},
    bookTitle = {Fleet Management and Logistics},
    year = {1998},
    publisher = {Springer US},
    address = {Boston, MA},
    pages = {33--56}
}

@article{kytojoki2007,
    title = {An efficient variable neighborhood search heuristic for very large scale vehicle routing problems},
    journal = {Computers \& Operations Research},
    volume = {34},
    number = {9},
    pages = {2743--2757},
    year = {2007},
    author = {Kyt{\"o}joki, J. and Nuortio, T. and Br{\"a}ysy, O. and Gendreau, M.}
}

@Inbook{Lourenco2003,
    title = {Iterated Local Search},
    author = {Louren{\c{c}}o, H. R. and Martin, O. C. and St{\"u}tzle, T.},
    bookTitle = {Handbook of Metaheuristics},
    year = {2003},
    publisher = {Springer US},
    address = {Boston, MA},
    pages = {320--353}
}

@article{zachariadis2010,
	author = {Zachariadis, E. E. and Kiranoudis, C. T.},
	title = {A Strategy for Reducing the Computational Complexity of Local Search based Methods for the Vehicle Routing Problem},
	journal = {Computers \& Operations Research},
	volume = {37},
	number = {12},
	year = {2010},
	pages = {2089--2105}
}

@article{dunn,
    author = {Dunn, O. J.},
    journal = {Journal of the American Statistical Association},
    number = {293},
    pages = {52--64},
    title = {Multiple Comparisons Among Means},
    volume = {56},
    year = {1961}
}

@article{wilcoxon,
     author = {Wilcoxon, F.},
     journal = {Biometrics Bulletin},
     number = {6},
     pages = {80--83},
     title = {Individual Comparisons by Ranking Methods},
     volume = {1},
     year = {1945}
}

@article{GLOVER1996223,
title = "Ejection chains, reference structures and alternating path methods for traveling salesman problems",
journal = "Discrete Applied Mathematics",
volume = "65",
number = "1",
pages = "223 - 253",
year = "1996",
note = "First International Colloquium on Graphs and Optimization",
issn = "0166-218X",
doi = "10.1016/0166-218X(94)00037-E",
url = "http://www.sciencedirect.com/science/article/pii/0166218X9400037E",
author = "Fred Glover"
}

@article{taillard1997tabu,
author = {Taillard, {\'{E}}ric and Badeau, Philippe and Gendreau, Michel and Guertin, François and Potvin, Jean-Yves},
title = {A Tabu Search Heuristic for the Vehicle Routing Problem with Soft Time Windows},
journal = {Transportation Science},
volume = {31},
number = {2},
pages = {170-186},
year = {1997},
doi = {10.1287/trsc.31.2.170},

URL = { 
        https://doi.org/10.1287/trsc.31.2.170
    
},
eprint = { 
        https://doi.org/10.1287/trsc.31.2.170
    
}
}

@inproceedings{Reinelt1994TheTS,
  title={The traveling salesman problem},
  author={Gerhard Reinelt and Giovanni Rinaldi},
  year={1994}
}

@article{MLADENOVIC19971097,
title = {Variable neighborhood search},
journal = {Computers \& Operations Research},
volume = {24},
number = {11},
pages = {1097-1100},
year = {1997},
issn = {0305-0548},
doi = {https://doi.org/10.1016/S0305-0548(97)00031-2},
url = {https://www.sciencedirect.com/science/article/pii/S0305054897000312},
author = {N. Mladenović and P. Hansen}
}

@article {Kirkpatrick671,
	author = {Kirkpatrick, S. and Gelatt, C. D. and Vecchi, M. P.},
	title = {Optimization by Simulated Annealing},
	volume = {220},
	number = {4598},
	pages = {671--680},
	year = {1983},
	doi = {10.1126/science.220.4598.671},
	publisher = {American Association for the Advancement of Science},
	issn = {0036-8075},
	URL = {https://science.sciencemag.org/content/220/4598/671},
	eprint = {https://science.sciencemag.org/content/220/4598/671.full.pdf},
	journal = {Science}
}

@article{vidal22,
author = {Vidal, Thibaut},
title = {Hybrid Genetic Search for the CVRP: Open-Source Implementation and SWAP* Neighborhood},
year = {2022},
issue_date = {Apr 2022},
publisher = {Elsevier Science Ltd.},
address = {GBR},
volume = {140},
number = {C},
issn = {0305-0548},
url = {https://doi.org/10.1016/j.cor.2021.105643},
doi = {10.1016/j.cor.2021.105643},
journal = {Comput. Oper. Res.},
month = {4},
numpages = {11}
}

@article{sisr,
author = {Christiaens, Jan and Vanden Berghe, Greet},
title = {Slack Induction by String Removals for Vehicle Routing Problems},
journal = {Transportation Science},
volume = {54},
number = {2},
pages = {417-433},
year = {2020},
doi = {10.1287/trsc.2019.0914},

URL = { 
        https://doi.org/10.1287/trsc.2019.0914
    
},
eprint = { 
        https://doi.org/10.1287/trsc.2019.0914
    
}
}

@article{ails2,
author = {M\'{a}ximo, Vin\'{\i}cius R. and Cordeau, Jean-Fran\c{c}ois and Nascimento, Mari\'{a} C. V.},
title = {AILS-II: An Adaptive Iterated Local Search Heuristic for the Large-Scale Capacitated Vehicle Routing Problem},
journal = {INFORMS Journal on Computing},
volume = {36},
number = {4},
pages = {974-986},
year = {2024},
doi = {10.1287/ijoc.2023.0106},

URL = { 
    
        https://doi.org/10.1287/ijoc.2023.0106
    
    

},
eprint = { 
    
        https://doi.org/10.1287/ijoc.2023.0106
    
    

}
}

@inproceedings{bentley1990k,
  title={K-d trees for semidynamic point sets},
  author={Bentley, Jon Louis},
  booktitle={Proceedings of the sixth annual symposium on Computational geometry},
  pages={187--197},
  year={1990}
}

@misc{cvrplib,
  author = {CVRPLIB},
  title = {Capacitated Vehicle Routing Problem Library},
  year = 2025,
  url = {http://vrp.galgos.inf.puc-rio.br/index.php/en/},
  urldate = {2025-09-24},
  note = {visited on 2024-09-24}
}

\end{document}